\documentclass[
reprint,
longbibliography,
eprint,
amsmath,
amssymb,
aps,
rmp
]{revtex4-1}

\usepackage{graphicx}
\usepackage{dcolumn}
\usepackage{bm}
\usepackage{multirow}
\usepackage{array}
\usepackage{soul}

\newcolumntype{H}{>{\setbox0=\hbox\bgroup}c<{\egroup}@{}}

\usepackage{comment}

\usepackage{hyperref} 

\usepackage{color}
\usepackage[dvipsnames]{xcolor}

\newcommand\marker[1]{{\color{Black}#1}}  
\newcommand\markertwo[1]{{\color{Black}#1}}

\newcommand{\fullbold}[1]{\bf{\boldmath{#1}}}

\begin{document}


\title{Topological ferroelectric  chirality}

\author{I.\,Luk'yanchuk}
\affiliation{University of Picardie\char`,{} Laboratory of Condensed Matter Physics\char`,{} Amiens 80039\char`,{} France}
\affiliation{Terra Quantum AG\char`,{} Kornhausstrasse 25\char`,{} CH-9000 St. Gallen\char`,{} Switzerland}

\author{A.\,Razumnaya}
\affiliation{Condensed Matter Physics Department\char`,{} Jožef Stefan Institute\char`,{} Jamova Cesta 39\char`,{} Ljubljana 1000\char`,{} Slovenia}

\author{S.\,Kondovych}
\affiliation{Institute for Theoretical Solid State Physics\char`,{} IFW Dresden\char`,{} Helmholtzstrasse 20\char`,{} 01069 Dresden\char`,{} Germany}
\affiliation{Life Chemicals Inc.\char`,{} Murmanska st. 5\char`,{} 02660 Kyiv\char`,{} Ukraine}

\author{Y.\,Tikhonov}
\affiliation{University of Picardie\char`,{} Laboratory of Condensed Matter Physics\char`,{} Amiens 80039\char`,{} France}

\author{V.\,M.\,Vinokur}
\email{vv@terraquantum.swiss}
\affiliation{Terra Quantum AG\char`,{} Kornhausstrasse 25\char`,{} CH-9000 St. Gallen\char`,{} Switzerland}
\affiliation{Physics Department\char`,{}
City College of the City University of New York\char`,{}
160 Convent Ave\char`,{} New York\char`,{} NY 10031\char`,{} USA}

\date{\today}

\begin{abstract}

 Chirality, an inherent property of most objects of the universe, is a dynamic research topic in material science, physics, chemistry, and biology. 
 The fundamental appeal of this extensive study is supported by the technological quest to manufacture materials with configurable chiralities for emerging applications ranging from optoelectronics and photonics to pharmaceutics and medicine. Recent advances put forth ferroelectrics as a host of chiral topological states in the form of Bloch domain walls, skyrmions, merons, and Hopfions, offering thus a unique ground for making chirality switchable and tunable. Here we review current developments, milestones achieved, and future routes of chiral ferroelectric materials. We focus on insights into the topological origin of the chirality in the nanostructured ferroelectrics, bringing new controllable functionalities.
 We pay special attention to novel developments enabling tunability and manipulating the chiroptical response and enantioselectivity, leading to new applications in nano-optoelectronics, plasmonics, pharmaceutics, and bio-medical industries. 
\end{abstract}




\maketitle

\tableofcontents

\newpage



\marker{In this paper, we develop a description bringing in novel knowledge of chirality in ferroelectric systems. Our approach is built on the topological methods and explores the field-tuned topological formations in nanostructured ferroelectrics. 
This description enables a breakthrough in realizing essential topological features via specifics of the ferroelectric chirality bringing in novel practical applications.
Our paper has been invited to the Review of Modern Physics journal on 20 October 2021, submitted on 12 March 2023. It represents numerous pioneering results establishing new physics of ferroelectricity. Unfortunately, after about a year of consideration, our paper was rejected. The reasons for the rejection given by the journal were that during their consideration period they have published other review touching on some aspects of topology of ferroelectrics. 
Importantly, our article not only brought in new topics that have not been touched upon in the previous reviews but does create the novel successful research methods and outlines the areas for the future cross-disciplinary development of chiral physics employing ferroelectrics.
We feel strongly that given the novelty of our pioneering results, the community will greatly benefit from their immediate publication in arXiv before they appear as a series of new extended papers in refereed journals.


  }
\section{\label{sec:intro}Introduction}

\marker{The chirality, or handedness, introduced by Lord Kelvin in 1894, is defined as a fundamental asymmetry property describing systems, distinguishable from their mirror images}. The concept of chirality provides a remarkable addition to the foundational pillars of our world that applies to a wide multiplicity of sciences ranging from particle physics and life science to the structure of the universe\,\cite{Mason1984,Monastyrsky2007,Hegstrom1990,Wagniere2008}.
The chirality of the materials has become one of the focuses of modern material science, focusing on the materials manifesting exceptional optical, plasmonic, biochemical, and pharmaceutic properties. 
So far, the handedness has been appearing as a built-in property that cannot be changed once the material is synthesized.  

Recent progress in exploring the field-tuned topological formations in nanostructured ferroelectrics, vortices\,\cite{Yadav2016}, Bloch domain walls\,\cite{Chauleau2020}, skyrmions\,\cite{Nahas2015,Das2019,Tikhonov2020,Das2021,Yin2021}, merons\,\cite{Wang2020}, and Hopfions\,\cite{Lukyanchuk2020}, revealed a new class of chirality, the controllable topological chirality. 
This enables a breakthrough in realizing emerging ferroelectric chiral systems for practical applications. In this Review, we systematize our current knowledge of chirality in ferroelectric systems. 

In Section\,\ref{sec:Meet}, we lay out the fundamental symmetric and topological foundation of chirality in ferroelectrics which enables us to systematically understand the recent experimental and theoretical findings where chirality emerges as an inherent property of the topological states. We present the typical examples of local chiral structural organization, from the organic and inorganic chiral molecules to solids, including ferroelectrics, with chiral crystallographic symmetry. Then, we 
describe the topological type of chirality emerging due to the nonuniform distribution of the order parameter at the scales extended to nanometers and even microns. After brief introductions to the applied aspects of topology, we highlight that the origin of the discovered topological states in nanostructured confined ferroelectrics is the long-range electrostatic forces and describe these states. 
This makes the fundamental difference between the ferroelectric and magnetic systems in which the origin of the extended topological states is usually the local Dzyaloshinskii–Moriya interaction (DMI).  
We establish the understanding of the chirality in ferroelectrics highlighting its fundamental yet underexplored topological foundations.

In Section\,\ref{sec:ExtTheor}, we describe the state-of-the-art of research on chirality in ferroelectrics, 
focusing attention on the explosive growth of publications on topological chirality over the past 2-3 years. We discuss the emergent but not yet sufficiently explored practical methods of chirality testing and manipulation suitable for ferroelectrics. 
We review the progress in the detection and measurements of chirality in ferroelectric materials as well as the advanced methods used for other emergent chiral nanosystems and highlight the opportunities for employing the same methods for nanostructured ferroelectrics. 
We further review the efficient methods for chirality switching and reconfiguration and outline the design of systems possessing switchable chirality.
Along with the discussion of the recently emerged methods of electrical switching, we offer other methods for manipulating chirality which are of particular importance for future chiral nanotechnologies. 
\marker{Quite often, the available experimental data are restricted mostly to the observation of polarization textures in the vicinity of the surface. The corresponding analysis may not provide an unambiguous identification of the detailed topological structure of the polarization states in the depth of the systems. 
In order to present an exhaustive 3D panorama picture of the emerging states, we complement the sets of available data by matching them with the phase-field simulations of the ferroelectric systems. The details of the simulations are presented in Supplemental Material,\,(\citeyear{SM}).}

In Section\,\ref{sec:Perspectives}, we discuss how the switchable and tunable chirality of the nanostructured ferroelectrics may advantageously be utilized and optimized in specific devices. We lay out the main avenues and horizons for applications of the switchable ferroelectric chirality, including laser-manipulation capabilities, chiral sensing, optical activity, and circular dichroism that finds diverse use in optic and plasmonics, and enantioselective technologies for pharmaceutics and bio-medical industries. This shapes the prospective developments of this promising field in the coming years.


\section{\label{sec:Meet}Symmetry meets topology}


\begin{figure*}
\begin{center}
\includegraphics [width=1\linewidth] {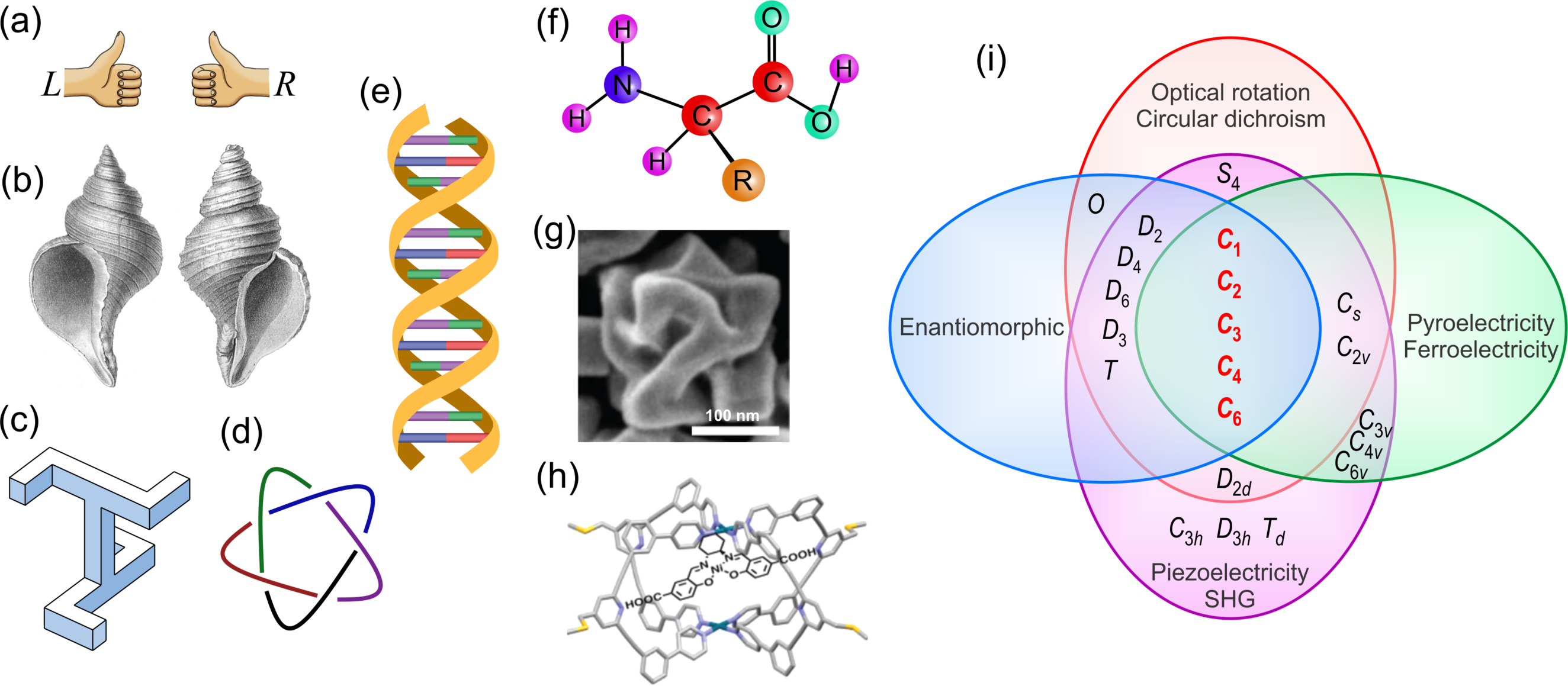}
\caption{ \textbf{Chirality in nature.} 
\textbf{(a)}, Images of the left (L) and right (R) hands that introduce the basic essence of chirality, the objects that cannot be overimposed by mirror reflections and rotations.
\textbf{(b)}, \textit{Neptunea angulata} and \textit{Neptunea despecta} are the left- and right-handed modifications of the sea snail representing yet another chiral object of nature illustrating the shape-induced chirality.
\textbf{(c)}, Chiral geometrical object: the basic illustration of structural chirality.
\textbf{(d)}, Chiral mathematical knot introducing the topological type of chirality. 
\textbf{(e)}, The DNA molecule possessing left-handedness.
\textbf{(f)}, Amino acid molecule having the structural chirality. 
\textbf{(g)}, The SEM images of chiral gold nanoparticle where chirality is the geometry-induced one. From Ref.\,\cite{Lee2018}.
\textbf{(h)}, Guest salen-Ni complex embedded into the metal-organic molecular cage possessing the topological chirality. From Ref.\,\cite{Tian2021chi}. 
\textbf{(i)}, \marker{Diagram suggested by Halasyamani and Poeppelmeier} \,\cite{Halasyamani1998}, illustrating \marker{enantiomorphic symmetry groups}\,\cite{SohnckeBook}  for materials possessing chirality and their interrelation with the symmetry groups of other functional materials, pyro/ferroelectrics, piezoelectrics that demonstrate the second-harmonics generation (SHG), and optically active materials demonstrating optical rotation and circular dichroism. The symmetry groups of chiral ferroelectric materials are highlighted in red. 
}
\label{FigNature}
\end{center}
\end{figure*}

\subsection{\label{sec:Nature}Chirality in nature}
Figure\,\ref{FigNature} displays the graphic introduction to the concept of chirality. The most widespread example used to describe the concept is the images of the left and right hands that cannot be overimposed by mirror reflections and rotations \marker{as} shown in Fig.\,\ref{FigNature}(a). 
The next illustration of chirality in live nature is the mirror-reflected left and right pieces of the sea snail, \textit{Neptunia}, see Fig.\,\ref{FigNature}(b).
Figure\,\ref{FigNature}(c) presents a simple geometric object possessing chirality, and Fig.\,\ref{FigNature}(d) demonstrates a mathematical chiral knot. 
Chirality appears to be one of the basic features of organic molecules. Figure\,\ref{FigNature}(e) depicts the chirality of one of the most important in the life DNA molecule. 

Figures\,\ref{FigNature}(f)-(h) demonstrate the different types of chirality, structural, geometric, and topological, appearing in material science which, in their practical realizations, can be viewed as an appropriate expansion of these simple but fundamental examples.  

The structural chirality, provided by the chiral structural arrangement of the atoms in molecules and crystals, is shown in Fig.\,\ref{FigNature}(f) using the example of the amino acid molecule. This type of chirality is appropriately described by the crystallographic groups, known as the Sohncke groups\,\cite{SohnckeBook}, similar to chiral geometrical objects exemplified in Fig.\,\ref{FigNature}(c). 
The relation of the Sohncke point symmetry groups with other functional symmetry groups in materials is illustrated by the \marker{diagram suggested by Halasyamani and Poeppelmeier}\,\hbox{\cite{Halasyamani1998}} in Fig.\,\ref{FigNature}(i), and is discussed in more detail in Sec.\,\ref{sec:symetry}. 
There are several ferroelectric materials that belong to the Sohncke symmetry groups and hence possess chirality. Remarkably, such well-known ferroelectrics, like the first discovered ferroelectric, Rochelle salt and triglycine sulfate (TGS) belong to this family. We review the structurally chiral ferroelectrics in Sec.\,\ref{sec:Struct}.

The next type of chirality, geometric chirality, is exemplified in Fig.\,\ref{FigNature}(g), which demonstrates the geometrical shape-induced chirality for the plasmonic nanoparticle of gold\,\cite{Lee2018}. It is analogous to the shape-induced chirality of the snail (Fig.\,\ref{FigNature}(b)). This geometric shape-induced chirality in nanomaterials emerges mostly due to the competition between the surface and bulk atomic forces or the specific atomic shaping\,\cite{Gregorio2020}. The typical examples are the metallic nanoparticles\,\cite{Xia2011,Lee2018,Gregorio2020} and artificial nanostructures\,\cite{Li2020,Liu2021,Wu2022}, fabricated for plasmonic applications\,\cite{Gregorio2020,Neubrech2020,Long2020}.

Figure\,\ref{FigNature}(h) illustrates an essential type of chirality, appearing mostly in the complex organic molecules\,\cite{Tian2021chi}. This topological chirality is provided by the nontrivial linking of the molecular chains. On the mathematical level, this type of chirality is described by the knots theory, similar to that describing the mathematical knots exemplified in Fig.\,\ref{FigNature}(d).  The concept of topological chirality brings in far-reaching generalizations, in particular in the description of the chiral topological linking of the streamlines of the electromagnetic field in the plasma magneto-hydrodynamics and astrophysics\,\cite{Moffatt1992}, and of the fluxes in streaming liquids\,\cite{Arnold2021}. 
This key chirality is remarkably realized in nanostructured ferroelectrics in the form of the knotted streamlines of the vector field in the nonuniform chiral topological states extended to the nanometer spatial scale. 
Importantly, the emergence of topological chirality in ferroelectrics is not related to the background structural crystallographic group of the material but is provided by the long-range electrostatic forces in the confined volume of the nanomaterial. We discuss this feature, central to our Review, in Sec.\,\ref{sec:Topology}.

Before turning to a general review of the chiral properties of ferroelectrics, we have to mention that the equivalent terms ``chirality" and ``handedness" are used on equal footing. The chiral objects, also called enantiomers, are distinguished as left (L) and right (R) pieces. In the figures, we present the left objects in blue and the right objects in red. The achiral species are drawn in green. 

\subsection{\label{sec:symetry}Symmetry approach}

In group theory, the chiral object is characterized by the absence of symmetry elements related to mirror reflections, such as rotation-reflections (also called improper rotations) in the point symmetry group of the molecule or crystal. In particular, such an object does not possess either the inversion center or the reflection planes, which are the particular cases of the rotation-reflections. A chiral object and its mirror image are called enantiomers. 
There exist 11 point symmetry groups, $C_1$, $C_2$, $C_3$, $C_4$, $C_6$, $D_2$, $D_3$, $D_4$, $D_6$, $T$, and $O$, corresponding to the 65 space Sohncke groups\,\cite{SohnckeBook}, enumerating all possible types of chiral crystals. The interrelation between the chirality and other functional properties of crystals, also defined by some types of symmetry, is \marker{ given in Table\,I and well illustrated by the diagram suggested by Halasyamani and Poeppelmeier} \,\cite{Halasyamani1998}, shown in Fig.\,\ref{FigNature}(i). 
\marker{The possibility of discerning the orientational domain states of the low-symmetry phase by the vector-like order parameters associated with these symmetries was given in}\,\cite{Erb2020}.

\begin{table}[t!]
\marker{
 \caption{The classification of non-centrosymmetric point groups according to their physical properties.  The  data is collected from\,\cite{Halasyamani1998}}
{
\centering
\renewcommand{\arraystretch}{1.2}
\begin{tabular}{ c c c c c c } 
\hline \hline
  Crystal 
& Point 
& Pyro-, 
& Enantio- 
& Optical 
& Piezo- \\

  system 
& group 
& Ferro-
& morphic 
& rotation
& electric \\

&  
& electric
& (chiral)
& 
&  \\

\hline 
  Triclinic
& $C_{1}$ (1)
& $\times$
& $\times$
& $\times$
& $\times$ \\
\hline 
  Monoclinic
& $C_{2}$ (2)
& $\times$
& $\times$
& $\times$
& $\times$ \\
  
& $C_{s}$ (m)
& $\times$
& {}
& $\times$
& $\times$ \\
\hline 
  Orthorhombic
& $D_{2}$ (222)
& {}
& $\times$
& $\times$
& $\times$ \\
  
& $C_{2v}$ (mm2)
& $\times$
& {}
& $\times$
& $\times$ \\
\hline 
  Trigonal
& $C_{3}$ (3)
& $\times$
& $\times$
& $\times$
& $\times$ \\
  
& $D_{3}$ (32)
& {}
& $\times$
& $\times$
& $\times$ \\
  
& $C_{3v}$ (3m)
& $\times$
& {}
& {}
& $\times$ \\
\hline 
  Hexagonal
& $C_{6}$ (6)
& $\times$
& $\times$
& $\times$
& $\times$ \\
  
& $C_{3h}$ ($\overline{6}$)
& {}
& {}
& {}
& $\times$ \\
  
& $D_{6}$ (622)
& {}
& $\times$
& $\times$
& $\times$ \\
  
& $C_{6v}$ (6mm)
& $\times$
& {}
& {}
& $\times$ \\
  
& $D_{3h}$ ($\overline{6}$2m)
& {}
& {}
& {}
& $\times$ \\
\hline 
  Tetragonal
& $C_{4}$ (4)
& $\times$
& $\times$
& $\times$
& $\times$ \\
  
& $S_{4}$ ($\overline{4}$)
& {}
& {}
& $\times$
& $\times$ \\
  
& $D_{4}$ (422)
& {}
& $\times$
& $\times$
& $\times$ \\
  
& $C_{4v}$ (4mm)
& $\times$
& {}
& {}
& $\times$ \\
  
& $D_{2d}$ (4$\overline{2}$m)
& {}
& {}
& $\times$
& $\times$ \\
\hline 
  Cubic
& $T$ (23)
& {}
& $\times$
& $\times$
& $\times$ \\
  
& $O$ (432)
& {}
& $\times$
& $\times$
& {} \\
  
& $T_d$ ($\overline{4}$3m)
& {}
& {}
& {}
& $\times$ \\
\hline \hline
\end{tabular}
}
}
\label{TableGroups}
\end{table}


Importantly, there exist 5 types of chiral ferroelectrics possessing the uniaxial point group symmetries $C_1$, $C_2$, $C_3$, $C_4$, and $C_6$; ferroelectrics having other point symmetry groups are achiral. 
Another important property called optical activity and often associated with chirality is the crystal's ability to rotate the plane of polarization of a light beam passing through the material. From the symmetry viewpoint, the optical activity and associated circular dichroism (different absorption of the right- and the left-polarized light beams) emerge in the non-centrosymmetric crystals. 
Therefore, all the chiral materials are optically active, whereas there exist optical active materials which are not chiral. Those are the crystals with the $C_{s}$, $C_{2v}$, $D_{2d}$, and $S_4$ point symmetry groups. Notably, the crystals having two last symmetries are ferroelectrics.  Finally, the majority of chiral materials, \marker{except that possessing the $O$ point symmetry,} also demonstrate the piezoelectric and the second-harmonic generation (SHG) effects. 
Hence, optical activity, circular dichroism, and, to a certain extent, piezo-electric effect and SHG can be used to test the chirality of the materials, in particular in the case of ferroelectricity. We discuss the corresponding experimental activities in detail in Sect.\,\ref{sec:Sensing}.    

There are quite a few chiral ferroelectrics in which the handedness arises from their structural properties. We discuss their properties in Sec.\,\ref{sec:Struct}. We turn now to the topological source of chirality in ferroelectrics, the property not directly related to the specific crystallographic symmetry and emerging at the mesoscopic scales due to specific long-range electrostatic interactions.

\subsection{\label{sec:Topology}Topology and topological states}

The beginning of the XXI century can be viewed as a new epoch in ferroelectricity, which has been marked by the discovery of the so-called topological states emerging in the nano-structured ferroelectric materials. 
A plethora of topological formations, stable domain walls (DWs)\cite{Chauleau2020,Tagantsev2010book}, vortices\cite{Yadav2016}, skyrmions\cite{Nahas2015,Tikhonov2020,Das2021,Yin2021}, merons\cite{Wang2020}, and Hopfions\cite{Lukyanchuk2020} were predicted and experimentally found in ferroelectric films, nanoparticles, nanodots, and layered ferroelectric heterostructures.  
These discoveries are in-line with the ongoing paradigmatic shift in physics, marking the change from deriving properties of the matter out of the local structural symmetries to using the global topological frame, which describes the fundamental properties of matter on the basis of their invariance under continuous transformations or perturbations.

There have been several nice reviews on ferroelectric topological states\,\cite{Zheng2017,Das2018,Ramesh2019,Das2020,Tian2021,Tang2021,Chen2021,Guo2022,Wang2022,Grunebohm2021,Fernandez2022,Govinden2023rev} describing them via references to their more familiar magnetic counterparts. Deriving from those, topological states in ferroelectrics have been classified in accord with the structural features of the vector field in real space. 
\marker{A comprehensive review of the application of the real space topological approach for ferroelectrics was recently given in\,\cite{Junquera2023}.}
Here, before addressing advanced concepts of topological chirality, we briefly make a pedestrian overview of the basic concepts of topology and their common use in ferroelectrics.

\subsubsection{\label{sec:TopApproach}Topological approach for the vector field}
 Topological states emerging in ferroelectric systems are described by the spatial distribution of the order parameter, which is the polarization vector $\mathbf P$. 
Analogously to magnetic systems in which the  order parameter is the magnetization vector $\mathbf{M}$, 
the topological consideration of polarization textures in ferroelectrics is usually based on comparing two fundamental spaces (manifolds)\,\cite{Mermin1979,Mineev1998,Volovik2019,Saxena2020}. 

The first fundamental manifold, $\mathcal M$, is the coordinate space in which the distribution of the order parameter is considered. In general, $\mathcal M$ is a subspace of the $n$-dimensional Euclidean system space $\mathbb{R}^n$, which, for example, is the one-dimensional (1D) space in the case of nanorods, 2D space in the case of thin films or layers in heterostructures, and 3D space in the case of the bulk systems. 
In most common cases, it is assumed that outside of topological states themselves, the order parameter is uniform. In the topological language, this means that at infinity, all the points of the considered topological state are identical. Then, the space  $\mathcal M$ is described by the closed manifolds, equivalent to the n-dimensional spheres: S$^1$, S$^2$, and S$^3$, respectively.  

The second fundamental manifold is the degeneracy space D$'$, of the internal states of the vector order parameter $\mathbf{V}$, 
which is $\mathbf{M}$ for magnetic systems and $\mathbf{P}$ for ferroelectrics.
This manifold consists of the states having the same or approximately the same minimal free energy (we use the prime superscript for the manifolds related to the degeneracy manifold). 
Importantly, by analogy with magnetism, in a ferroelectric system, the anisotropy energy is neglected or assumed to be small compared to other energy contributions. Hence, the isoenergy manifold D$'$ is degenerate with respect to rotations of the order-parameter space and is presented by the sphere S$'^2$ in the fully isotropic systems or by the circle S$'^1$ in the case of the uniaxial anisotropy with the easy-plane, where the order parameter vector is confined in the plane perpendicular to the anisotropy axis.  In the case of the easy-axis anisotropy, the degeneracy space reduces to two points in the order parameter space, $\mathbb{Z}'_2$, corresponding to the up and down-directed stable states of the order parameter.  


\begin{figure*}
\begin{center}
\includegraphics [width=0.9\linewidth] {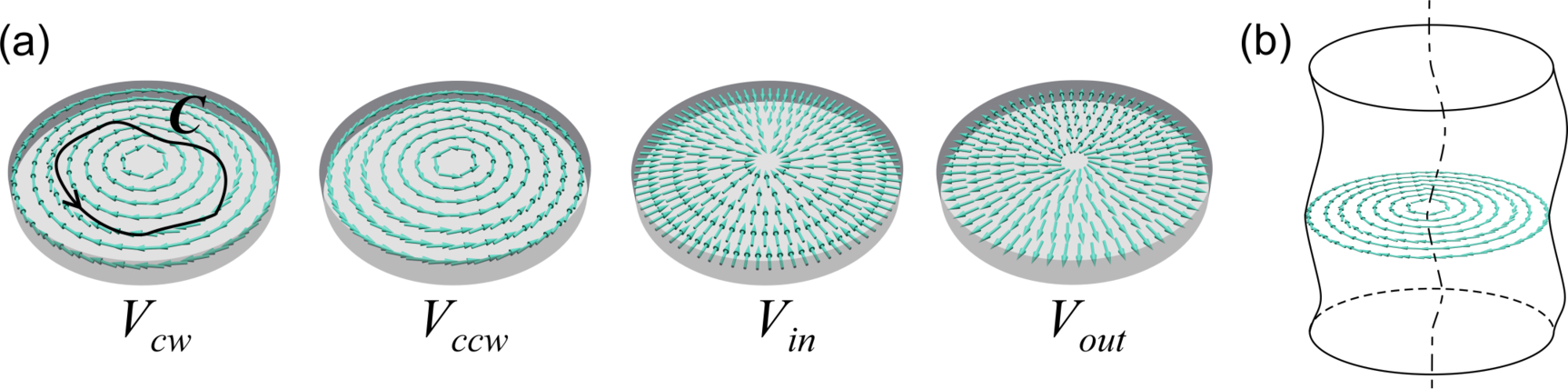}
\caption{ \textbf{Vortices and vortex tube.}  
\textbf{(a)}, Different types of vortices in two-dimensional space with the clockwise, $V_{cw}$, counterclockwise, $V_{ccw}$, incoming, $V_{in}$, and outcoming, $V_{out}$, distributions of the vector field. Closed contour $C$ encircling vortex core is illustrated for vortex $V_{cw}$. 
\textbf{(b)}, The illustration of the vortex tube.   
}
\label{FigVort}
\end{center}
\end{figure*}

 \begin{figure*}
\begin{center}
\includegraphics [width=0.9\linewidth] {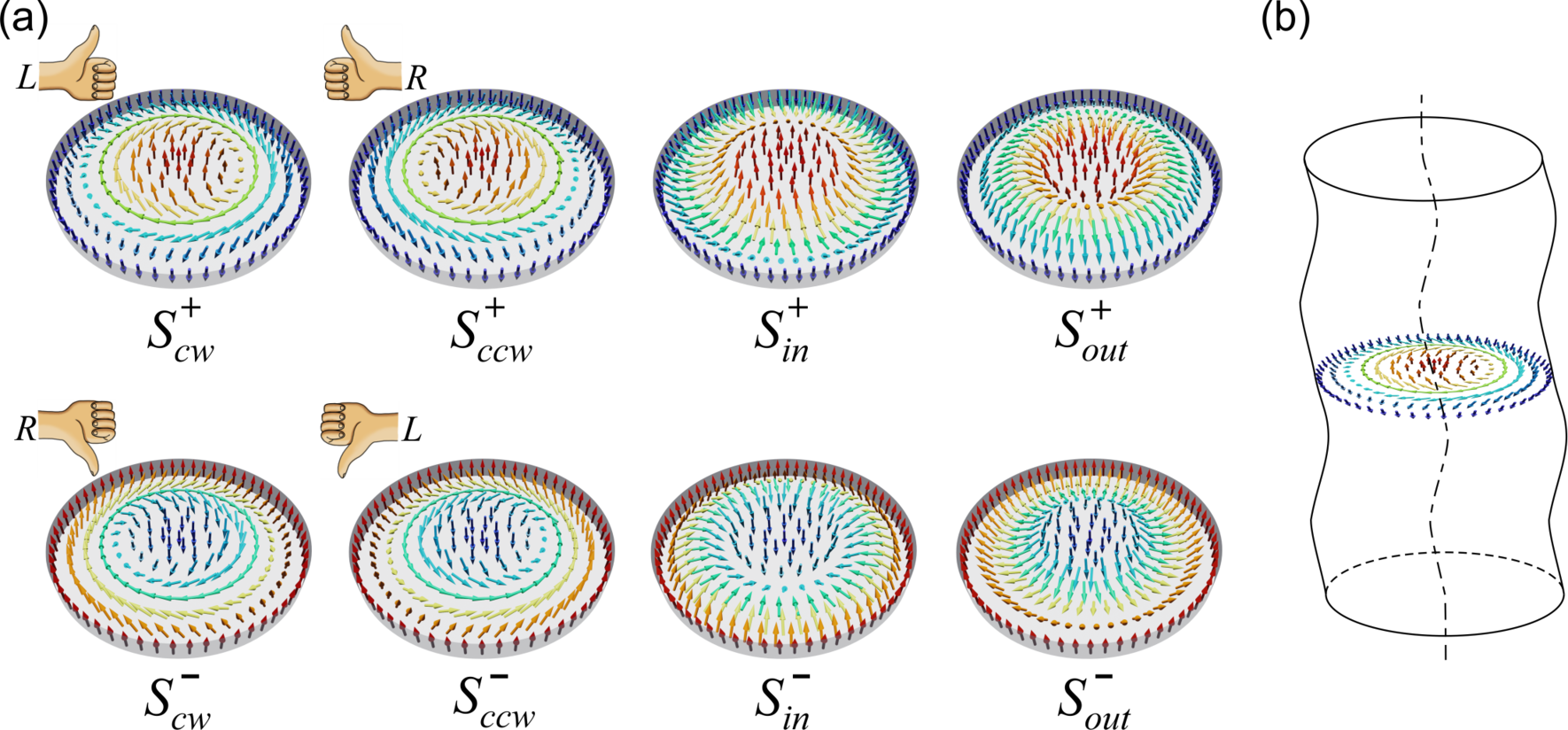}
\caption{ \textbf{Skyrmions and skyrmion tube.}  
\textbf{(a)}, Different types of skyrmions in two-dimensional space, either Bloch skyrmions, with the left-handedness, $S_{cw}^+$, $S^-_{ccw}$, and right-handedness, $S_{ccw}^+$, $S^-_{cw}$, or N\'eel skyrmions, with incoming, $S_{in}^+$, $S_{in}^-$, and outcoming, $S_{out}^+$, $S_{out}^-$, distribution of the vector field.  The superscript index denotes the polarity of the skyrmion. 
\textbf{(b)}, The illustration of the skyrmion tube.  
}
\label{FigSkyrm}
\end{center}
\end{figure*}

The classification of the topological states of the order parameter vector field is based on the mapping of fundamental manifolds, $\mathcal M \to $ D$'$. 
This implies that the order parameter vector $\mathbf{V}$ in each point of $\mathcal M$ has a definite value from D$'$. However, the correspondence is not necessarily one-to-one. For instance, the order parameter vectors from all the points of  $\mathcal M$ can map to only one point in D$'$ (or a small region nearby), which corresponds to the (almost) uniform distribution of the order parameter.  Otherwise, the image of the space $\mathcal M$ can completely sweep the manifold D$'$ one or several times. The latter cases are attributed to the non-trivial topological states of the vector field $\mathbf{V}$. The degree of sweeping,  
\marker{indicating how many times the image of $\mathcal M$ covers D$'$  defines the term topological charge}, also called the winding number, $N$, of the state. Because the degree of sweeping is an integer and can not be changed by the small continuous deformations of the vector field $\mathbf{V}$, say that the topological states with given $N$ are stable or topologically protected. For the described most common types of mapping, S$^n\to$ S$'^m$ (with $n=1,2,3$ and $m=1,2$), the calculation of the degree of the mapping is done within the so-called formalism of homotopy groups of spheres, $\pi_n($S$'^m)$, which is the well-established procedure. For other types of mapping  $\mathcal M \to $D$'$, the finding of the topological charges can be rather challenging.

\subsubsection{\label{sec:TopStates}Topological states. Vorticity and chirality}
\marker{We now describe the topological states of the vector field commonly discussed in literature}

\medskip

\noindent \textit{a) Vortices} 

Figure\,\ref{FigVort}(a) shows several types of vortices forming in 2D films in the case where the vector order parameter is confined within this film. They have the clock- or counterclockwise vorticity 
(vortices $V_{cw}$ or $V_{ccw}$, respectively) or star-like structure with the incoming to- or outcoming from the center vector field (vortices $V_{in}$ or $V_{out}$, respectively). 
These vortices are characterized by the S$^1 \to $S$'^1$ mapping where the real space manifold, S$^1$, is equivalent to the contour $C$ (illustrated for the vortex $V_{cw}$ in Fig.\,\ref{FigVort}(a)), encircling the vortex core and the degeneracy space S$'^1$ is the circle of all possible orientations of the vector $\mathbf{V}$ in the vector space plane. Traveling around the vortex core along the contour $C$ induces the complete turn of the order parameter vector $\mathbf{V}$ in the degeneracy space, which means that the topological charge of a vortex is $N_1=1$. Since the winding number is the same for all the vortices shown in Fig.\,\ref{FigVort}(a), they all belong in the same topological class, which means that they can be transformed into each other by continuous deformation of the vector field, in particular by a continuous rotation of each order parameter vector over the same angle. 

There exist other types of vortices with other integer winding numbers, for instance, the double-quanta vortex with $N_1=2$ obtained by unifying two single quanta vortices with $N_1=1$ together or antivortex with $N_1=-1$ which is the saddle-point of the order parameter vector field. The topological charge $N_1=0$ corresponds to the uniform state.
\marker{Each group of vortices remains stable preserving its total winding number $N_1$.} The states corresponding to the respective kinds of vortices are separated by huge energy barriers. In particular, the barriers separate the uniform state from the vortex-carrying states.

Importantly, the vector textures are achiral since they can be superimposed by the mirror reflection \marker{with respect to} the film plane. The 2D vortex textures emerging in the films can also extend in the third dimension in the case of uniaxial crystals with the easy-plane anisotropy, forming the vortex tubes aligned along the anisotropy axis, as shown in Fig.\,\ref{FigVort}(b).

\medskip


\noindent \textit{b) Skyrmions}

Other topological states emerging in the 2D systems, in particular in magnetic films are skyrmions\,\cite{Nagaosa2013,Fert2017,Liu2017book,Gobel2021}. They differ from vortices in such a way that their order parameter vector $\mathbf{V}$ is not confined within the film but can freely rotate in space. The end of the order parameter vector sweeps, therefore, the sphere S$'^2$, which constitutes the degeneracy space of the system. The different types of skyrmions are shown in Fig.\,\ref{FigSkyrm}(a). The specific feature of skyrmion states is that they are localized in a particular region of 2D space, whereas the vector field outside this region, including the infinite points, is supposed to be uniform and be directed perpendicular to the film plane, i.e., ``up" or ``down". Accordingly, the coordinate space of the system is presented by the plane with the identified infinite points, which is equivalent to the sphere, S$^2$. 
 Therefore the skyrmion states are defined by the mapping S$^2\to$S$'^2$. All skyrmions shown in Fig.\,\ref{FigSkyrm} have the topological charge $N_2=1$ since the order parameter vector sweeps the sphere S$'^2$ only one time when passing all the points in the 2D coordinate space. They differ, however, by the symmetry resulting from the intrinsic arrangement of the order parameter vector. 

 The Bloch-type skyrmions, $S^\pm_{cw}$ and $S^\pm_{ccw}$ are characterized by the clock- or counterclockwise 
 vorticity in their middle parts, denoted by the subscripts {\it cw} and {\it ccw}. They are also characterized by the up- or downward directions of the order parameter vector $\mathbf{V}$ in their core regions (which is opposite to the direction of $\mathbf{V}$ in the periphery regions), denoted by the polarity superscript index, ``$+$" or ``$-$", respectively. Bloch skyrmions possess handedness. 
 Indeed, the reflection of the Bloch-type skyrmions in the plane of the film gives the skyrmions of the same vorticity but of opposite polarities, whereas their reflection in the plane containing the axis gives the skyrmions of the same polarity but of opposite vorticity. These images can not be superimposed on their preimages by any rotation of the coordinated space. The skyrmions $S^+_{ccw}$ and $S^-_{cw}$ are right-handed, whereas the skyrmions $S^+_{cw}$ and $S^-_{ccw}$ are left-handed, which is illustrated by the hand pictograms in Fig.\,\ref{FigSkyrm}(a).
 
 The N\'eel-type skyrmions $S^\pm_{in}$ and $S^\pm_{out}$ are characterized by the in- and outcoming directions of the order parameter vector of their middle parts, denoted by the subscripts {\it in} and {\it out}. They are also characterized by the $\pm$-polarity like the Bloch-type skyrmions. 
 However, N\'eel-type skyrmions do not possess definite handedness.
 For instance, the N\'eel-type skyrmions, after reflection in the plane of the film, can be superimposed on themselves by the $180^\circ$ rotation around any axis lying in this plane. 
 
 Similar to vortices, skyrmions can form extended skyrmion tubes in the 3D coordinate space of the crystals with strong uniaxial easy-plane anisotropy, see Fig.\,\ref{FigSkyrm}(b).

\medskip

 \noindent \textit{c) Merons}

\begin{figure}
\begin{center}
\includegraphics [width=0.7\linewidth] {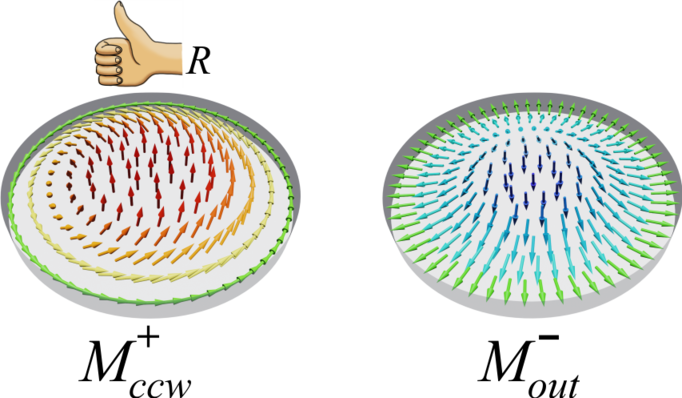}
\caption{ \textbf{Merons.}  Exemplary illustration of the Bloch meron $M_{ccw}^+$, and N\'eel  meron $M_{out}^-$.}
\label{FigMeron}
\end{center}
\end{figure}

 The merons represent another topological state dwelling in 2D films and having a structure analogous to that of skyrmions\,\cite{Gobel2021}. Similarly, they can be of the Bloch type, $M^\pm_{cw}$  and $M^\pm_{ccw}$, or of the N\'eel type, $M^\pm_{in}$ and $M^\pm_{out}$. Figure\,\ref{FigMeron} shows the typical representatives, merons  $M^+_{ccw}$  and $M^-_{out}$.
 The distinction from skyrmions is that the order parameter vector field in skyrmions at infinity is supposed to be either up or down-directed with respect to the film, whereas, in merons, it remains confined within the plane of the film. Accordingly, the topological structure of merons presents a ``half-skyrmion" structure. It means that meron's vector field texture results from the cutting of the inner region of the skyrmion in which the vector sweeps only the upper half of the S$'^2$ sphere in the degeneracy space. At its periphery, the  vector order parameter lies in the film plane and makes a full $360^\circ$ turn when going around the meron center. Therefore, merons can also be globally viewed as vortices in which the order parameter vector escapes from the film plane at the core, avoiding hence the singularity.

To conclude here, these are the planar or tube-like topological states with non-vanishing vorticity,  skyrmions, $S^\pm_{cw}$ and $S^\pm_{ccw}$, and merons, $M^\pm_{cw}$ and $M^\pm_{ccw}$, which carry the handedness of the vector field. The zero-vorticity skyrmions, $S^\pm_{in}$ and $S^\pm_{out}$, and merons, $M^\pm_{in}$ and $M^\pm_{out}$, and all types of the vortices, $V_{cw}$, $V_{ccw}$, $V_{in}$ and $V_{out}$, and vortex tubes possess no handedness. 
More complex, 3D chiral units, Hopfions\,\cite{Rybakov2022,Liu2022}, will be discussed in the next subsection. 
Aggregating these basic chiral building blocks into the space-extended textures like lattices or disordered networks enables the construction of structures with topological spatially-nonuniform chirality.

\begin{figure*}
\begin{center}
\includegraphics [width=0.9\linewidth] {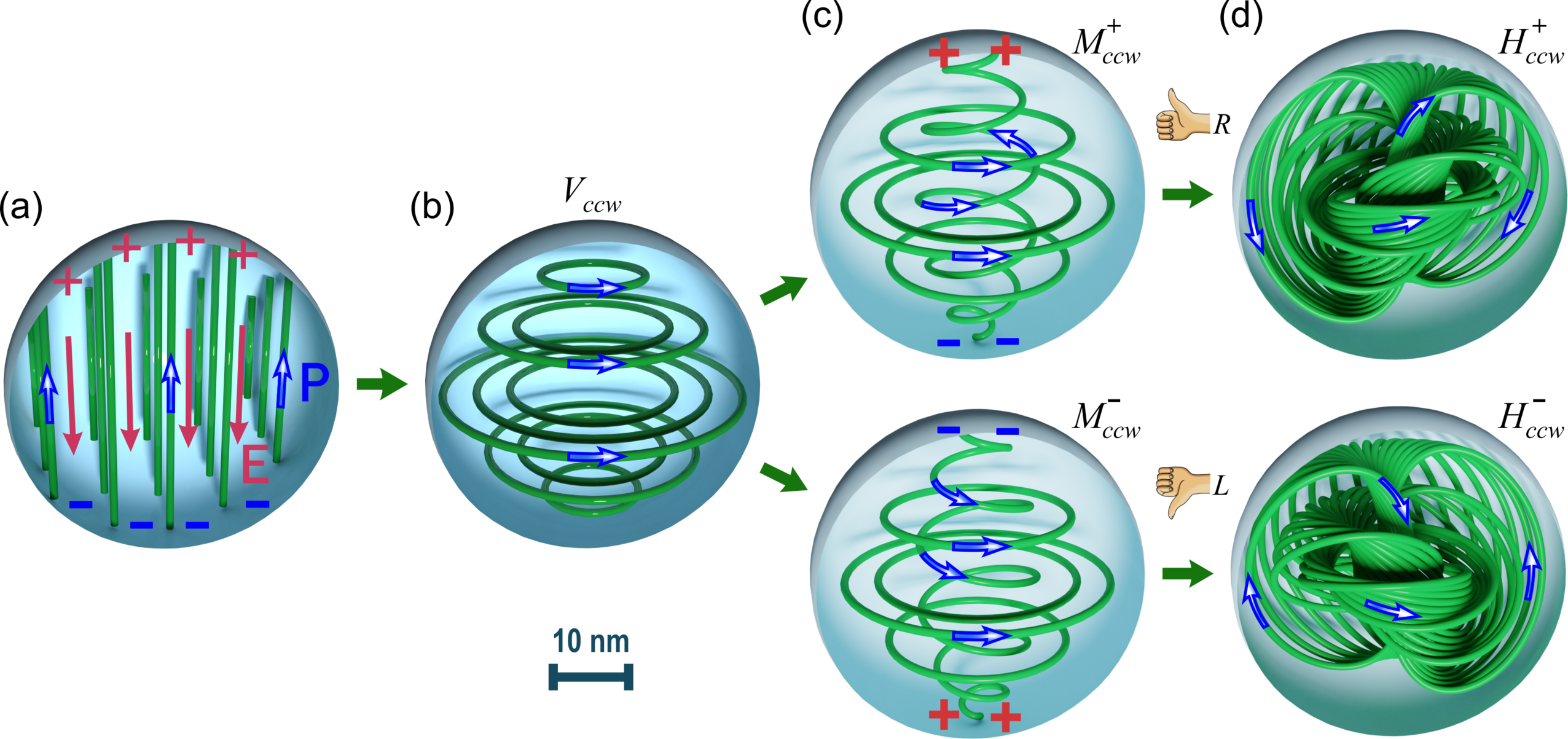}
\caption{ \textbf{Hopfion formation in a spherical nanoparticle.} 
\textbf{(a)}, Uniform distribution of the polarization, $\mathbf{P}$, is shown by green streamlines. Blue arrows show the polarization direction. Positive and negative bound charges at the surface induce energetically undesirable depolarization electric field, $\mathbf{E}$, shown by arrows.
\textbf{(b)}, Formation of polarization vortex $V_{ccw}$ eliminates bound charges but requires additional energy for suppressing polarization at the core.
\textbf{(c)}, The escape of the polarization vortex streamlines into the third direction, allowing to avoid the singular core formation, is possible in two ways. It is either parallel to the vortex axis and results in the formation of the right-handed meron tube $M^+_{ccw}$, or it is anti-parallel to the vortex axis and leads to the formation of the left-handed meron tube $M^-_{ccw}$. However, this again results in appearing the positive and negative bound charges at the surface.
\textbf{(d)}, Spreading of the polarization streamlines into a back-flow along the sphere’s surface eliminates the energetically undesirable onset of depolarization charges. This results in the formation of new topological states, Hopfions, $H^+_{ccw}$ and $H^-_{ccw}$, possessing right-handed and left-handed chiralities, respectively.
Adopted from Ref.\,\cite{Lukyanchuk2020}.
}
\label{FigFormation}
\end{center}
\end{figure*}

\subsection{\label{sec:Foundations} Foundations of topological chirality in ferroelectrics}

\subsubsection{\label{sec:Origin} Origin of topological chirality in ferroelectrics}

Despite the seeming similarity, the origin of chiral topological structures is very different in magnetic and ferroelectric systems. 
In the majority of the chiral magnetic systems, it is the local built-in Dzyaloshinskii–Moriya interaction (DMI) that induces long-range chiral ordering and, in particular, stabilizes the skyrmion and meron states. Therefore, the chirality of the magnetization texture in magnetic systems is predefined by the sign of the DMI interaction.

At variance, a distinct feature of ferroelectrics is that the chirality appears as a result of the spontaneous symmetry breaking due to an interplay of the confinement and depolarization effects where the topological states are generated by the long-range interaction of the bound charges. 
To detail this, we highlight the essential physical properties of ferroelectrics resulting in the formation of the chiral topological states. 
 
(i) The considerable anisotropy effect removes the degeneracy of the ferroelectric energy with respect to the rotation of the polarization order parameter $\mathbf{P}$. This complicates the introduction of the degeneracy space D$'$, which was a core element of the described above topological approach.
 
(ii) \marker{The characteristic value of the coherence length in ferroelectrics, the length scale over which the disrupted order parameter reverts to its equilibrium value, $\xi_0\simeq 1$-$3$\,nm, is much smaller than the corresponding length in magnetic systems, where it is of the order of tens of nanometers. Therefore, ferroelectric nanostructures with a size exceeding a few nanometers should be considered as 3D systems with the coordinate space $\mathcal{M}\subset\mathbb{R}^3$. Importantly, for the atomic-scale sample of the size below $\xi_0$, in particular for the atomically thin ferroelectric films\,\cite{Guan2020,Qi2021,Wang2023}, the fluctuations of atomic ordering, surface effects, and imperfection of the crystal lattice may play the decisive role in establishing the ferroelectric order. These effects 
are very material-dependent, and the design of the particular chiral topological states requires an advanced utilization of the material chemistry methods on the atomistic level. Currently, this direction is undergoing impressively rapid development. We then reserve a detailed consideration of the atomic-scale chiral ferroelectrics for future publications. }  
 
(iii)  Depolarization effects, not accounted for by the described above vector field topological approach, are pivotal in the formation of the chiral topological structures in ferroelectrics. In the non-uniform polarization state, the bound charges emerge in the bulk with the volume charge density $\rho=-\mathrm{div}\mathbf{P}$ and at the surface with the surface charge density $\sigma=-P_\perp$, where $P_\perp$ is the polarization vector component normal to the surface.  
The energy of the depolarization field ${E}_{dep}$, induced by these charges, $\mathcal{E}_{dep}=(\varepsilon_0/2)\int {E}_{dep}^2 \,dV $, is huge (here $\varepsilon_0$ is the vacuum dielectric permittivity). Therefore, the system tries to find the configuration of $\mathbf{P}(r)$ that does not result in the formation of bound charges and the related intrinsic depolarization fields, see the analysis below in Section\,\ref{sec:Extensions}.   
Notably, the situation in ferroelectrics differs from that in magnetic systems, where the demagnetization energy constitutes only a small part of the energy of the system. 

Accordingly, in the ground state, the polarization field within the confined ferroelectric endeavors to swirl in order to become tangent to the confining surface while maintaining its divergenceless character. This strong electrostatic constraint of the nanostructured ferroelectrics sets the topological states possessing finite vorticity and chirality.

\begin{figure*}
\begin{center}
\includegraphics [width=\linewidth] {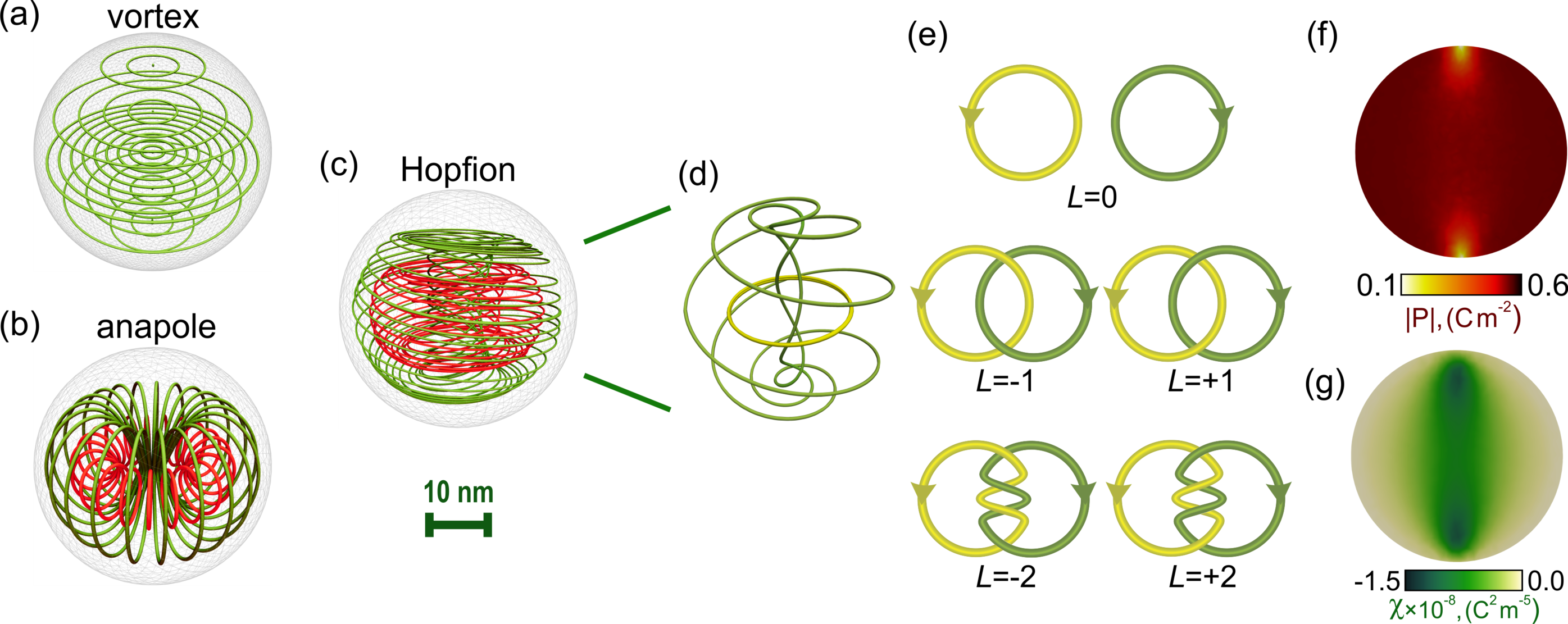}
\caption{ \textbf{Hopfion properties.} 
\textbf{(a)}, A vortex, the toroidal component of a Hopfion.
\textbf{(b)}, An anapole, the poloidal component of a Hopfion. 
\textbf{(c)}, A Hopfion, the combination of the toroidal and poloidal components.
\textbf{(d)}, Exemplary polarization streamlines belonging in the tori shown in $\mathbf{c}$ that illustrate linking of streamlines loops in the Hopfion.
\textbf{(e)}, Illustrations of the topological linking number, \marker{quantifying how many times the two lines wind around each other}. 
\textbf{(f)}, The color map of the distribution of the amplitude of the polarization, $|\mathbf{P}|$.
\textbf{(g)}, The colour map of the distribution of the chirality density $\chi= \mathbf{P\cdot }\left(\mathbf{\nabla}\times\mathbf{P}\right)$.
Panels \textbf{(c)},\textbf{(d)}, \textbf{(f)}, and \textbf{(g)} are from Ref.\,\cite{Lukyanchuk2020}.
}
\label{FigHopfStruc}
\end{center}
\end{figure*}

\subsubsection{\label{sec:Hopfions} Hopfions. The emergence of topological chirality in a ferroelectric nanoparticle}

We present an illustrative example of the formation of chiral topological states in a spherical ferroelectric nanoparticle, assuming that the emerging polarization structures are axially symmetric\,\cite{Lukyanchuk2020}. 
As we mentioned above, a uniform polarization state in the nanoparticle is not energetically favorable because of the formation of the surface-bound charges located at the termination points of polarization lines, see Fig.\,\ref{FigFormation}(a).  A more advantageous configuration is the divergenceless texture of a vortex tube with the polarization vector tangential to the surface, see the vortex $V_{ccw}$ in Fig.\,\ref{FigFormation}(b). However, the creation of a vortex costs additional energy because of the existence of the vortex core in which the ferroelectricity is suppressed. 

A singularity at the vortex core can be eliminated by the escape of the vector field into the third dimension along the vortex axis. Then, the chiral meron-like structure forms, and the meron tubes, $M_{ccw}^+$ and $M_{ccw}^-$, with left- and right-handedness, emerge with equal probability, see Fig.\,\ref{FigFormation}(c).  However, because of this axial polarization component, the bound charge will emerge at the spots where the vortex axis hits the nanoparticle boundary, creating an unfavorable depolarization field again. Spreading into a back-flow over the sphere’s surface maintains the polarization tangent to the surface eliminating this undesirable onset of depolarization charges. 
The described structure is a fundamental chiral topological state called Hopfion\,\cite{Hopf1931}. The chirality of the vector field is provided by the counterplay of the 
upcoming helix-like central flux and downcoming peripheral helix-like fluxes. The central part usually prevails and defines the total handedness of the nanoparticle.

The left and right modifications of Hopfions, $H_{ccw}^+$ and $H_{ccw}^-$, are shown in Fig.\,\ref{FigFormation}(d). 
The Hopfion $H_{ccw}^-$ is obtained from $H_{ccw}^+$  by the mirror reflection in the equatorial plane.
According to the described above convention, the superscript $\pm$ describes polarization polarity at the central axis of Hopfion, and the subscript {\it ccw} stands for the polarization vorticity in the middle of the equatorial plane of the Hopfion. 
The Hopfions possessing the same handedness can be transformed into each other by rotation. For example, the right-handed Hopfions, $H_{ccw}^+$ and $H_{cw}^-$,  can be obtained from each other by $180^\circ$ rotation around any axis, lying in the equatorial plane.  

Figure\,\ref{FigHopfStruc} displays the structure and properties of a Hopfion topological state emerging in a spherical ferroelectric nanoparticle. For simplicity, we assume that the system is isotropic and Hopfion has an axisymmetric structure. Then the superposition of the vortex and anapole components, shown in Fig.\,\ref{FigHopfStruc}(a) and Fig.\,\ref{FigHopfStruc}(b), respectively gives the Hopfion structure, see Fig.\,\ref{FigHopfStruc}(c).  This is also called a toroidal-poloidal decomposition\,\cite{Moffatt1992,Chandrasekhar2006}. 
The vortex state is a vortex tube with a straight axis coinciding with the vertical nanoparticle axis. 
The anapole (this term was first introduced in the physics of elementary particles by Ia.\,B.\,Zel'dovich\,\cite{Zeldovich1957} after the suggestion of B.\,L.\,Ioffe and A.\,S.\,Kompaneets) is the vortex ring state with the circular vortex axis entwining the nanoparticle axis. \marker{In ferroelectric literature the anapoles also quoted as electrical skyrmion bubbles, skyrmion bubbles, polar
bubble skyrmion, polar skyrmion bubbles\,\cite{Govinden2023rev}, and torons\,\cite{Lich2023}.}

In spherical coordinates 
 $\left(r,\theta,\varphi \right)$ the toroidal-poloidal decomposition of the divergenceless vector field of the Hopfion is\,\cite{Moffatt1992,Chandrasekhar2006}
\begin{equation}
\mathbf{P}
=\mathbf{P}_{\mathrm{tor}}+\mathbf{P}_{\mathrm{pol}}
= \alpha_t \, \eta \mathbf{e}_{\varphi }
+\alpha_p \, \mathbf{\nabla }\times \left( \zeta \mathbf{e}_{\varphi }\right)\,,
\label{EqPolTor}
\end{equation}
 where $\eta=\eta(r,\theta) $ and $\zeta=\zeta(r,\theta)$ are scalar potentials and $\alpha_t$ and $\alpha_p$ are the corresponding weights of the toroidal and poloidal components.   

For illustrative purposes, we select a simple form of the $\eta$-  and $\zeta$-potentials, reproducing qualitatively the topology of the toroidal and poloidal components of streamlines, maintaining the polarization tangential to the nanoparticle surface, and enabling analytical calculations as
\begin{equation}
\eta  =  \sin \left(\frac{\pi r}{2 R}\right)\sin \theta\, , \hspace{10pt}
\zeta  =  R\,\frac{\sin \left( {\pi r}/{R}\right) }{r/R}\sin \theta\, .
\label{EqPolTorPot}
\end{equation}

In Hopfion, polarization lines form the upcoming flux in the central region and the compensating down-coming flux at the periphery, making the total flux equal to zero. Herewith the polarization lines turn out to be wound on a system of tori concentrically nested around the central ring encircling the vertical nanoparticle axis. The important property of the polarization vector field of Hopfions is the entanglement of the streamlines, which provides the handedness of the structure.  As illustrated in Fig.\,\ref{FigHopfStruc}(d), any two streamlines, winding either around the same torus or around the different tori, are knotted with each other. The situation is different for the vortices or anapoles in which the streamlines form the rings that are not linked. 

The exemplary types of knots, having different linking numbers $L$,  \marker{quantifying how many times the two lines wind around each other} are illustrated in Fig.\,\ref{FigHopfStruc}(e).  
This property is intimately related to the handedness of the vector field, as we explain more in detail in Sec.\,\ref{sec:Quantification}. The knotting of the streamlines is very similar to the structure of the interlocked molecules, in which the linking between different parts of the molecular chain provides the topological handedness of the structure, see Sec.\,\ref{sec:Nature} and Fig.\,\ref{FigNature}(h) therein. 

Figure\,\ref{FigHopfStruc} further outlines the principal properties of the polarization vector field in a Hopfion described in\,\cite{Lukyanchuk2020}. The distribution of the amplitude of the  polarization is shown in Fig.\,\ref{FigHopfStruc}(f). The distribution of the chirality density $\chi= \mathbf{P\cdot }\left(\mathbf{\nabla}\times\mathbf{P}\right)$, calculated below, see Sec.\,\ref{sec:Quantification}, is presented in Fig.\,\ref{FigHopfStruc}(g).  As we have already mentioned, the core region of the Hopfion gives the dominant contribution to the total chirality of the nanoparticle.

\subsubsection{\label{sec:Arnold} Topological hydrodynamics and Arnold theorem}

Using an illustrative example of a spherical nanoparticle, we have demonstrated above how the polarization inside a restricted volume swirls to either achiral vortices and anapoles or chiral Hopfions to keep its divergenceless structure, which is the principle topological constraint for confined ferroelectrics. Understanding the polarization states in the systems having an arbitrary geometry and size requires a general topological approach for the description of the confined divergenceless vector fields. This approach, called topological hydrodynamics, was introduced in pioneering works by Moffat and Arnold\cite{Moffatt1992,Arnold2021} for a specific example of the hydrodynamic flow of an incompressible liquid in confined volume.

In topological hydrodynamics, the classification scheme of the topological states of the divergenceless vector field is built on Arnold theorem. Arnold theorem establishes that the interior of the bounded manifold holding within a divergenceless vector field splits into a finite number of cells of two types. The streamlines in the cells of the first type are winded around the set of concentric cylinders and all closed. The streamlines in the cells of the second type are winded around the set of nested tori and are either all closed or everywhere dense. 

This fundamentally important statement enables us to straightforwardly classify topological states emerging in ferroelectrics, associating the topological states in the cells of the first type with vortices and the topological states in the cells of the second type with Hopfions. 

\marker{Which \markertwo{of the states} appear most stable depends on the tiny energy balance between the different contributions to the ferroelectric energy, in particular the Ginzburg-Landau energy, including the anisotropy terms, \markertwo{ the gradient energy}, the elastic energy, and also \markertwo{depends}  on the shape of the cell.}

The emergence of Hopfions provides the chirality of the confined ferroelectric systems.  
It appears that the resulting field configurations in ferroelectrics are of a general character, and the hydrodynamic approach works 
perfectly even in highly-anisotropic ferroelectrics\,\cite{Tikhonov2022}. 
Furthermore, the topological states in ferroelectrics and similar to those in a variety of physical systems ranging from the whirlpools to the interiors of the stars and plasma devices, tokamaks, and stellarators, are described by the topological hydrodynamics.

\subsubsection{\label{sec:Quantification} Quantification parameters}
\noindent a)\,\textit{Numerical indicators} 

Characterization of the topological states in confined ferroelectrics requires numerical indicators, allowing to quantize the degree of emerging polarization swirling and handedness in the system of interest. 

The swirling of the polarization can be characterized by parameter $\mathbf{\Omega}$ expressing the integral of the quantified vorticity $\boldsymbol{\omega}$\,\cite{Moffatt1992} over the whole volume $V$ of the coordinate space $\mathcal{M}$ of the system
\begin{equation}
\mathbf{\Omega}=\int_{\mathcal{M}}\boldsymbol{\omega}\,dV, \qquad  \boldsymbol{\omega}= \mathbf{\nabla }\times \mathbf{P}\,. 
\label{EqVorticity}
\end{equation}
Nonzero $\mathbf{\Omega}$ indicates the presence of vortices. 
However, its sign does not necessarily reflect the direction of the rotation, CW, or CCW. 

\markertwo{
To illustrate this, let us consider the polarization vortex confined in the cylinder of the radius $R$ and length $L$ with an exemplary radial dependence of the polarization $\mathbf{P}(\rho) = P(\rho) \,\mathbf{e}_\varphi$ in cylindrical coordinates $\rho$ and $\varphi$.  The ferroelectric cylinder is surrounded by the dielectric material with zero spontaneous polarization, resulting in the suppression of the polarization amplitude in the vicinity of the cylinder surface, hence the non-monotonous behavior of ${P}(\rho)$ through the whole vortex.  
Then, the radial distribution of the vortex polarization in the near-surface region, $\mathbf{P}(\rho)\simeq(P(R)+C_1(R-\rho))\,\mathbf{e}_\varphi$, provides the vorticity $\boldsymbol{\omega}=-C_1 \,\mathbf{e}_z$ with the sign, opposite to that realized in the vortex core, where $\mathbf{P}(\rho)\simeq C_0\rho\,\mathbf{e}_\varphi$ and 
$\boldsymbol{\omega}=C_0 \,\mathbf{e}_z$. Here both, the magnitude of the vortex polarization at the surface, $P(R)$, and numerical coefficients, $C_0$ and $C_1$, are either all positive, in the case of the CCW vortices, or all negative, in the case of the CC vortices.
Furthermore, the decrease of the order parameter far from the vortex core can be facilitated by the 
elastic coupling of the polarization with crystal lattice \cite{Kondovych2023}. 

In general, the swirling parameter of the whole system, $\mathbf{\Omega}$, is equal to zero. According to the Stokes theorem, Eq.\,(\ref{EqVorticity}) transforms to polarization circulation around the remote circular contour $C$, encircling the cylinder and lying far away in the dielectric region, 
$\mathbf{\Omega}= \mathbf{e}_z L\oint_C\, \mathbf{P} d \mathbf{l}$\, which vanishes because of the suppression of the polarization in the external dielectric media. Here $d\mathbf{l}$ is the differential length along $C$.  The equality to zero of the swirling parameter implies that within the vortex area, having the definite direction of the polarization rotation, CW or CCW, the sign of the vorticity is not homogeneously defined. Namely, the vorticity of the vortex core region is compensated by the vorticity of the outer region, having the opposite sign.   

As we describe further in Section\,\ref{sec:Films}, the inversion of the sign of vorticity is also observed for the polarization vortices packed in the ferroelectric/paraelectric superlattices in the vicinity of the interfaces between these materials. Hence, the sign of the vorticity cannot be used as an unambiguous indicator of the vortex rotation in this case as well.}

The vortex structure can also be characterized by the toroidal, ${\mathbf{T}}$, and hypertoroidal, $\mathbf{G}$, moments\,\cite{Prosandeev2008}, 
that we define as the integrals of the corresponding toroidal, ${\mathbf{\mathcal{T}}}$, and hypertoroidal, ${\mathbf{\mathcal{G}}}$, moment densities
\begin{eqnarray}
\mathbf{T}&=&\int_{\mathcal{M}}\mathbf{\mathcal{T}}\,dV, \qquad  
\mathbf{\mathcal{T}}= \frac{1}{2}\,\mathbf{r}\times (\mathbf{P}-\mathbf{\bar{P}})\,, 
\label{Toroidal}
\\
\mathbf{G}&=&\int_{\mathcal{M}}\mathbf{\mathcal{G}}\,dV, \qquad  
\mathbf{\mathcal{G}}= \frac{1}{4}\,\mathbf{r}\times \left( \mathbf{r}\times (\mathbf{P}-\mathbf{\bar{P}})\right)\,. 
\label{EqHyperToroidal}
\end{eqnarray}
To reflect the absence of the depolarization field in the confined system, we have to put the condition $\mathbf{\bar{P}}=0$ for the average polarization.  Unlike the swirling parameter $\boldsymbol{\Omega}$, the signs of toroidal and hypertoroidal moments discriminate between CW and CCW rotation of the polarization vector.

Note, however, that none of the described indicators can be used for the detection of the handedness of the system since their mirror images can be overimposed onto the originals after the proper turn.
The handedness of topological states can be quantified by the chirality, $X$, as an integral of the corresponding chirality density, 
\begin{equation}
X=\int_{\mathcal{M}}{\chi}\,dV, \qquad  \chi= \mathbf{P\cdot }\left(\mathbf{\nabla}\times\mathbf{P}\right)\,. 
\label{EqChiraliyy}
\end{equation}

The introduced scalar indicator, chirality, has, however, the same shortcoming as vorticity $\boldsymbol{\Omega}$.
While being capable of detecting the presence of the chirality itself, it cannot unambiguously determine the sign of the handedness because the differential vorticity density ${\nabla }\times\mathbf{P}$ does not univocally determine 
the direction of rotation of the polarization vector. 
\markertwo{Let us consider, for instance, the right-handed helicoidal structure formed by the bounded CCW vortex, nucleating within the uniformly polarized state,  $\mathbf{P}=P_z\mathbf{e}_z+P(\rho)\mathbf{e}_\varphi$, where the uniform $z$-directed component of polarization, $P_z=$const and the perpendicular vortex component, $P(\rho)$, vanishes at the vortex core and also at the infinity, as we described above. The total chirality can be calculated using again the Stokes theorem, $X= L P_z  \oint_{C} \mathbf{P}\,d\mathbf{l}$. Then, similar to the swirling parameter, the total chirality vanishes for the remote enough contour $C$, not allowing to determine the handedness of the system.
}

\medskip

\noindent b)\,\textit{Topological charges}~~\\ 
Other types of parameters quantifying the properties of the topological states are the so-called topological charges, characterized by their invariance with respect to small continuous deformation of the system, in particular with respect to the volume-preserving diffeomorphisms, keeping the divergenceless character of the vector fields.

Although, as we mentioned in Sec.\,\ref{sec:Origin}, the coordinate space of the nanostructured ferroelectrics is a 3D space, certain polarization textures, like, for instance, vortex-, skyrmion-, or meron tubes, are, in fact, quasi-2D objects. In this case, for quantitative characterization of the emerging topological states, the formalism of the winding numbers expressing the topological charges that usually applies to describe objects confined in the 2D space can be used. In particular, vortex tubes can be characterized by the integer vortex winding number\,\cite{Zang2018}
\begin{equation}
  N_1=\frac{1}{2\pi}\oint_C \frac{\partial \varphi}{\partial l} dl\,, 
    \label{vort}
\end{equation}
\marker{where} the  integration is carried out over the contour $C$, encircling the vortex core, $dl=|d\mathbf{l}|$ is the element of the length of this contour and $\varphi$ is the azimuthal angle of the polarization vector $\mathbf{P}$.  

In the case of the skyrmion- or meron-like tube, the corresponding winding number also called the Pontryagin index, is\,\cite{Zang2018} 
\begin{equation}
    N_2=\frac{1}{8\pi}\int_{\mathrm{S}^2} 
     \epsilon_{ijk} \, \mathbf{n} \cdot 
    \left[
    \frac{\partial \mathbf{n}}{\partial x_j}
     \times
     \frac{\partial \mathbf{n}}{\partial x_k}
     \right]
    \, 
    d{\marker{s_i}}\,,
    \label{skyrm}
\end{equation}
where $\mathbf{n}=\mathbf{P}/P$ is the unit vector, indicating the direction of polarization. The integration is done in the plane perpendicular to the \marker{tube axis}, $d\mathbf{s}$ is the vector area element of the integration plane,  $\epsilon_{ijk}$ is the antisymmetric Levi-Civita symbol, and $i,j,k$ run through the $1,2,3$ values. Index $N_2$ indicates the number of $180^\circ$ half-turns of the vector $\mathbf{P}$ accomplished when $\mathbf{P}$ goes from the tube center to the periphery. It assumes integer values in the case of skyrmions, semi-integer values in the case of merons, and arbitrary values in the case where the direction of $\mathbf{P}$ at the periphery of the tube is not fixed. Note that neither $N_1$ nor $N_2$ reflect the handedness of the system.  

It is of great fundamental importance to introduce another characteristic value, the vector potential $\mathbf{A}$ of the polarization vector field. 
Defining, in an analogy with electromagnetism,  $\mathbf{A}$ via $\mathbf{P}= \mathbf{\nabla }\times \mathbf{A}$, we straightforwardly obtain $\mathrm{div}\,\mathbf{P}=0$. Note that in a particular case of the uniform distribution of polarization,  the vector potential can be presented as $\mathbf{A}=(1/2)\left( \mathbf{r}\times \mathbf{P}\right)$ that coincides with the density of the toroidal moment. 
For the axisymmetric Hopfion, described by Eqs.\,(\ref{EqPolTor},\ref{EqPolTorPot}) the vector potential is
\begin{equation}
\mathbf{A}=\alpha_t \, \xi \mathbf{e}_{r}+\alpha_p \,\zeta \mathbf{e}_{\varphi}\,,
\label{EqAHopf}
\end{equation}
where the function  $\xi$ is calculated from $\eta =-r^{-1}\partial \xi /\partial \theta $, and for the particular choice of $\eta$ (\ref{EqPolTorPot}) is given by 
\begin{equation}
\xi = r \sin \left(\frac{\pi r}{2 R}\right) \cos \theta \,.
\label{EqXiAxial}
\end{equation}

The use of the vector potential $\mathbf{A}$ allows for the introduction of another topological characteristic, helicity,  emerging in topological hydrodynamics and electromagnetism and characterizing the handedness of the system\,\cite{Arnold2021}.
\begin{equation}
H=\int_{\mathcal{M}} \mathcal{H}\,dV, \qquad  \mathcal{H}= \mathbf{P}\cdot\mathbf{A}\,. 
\label{EqHelicity}
\end{equation}
 This scalar quantity changes the sign when passing from the left- to the right-handed coordinate system, hence as well as chirality, $X$, specifies the handedness. 
In the particular case of the axisymmetric Hopfion Eqs.\,(\ref{EqPolTor}) the calculation of helicity is simplified to\,\cite{Finn1985} 
\begin{equation}
H=2\,\alpha_p \alpha_t\int_{\mathcal{M}} \zeta \eta\,dV\, .
\label{EqChiralityAxial}
\end{equation}

The topological nature of the helicity follows from the statement that it conserves its value under any volume-preserving diffeomorphism of the system, \marker{for instance under volume-preserving deformations. Notably, the description of the full group of the volume-preserving transformations for ferroelectrics including dynamical ones under which the helicity is conserved, remains a challenging question.}

Moreover, helicity characterizes the important geometric parameter of the divergenceless vector field,  the average of the linking number of the streamlines, $L$\,\cite{Arnold2021}. 
This topological attribute of the streamlines quantifies the number of times that each line, on average, winds around the other, see Fig.\,\ref{FigHopfStruc}(d). 
The important result of the topology is that the average linking number of the streamlines is equivalent to the helicity of the topological state. The sign of $L$ is different for the left- and right-handed modifications of the knots of the winding streamlines, as illustrated in Fig.\,\ref{FigHopfStruc}(e).  This explains -- from another, highly demonstrative perspective –– why helicity is related to handedness.    


\marker{Note that sometimes another definition of helicity, coinciding with the introduced in Eq.\,(\ref{EqChiraliyy}) chirality parameter is used for characterization of the handedness of the ferroelectric topological states, see e.g.,\,\cite{Shafer2018}. 
An ambiguity in the definition of $H$ appears due to the existence of two different approaches in hydrodynamics. In one approach, the velocity field of the liquid, $\mathbf{u}(r)$, is used as a principal vector field, characterizing the system. Then, the helicity, defined as $H$$=$$\int_{\mathcal{M}}\mathbf{u}\cdot\mathbf{A}\,dV$, where the vector potential $\mathbf{A}$, coming from the relation $\mathbf{u}$$=$$\mathbf{\nabla }\times \mathbf{A}$, characterizes the average linking number of the streamlines of the field $\mathbf{u}(r)$\,\cite{Arnold2021}. 
In another approach, it is vorticity field, $\boldsymbol{\omega}$$=$$\mathbf{\nabla}\times\mathbf{u}$, that is considered as a principal vector field. The corresponding helicity, defined as $H$$=$$\int_{\mathcal{M}}\boldsymbol{\omega}\cdot\mathbf{u}\,dV$, characterizes the average linking number of the vortex tubes, formed by the streamlines of the field $\boldsymbol{\omega}(r)$\,\cite{Moffatt1992}. 
We follow the first approach taking into account that 
this is
the polarization vector field $\mathbf{P}(r)$ in ferroelectricity
 which is an analog of 
the divergenceless velocity field $\mathbf{u}(r)$ in hydrodynamics. 
Then, the helicity defined by Eq.\,(\ref{EqHelicity}) provides 
a distinct topological characterization of the handedness and the linking number of the streamlines of the field $\mathbf{P}(r)$. Another definition of helicity operates with the streamlines of the vorticity field  $\boldsymbol{\omega}$$=$$\mathbf{\nabla}\times\mathbf{P}$ which do not have a clear physical meaning and, hence, cannot be used for judging about the total handedness of the system, as we have already discussed above.  
}

\medskip
\noindent\,c)\,\textit{The application of quantification parameters}

The parameters represented above well describe the variance of the properties of the polarization field in nanostructured systems and are widely used to treat the experimental and simulation data. 
Yet, for the most efficient use of these parameters, one has to set up their respective areas and conditions of applicability. 
To illustrate this, we consider the model Hopfion, in which the distributions of polarization and vector potential are given by Eqs.\,(\ref{EqPolTor},\ref{EqPolTorPot}), and (\ref{EqAHopf},\ref{EqXiAxial}) respectively.  The quantitative parameters are calculated using the corresponding equations (\ref{EqVorticity}-\ref{EqChiraliyy}),
and\,(\ref{EqHelicity}) (the same parameters per unite volume can also be used). The results are given in Table\,II.
\begin{table}[h!]
{
\centering
 \caption{Parameters, quantifying the swirling and handedness of the model Hopfion.}
 \renewcommand{\arraystretch}{1.2}
\begin{tabular}{l l }
\hline
\hline
Swirling parameter,        & $\mathbf{\Omega }\,=\,8.38\, \alpha_t R^{2}\, \mathbf{e}_{z}$
\\ 
Toroidal moment,          & $\mathbf{T}\,=\,0.96\,\alpha_t R^4 \, \mathbf{e}_{z}$ 
\\ 
Hypertoroidal moment, \;    & $\mathbf{G}\,=\,1.19\,\alpha_p R^{5}\,\mathbf{e}_{z}$ 
\\ 
Chirality,               & $X\,=\,43.72\, \alpha_p \alpha_t R^{2}$
\\ 
Helicity,               & $H\,=\,4.09\, \alpha_p \alpha_t R^4 $ 
\\ 
\hline
\hline
\end{tabular}
}
\label{TableParam}
\end{table}

As follows from Table\,II, the swirling parameter and toroidal moment are sensitive to the toroidal component of the Hopfion, i.e., it is effective for vortices where $\alpha_p=0$, $\alpha_t\neq 0$ and vanishes for the anapoles where $\alpha_p\neq 0$, $\alpha_t=0$. The hypertoroidal moment, in contrast, is sensitive to the poloidal component, hence to the emergence of anapoles.
\marker{Note that neither toroidal nor hypertoroidal moments can be used to discriminate between the chiral and achiral states since they remain non-zero in the states with no handedness, vortices, and anapoles, respectively.}
The handedness is characterized by the chirality and helicity parameters that both are non-zero in the chiral state, where both $\alpha_t, \alpha_p\neq 0$, and vanish in the achiral state of the vortex or anapole where one of the components, $\alpha_t$ or $\alpha_p$, is equal to zero. Note, however, that although the indicated parameters can be used for the characterization of particular properties of the polarization distribution, their full relevance realizes when one does associate the measurable physical parameter with them. For instance, having similar integral expressions in Table\,II, the chirality and helicity may be very different in terms of the distribution of their densities over the sample volume. 
Which precisely parameter becomes most significant depends on the concrete experimental realization of the experiment that reveals handedness, for instance, on the type of the chiral fields testing the handedness\,\cite{Mackinnon2019}. 
To describe the distribution of chirality inside nanostructured ferroelectrics, we will use the parameter $\chi$, which is easy to calculate. 

Note, finally, that the topological charges (\ref{vort}) and (\ref{skyrm}), conventionally defined for the low-dimension coordinate spaces, can, to some extent, be used for the 3D nanostructured ferroelectrics. For instance, the vortex winding number $N_1$ can be helpful in detecting singularities of the vortex core lines extended in 3D space, \marker{whereas the Pontryagin index $N_2$ was successfully used in\,\cite{Das2019} to test the behavior of the vector field in the cross-sectional areas of the chiral bubbles, called polar skyrmions.}

\marker{

\subsubsection{\label{sec:Extensions} Extensions of hydrodynamic approach}

Here we discuss the peculiarities of applications of topological hydrodynamics, in particular, the Arnold theorem, to the intriguing subtleties appearing in ferroelectric systems and promising new routes for remarkable advancements in the field. 

\medskip

\noindent a)\,\textit{Characteristic energies and bound charges} 

In Sections\,\ref{sec:Origin}-\ref{sec:Quantification} we have provided a systematic hydrodynamic approach for the description of topological properties of nanostructured ferroelectrics based on the selection of specific low-energy polarization states $\mathbf{P}(r)$ in which the strong depolarization effects arising from the emerging bound charges $\rho=-\mathrm{div}\,\mathbf{P}$ are essentially reduced. The key aspect of our approach is focusing on a subset of polarization fields that satisfy the condition $\mathrm{div}\,\mathbf{P}=0$. This constraint is fundamentally different from the constraint imposed by the spatial constancy of the order parameter amplitude, usually utilized in the real-space vector field approach in magnetism for the magnetization vector  $|\mathbf{M}(\mathbf{r})|=\mathrm{const}$. 

We discuss now to what extent the imposed constraint, $\mathrm{div}\,\mathbf{P}=0$, is effective. Notably, we employ the standard approach in physics, where, first, the major effect, which, in our case, is the impact of depolarization fields, is explored, and then the effects bringing in small corrections are taken into account. 
A topological consideration that we employ is based exactly on this approach. We first find the divergenceless fields corresponding to the given geometry of the problem and then consider the interplay of other energies, like Ginzburg-Landau energy, gradient energy, and elastic energy perturbations that define the resulting state of the system.  See\,(\citeyear{SM}) for the full energy functional of the system.

To put the introduced energy hierarchy on a quantitative basis, we consider an exemplary Ginzburg-Landau functional, describing the emergence of the uniform one-component polar state, characterized by the  order parameter $\bf P$ in ferroelectric and its interaction with an intrinsic electric field $\bf E$, 
\begin{equation}
F=\frac{1}{2} a (T-T_c) {\bf P}^2 + \frac{1}{4} B \left({\bf P}^2\right)^2-\mathbf{EP}\,. 
\label{GLsimpl}
\end{equation}
Here, $T_c$ is the ferroelectric transition temperature. The phenomenological constants $a$ and $B$ are expressed through the measurable parameters, Curie constant, $C$, and polarization magnitude $P_0$ at $T=0$, and $\mathbf{E}=0$ as: $a=1/(\varepsilon_0 C)$,\cite{StrukovBook}, $B=aT_c/P_0^2=T_c/(\varepsilon_0 C P_0^2)$, where $\varepsilon_0$ is the vacuum dielectric permittivity. The Ginzburg-Landau energy density of the uniform state at $T=0$ is $F_{GL}=- 1/(4\varepsilon_0) (T_c/C) P_0^2$. This energy is negative, which reflects the instability of the system with respect to the formation of the ferroelectric state. However, the termination of polarization at the surface of the sample results in the emergence of the surface bound charge $\sigma_0=P$ which, in turn, produces the depolarization electric field, oriented oppositely to $\bf P$, $E_{dep}=-\sigma_0/\varepsilon_0=-P_0/\varepsilon_0$. The associated with $E_{dep}$ depolarization energy density, $F_{dep}=\varepsilon_0E_{dep}^2/2=P_0^2/2\varepsilon_0$, is positive and by the factor $2C/T_c$ larger than the absolute value of $F_{GL}$. In the most functional oxide ferroelectrics that are displacive ferroelectrics, this factor is especially large\,\cite{StrukovBook}. Substantial prevailing of the positive depolarization energy over the negative Ginzburg-Landau energy strongly destabilizes the uniform ferroelectric state with $\mathbf{P}=\mathrm{const}$ through the polarization-field coupling term in (\ref{GLsimpl}) and finally results in the strongly non-uniform reconstructed state, in which the effective depolarization fields and producing them emerging bound charges are correspondingly reduced by at least the factor of $(2C/T_c)^{1/2}\simeq 20-30$.
Note that the factors $2C/T_c$ are so huge, for example, $2C/T_c\approx 10^3$, for PbTiO$_3$, and $0.5\times 10^3$ for BaTiO$_3$\,\cite{LandoldtOx}, that the specific oversimplified form of the functional\,(\ref{GLsimpl}) does not change the estimate much.

\markertwo{The material-specific functionals, given in\,(\citeyear{SM}), that account for the multi-component nature of the order parameter give rise to similar conclusions.  We consider the stable Hopfion topological state emerging in  PbZr$_{0.6}$Ti$_{0.4}$O$_3$ nanoparticle of a radius $R=10$\,nm at room temperature and present the relevant  Ginzburg-Landau, $\mathcal{F}_{GL}$, gradient, $\mathcal{F}_{grad}$, elastic, $\mathcal{F}_{elast}$, and depolarization, $\mathcal{F}_{dep}$, contributions to the total energy, $\mathcal{F}_{tot}$ in the Table\,III. We present also the corresponding energy values for the unstable uniform states. As explained above, the positive depolarization energy of the uniform state by two orders of magnitude exceeds the negative Ginzburg-Landau energy, making the uniform state highly unstable towards swirling to the almost divergenceless polarization texture. In the final  Hopfion state, the unfavorable for ferroelectrics $\mathcal{F}_{dep}$ almost vanishes, whereas the beneficial negative sum $\mathcal{F}_{GL}+\mathcal{F}_{elast}$ increases only slightly. Importantly, the positive gradient energy is concentrated in the region of the pole singularities, giving a contribution small by a factor $\xi/R$. However, together with the Ginzburg-Landau energy and elastic energy, (the magnitude of which is also smaller than $\mathcal{F}_{GL}$), it plays the decisive role in defining quantitatively, which of the almost divergenceless swirled states is the most stable one.  
}

\begin{table}[h!]
\markertwo{
{
\centering
 \caption{Ginzburg-Landau $\mathcal{F}_{GL}$, gradient $\mathcal{F}_{grad}$, elastic $\mathcal{F}_{elast}$, and depolarization $\mathcal{F}_{dep}$,  contributions to the total energy $\mathcal{F}_{tot}$, of the uniform and Hopfion states in PbZr$_{0.6}$Ti$_{0.4}$O$_3$ nanoparticle of radius $R=10$\,nm at room temperature.  
 The expressed in the units of $10^{-18}$J energies are found according to\,(\citeyear{SM}).}

\renewcommand{\arraystretch}{1.4}
\begin{tabular*}{\columnwidth}{@{\extracolsep{\stretch{1}}}*{7}{r}@{}}
\hline
\hline
  & $\mathcal{F}_{GL}$ & $\mathcal{F}_{grad}$ & $\mathcal{F}_{elast}$ & $\mathcal{F}_{dep}$ &  $\boldsymbol{\mathit{\mathcal{F}_{tot}}}$ 
\\
\hline
Uniform & -34.64 & 0 & -4.28 & 4540.14 & \textbf{4501.22}
\\
Hopfion & -15.49 & 4.79 & -10.22 & 0.011 & \textbf{-20.91}
\\ 
\hline
\hline
\end{tabular*}
}
\label{TableEnergies}
}
\end{table}

The most known example of the formation of the nonuniform state reducing the depolarization energy is the emergence of the alternatively polarized Landau-Kittel domains in ferroelectric slabs\,\cite{Landau1935,Kittel1946, Bratkovsky2000}. It was shown in\,\cite{Stephanovich2003,DeGuerville2005} that in thin ferroelectric films, the electrostatic energy of the Landau-Kittel domains is minimized even more if the surface depolarization charges, 
at points of domain termination, draw back into the domain volume. This results in the gradual profile of the out-of-plain polarization inside domains from almost zero at points of domain termination to almost equilibrium value $P_0$ in the bulk of the domain. 
Such a gradual distribution of the out-of-plane polarization component called in\,\cite{DeGuerville2005,Lukyanchuk2009} a soft polarization domain is completed by the field-induced in-plane polarization component. The overall polarization texture, looking like the periodic vortex structure\,\cite{Lukyanchuk2022} (Supplementary Information), was experimentally observed in\,\cite{Yadav2016}. We discuss this issue in detail in Sec.\,\ref{sec:Films}. 

Analysis of the experimental data from\,\cite{Yadav2019} provides a quantitative estimate of the magnitude of the depolarization effects emerging in the vortex phase realized in the PbTiO$_3$/SrTiO$_3$ superlattice. We compare the experimentally measured characteristic value of an internal electric field in the vortex phase, $E_i\simeq 2$\,MV/cm, with the characteristic value of the depolarization field that would exist in the uniformly-polarized slab, $E_{dep}=-P_0/\varepsilon_0 \simeq 8 \times 10^2$\,MV/cm where $P_0\simeq -70$\,$\mu$C/cm$^2$. Notably, the value of $E_i$ is four hundred times smaller than $E_d$; this indicates that the bound charges inducing the field $E_i$ are smaller than the bound charges expected for the uniformly polarized ferroelectric over approximately the same factor. 

The above theoretical and experimental evaluations do confirm that the bound charges, $\rho=-\mathrm{div}\,\mathbf{P}$, emerging in the polarization nonuniform states in nanostructured ferroelectrics, are extremely small. This justifies the topological constraint $\mathrm{div}\,\mathbf{P}=0$ and enables consideration of these states in the framework of the topological hydrodynamics. In fact, we construct the perturbation approach with respect to the small parameter $T_c/2C\simeq 10^{-3}$ or even less, in which the states with $\mathrm{div}\,\mathbf{P}=0$ correspond to the zero-order approximation. The next-order corrections will slightly modify the spatial distribution $\mathbf{P}(r)$, resulting, in particular, in the emergence of the small bound charges. However, such corrections will not change the overall structure of the described topological states and, in particular, the chiral properties of nanostructured ferroelectrics.

Unfortunately, experimentally available data do not enable access to the bound charges, in particular, to the surface bound charges related to the normal to the surface polarization component. Notably, the piezoresponse force microscopy (PFM) measures the average values within the finite penetration depth-thick, about 30nm\,\cite{Harnagea2004}, layer, which is larger than the characteristic coherence length $\xi$, constituting
several nanometers. From the instrumental perspective, it is important to have the possibility of extracting these data from direct measurements. 

\medskip 

\noindent b)\,\textit{Intrinsic and external charges and the electrostatic screening}

Another factor that perturbs the polarization streamlines is the electric charges, which either are externally introduced into the system or are inherent to the considered ferroelectric material.  

Although it seems that the screening of the emergent depolarization field by the short-circuited electrodes should destroy the domain pattern, accounting for the realistic situation, where the electrode-induced screening is not the perfect one, demonstrates that the topological states may survive even in the system with highly-conductive electrodes. The work\,\cite{Kornev2004} in which the partial screening by the electrodes was studied using the effective screening parameter $\beta$ demonstrated that the topological states exist in the vicinity of $\beta\simeq 90\%$, whereas the complete screening is characterized by $\beta=100\%$. Moreover, the structure of the emerging topological states can be tuned by the variation of the screening parameter $\beta$.  

The influence of the surface-deposited electrodes on the periodic vortex-domain structure in compressively-strained ferroelectric films was also discussed in\,\cite{Bratkovsky2009}. It was shown that for non-ideal electrode materials, the emergence of the monodomain state may hardly take place. The realistic metallic properties of the electrodes are usually characterized by the Thomas-Fermi screening length $\lambda_{s}$. Given in\,\cite{Bratkovsky2009} energetic consideration demonstrates that even for the very short $\lambda_{s}$ of the order of an {\AA}ngstr\"{o}m scale, which is characteristic for a good metal, the domain structure still remains stable. Furthermore, the electrode-ferroelectric junction may be also non-perfect and be characterized by the so-called buffer or dead dielectric layer in which the ferroelectricity is destroyed. The nanometer-scale buffer layer also stabilizes the domain pattern. In a similar way, the feasible semiconducting charges inside ferroelectric materials are not sufficient to screen the depolarization fields and do not modify the periodic polarization vortices. 

The situation, however, may change in the atomically thin ferroelectric layer. Because the ferroelectric layer thickness becomes smaller than the coherence length, the creation of a nonuniform texture requires considerable gradient energy. Essentially, the vortex domain sizes do not fit the ferroelectric film thickness. In such a situation, the monodomain state can become more beneficial, especially if the depolarization field is further screened by conducting electrodes\,\cite{Aguado-Puente2008}. 
Significantly, due to image forces, the electrostatics of a ferroelectric film confined between two electrodes is equivalent to that of a periodically repeating heterostructure. That is why a similar transition from a multi- to monodomain state was predicted for the ferroelectric/dielectric superlattices with the atomically small period.\,\cite{Stephanovich2005}.

Notably, the interaction of charges 
located within a few-nm scale dielectric or ferroelectric films sandwiched between two metallic or semiconducting electrodes is described by quasi-2D electrostatics. In particular, the 3D Coulomb electrostatic potential, decaying as $\varphi_\mathrm{3D}$$\propto$${1/r}$ would evolve into the logarithmic potential $\varphi_\mathrm{2D}\propto\mathrm{ln}\,(r/\Lambda)$, for $r<\Lambda$, that changes to exponentially-decaying potential $\varphi_\mathrm{2D}$$\propto$$\mathrm{exp}\,(-r/\Lambda)$ for $r$$>$$\Lambda$, where the screening length $\Lambda$ is determined by materials parameters and geometry of the system\,\cite{Kondovych2017}.  

In particular, the possibility of the existence of the Bloch-type skyrmion, producing substantial bound charge $\sim 0.6$e/uc, in the ultra-thin film of PbZr$_{0.4}$Ti$_{0.6}$O$_3$ with about $\beta=90\%$ screening of the polarization-induced surface charge resembling the effect of the electrode, was recently demonstrated by simulations\,\cite{Gao2023}. 

\markertwo{Furthermore, the residual bound charges, albeit of lesser magnitude $\sim 0.45$e/uc,  have been observed in simulations of the ferroelectric PbTiO$_3$ layer inserted in between realistic electrodes with  $\beta=80\%$ screening and packed within PbTiO$_3$/SrTiO$_3$ multilayers\,\cite{Lu2018}. These charges were associated with the vector field of the topologically protected states, dipolar waves, and disclination patterns.}
  
These systems are of high interest as they offer the possibility of observation of the crossover from the topological hydrodynamics to the topological vector field approach by variation of the screening parameters.  

The stability of the Hopfions confined in a spherical PbZr$_{0.6}$Ti$_{0.4}$O$_3$ nanoparticle of the typical diameter of 50\,nm in the presence of the semiconducting or impurity-induced free charges was investigated in\,\cite{Lukyanchuk2020} (Supplementary Information). It was found that at the typical for the PZT values of the screening length, about $\lambda_s\simeq$\,80–100\,nm, the screening does not influence the Hopfion texture. At higher densities of free charges, where the screening length decreases to around $\lambda_s \simeq$\,20\,nm, the polarization lines still conserve the tori-twinning structure.
However, the nested tori set transforms into the snail-like convolving configuration in which the polarization lines convolve or untwine along the spiral-like paths and creep from torus to torus. 
As the density of free charges continues to increase, resulting in  $\lambda_s \lesssim 5$ nm, the Hopfions are disrupted, and the polarization helical texture begins to form inside the nanoparticle. Achieving a complete unwinding of the polarization lines into a monodomain structure requires an extremely high concentration of free carriers typical for a good metal.  

Another interesting question that has not been thoroughly investigated yet is the emergence of localized charges resulting from impurities, ion vacancies, or surface-reconstruction defects. 
Recent work\,\cite{Govinden2023rev} has highlighted various factors that can contribute to the appearance of the remnant-bound charges at interfaces or within the interior. They include (i) epitaxial strains whose presence can induce stress in the material leading to the alignment of dipoles in specific orientations; 
(ii) interplay of different structural phases which can influence the arrangement of charges and dipoles, contributing to the presence of remnant-bound charges; 
(iii) the chemical composition of the ferroelectric materials, such as in PbTiO$_3$ or Ti-rich PbZr$_x$Ti$_{1-x}$O$_3$ solid solutions, favoring the tetragonal orientation of dipoles; and 
(iv) the partial screening of bound charges or the presence of a symmetry-breaking electric field (e.g., built-in bias) which can favor the Ising-like symmetry, where the dipoles point ``up" and ``down," preventing complete rotation of interfacial dipoles.
To minimize the electrostatic energy, polarization dipoles should align themselves to counterbalance the electric fields generated by these charges. This alignment naturally brings in the source and sink sets into our consideration of the effects of the divergenceless polarization field. While a small number of localized charges may not significantly disrupt the long-range polarization topological states, a high concentration of such charges can potentially disrupt this ordering.

To conclude here, the emergence of volume and surface charges, either of intrinsic, external, or electrode-induced origin, breaks down the condition for the field to be divergenceless, violating the conditions for the Arnold theorem. However, in realistic systems where the charge concentration is not extensively high, this breakdown only slightly changes the texture of the polarization field, hence being a perturbation. Such a perturbation does not influence the topological states but can result in the local re-commuting of polarization streamlines. This is another top aspect of the topology of ferroelectrics that is attractive to study.

\medskip

\noindent c)\,\textit{Beyond Arnold theorem}

It appears that the application of topological hydrodynamics to nanostructured ferroelectrics offers possibilities for extending and generalizing Arnold theorem. One of the conditions of the Arnold theorem is that the manifold, confining the divergenceless vector flux, is bounded and the polarization streamlines can not penetrate outside the confining surface. 
As a result, the vector field remains tangential to the surface, while any normal-to-the-surface component of the field would imply a nonzero divergence at the surfaces. In the case of ferroelectric nanoparticles in a vacuum (in which the polarization is zero), 
this implies the absence of the appreciable surface bound charges that would cost substantial energy. 
However, in the case where the ferroelectric is embedded into the polarizable dielectric medium, the polarization flux can overflow from the nanoparticle to the outside, making the nanoparticle surface semi-transparent.  The characteristic example of such overflow can be seen in ferroelectric/dielectric heterostructures, such as the PbTiO$_3$/SrTiO$_3$ superlattices. In these structures, one observes the slight escape of the polarization field of the periodic ferroelectric domains into the dielectric layers\,\cite{Lukyanchuk2009,Zubko2012,Yadav2016} (we describe such systems in more detail in\ref{sec:Films} below). 
This phenomenon appears as the fringing of the polar vortices beyond the boundaries of the ferroelectric layer. 
Such spreading out of the polarization field reduces the gradient energy of confined vortices, pushing out the antivortex-type singularities in the points where the interfaces of two oppositely winding vortices meet the ferroelectric surface. However, the additional polarization of the dielectric space costs a substantial energy. Accordingly, the amount of the emerging flux is small and does not exceed 8-10$\%$ of the total flux confined inside the ferroelectric layer.  
\markertwo{Notably, in the absence of free charges, the normal component of the electric displacement field $\mathbf{D}=\mathbf{P}+\varepsilon_0\mathbf{E}$ must be preserved at the ferroelectric-dielectric interface\,\cite{Landau8,Stengel2009}. In the case of the vanishing surface bound charges $\sigma$ this implies the continuity of the polarization flux through the interface. If some small surface bound charges yet emerge, they would produce the oppositely oriented normal field component, hence a correspondingly small discontinuity in the normal component of the polarization, $\Delta P_n = -\sigma$.}
 We foresee the possible escape of the polarization lines from the Hopfion-like textures in ferroelectric nanoparticles if one puts them into the high-$\varepsilon$ dielectric media. This flux emergence would help to avoid the pole singularities of the polarization field. However, this effect, again, is expected to be small. 

Although the effects of the polarization field emerging beyond the ferroelectric nanospecies are small and, according to our approach, can be considered perturbatively, they can also be included in the general topological consideration of Arnold hydrodynamics. 
In fact, the outcoming polarization streamlines are continuous (or slightly discontinuous in the case of the small bound charges) at the paraelectric-ferroelectric interfaces and are localized within the narrow layer close to them. The polarization vanishes when we are moving away from the interface. Otherwise, polarizing the external media would cost substantial energy. Placing the bounding surface slightly outside of the ferroelectric, where the induced polarization is zero, we satisfy the condition of the zero tangential polarization and come back to the Arnold classification of the topological states.    

We provide the state-of-the-art description of the available experimental ferroelectric systems carrying the topological chirality in Section\,\ref{sec:Sources}. Notably, certain geometries do not permit the complete bounding of ferroelectric media. Hence for these systems, the confinement condition of Arnold theorem is not fully satisfied.
Here, we give examples of how the hydrodynamic approach can be extended over more complex geometries. 
First, we mention the thin-film geometry in which the polarization texture is confined inside the ferroelectric layer by two planar interfaces. Such confinement is incomplete since the system propagates to infinity in transversal in-plane directions. In most cases, the topological states emerging in ferroelectric films present the periodic in-plane structures, vortices, or bubbles, which enable the restriction of their coordinate space to the cells with the imposed periodic boundary conditions. 
Topologically, such a configuration is equivalent to the quasi-2D toroidal shell, an exemplary formation whose topology differs from the considered so far topology of the confined nanoparticle by having multiconnectivity. 

\markertwo{Furthermore, in ferroelectric/paraelectric superlattices, the emerging polarization structures in ferroelectric layers may start to interact with each other when the thickness of the ferroelectric/paraelectric layers becomes comparable to or smaller than the period of the polarization domain structure, in particular, the period of the bubble arrays or vortex structures\,\cite{Stephanovich2003, Stephanovich2005}. In such a strong-coupling regime, the polarization flux of domains not only escapes from the original ferroelectric layer but also enters the next ferroelectric layer, propagating from one layer to another. The coherent polarization periodicity in the third, out-of-plane direction, may be established. In the limiting case of an ultrathin paraelectric layer, the domain period diverges leading to the appearance of the monodomain state. }

The established periodicity in all three directions corresponds to that of the 3D torus, which is another remarkable example of a multiconnected manifold. The coordinate space of nanorods and nanowires in which the ferroelectric is confined in two lateral directions of nanorods or nanowires but propagates along their axes, can be also mapped on the multiconnected manyfold of the solid torus provided the polarization structure is periodic along the nanorod or nanowire. 

The above examples demonstrate the possibility of the profound relation between the connectivity of the bounded system and the structure of the emerging topological states. For instance, the periodic chiral chain of coupled polarization vortices was simulated in nanorods of PbZr$_{0.5}$Ti$_{0.5}$0$_3$\,\cite{Naumov2004}; we discuss this structure in detail in Section\,\ref{sec:Nanorods}. Such a structure, appropriate to the solid torus topology of a nanorod, cannot be present as a combination of the independent vortex and Hopfion cells, appropriate for the single-connected bounded coordinate space of the system. 

This poses the challenge for the corresponding extension of topological hydrodynamics \markertwo{and requires the generalization of the Arnold topological approach, in particular, the notion of helicity} since Arnold theorem in its original formulation does not apply to multiconnected systems. \markertwo{This may lead to important consequences for the classification of the topological states in multiconnected systems which may go beyond the vortex and Hopfion states, as shows the quoted above example of the chiral chain in toroidal topology discovered in\,\cite{Naumov2004} }.

Another exciting possibility is extending the topological hydrodynamic approach onto the spatially nonuniform systems, like labyrinth structures emerging in ferroelectric heterostructures, see Section\,\ref{sec:Films}, composites and disordered bulk ferroelectrics, see Section\,\ref{sec:Composites}. Although in these systems, the coordinate space of a ferroelectric material is not bound, certain geometrical constraints are imposed. For example, the ferroelectric nanocomposite material is presented as an ensemble of ferroelectric nanoparticles that may touch each other. Then the polarization field, confined within each nanoparticle, may flow between the contacting nanoparticles through the hot spots of their junctions, forming the chiral helical streams. The topological states inside nanoparticles are similar to those emerging in a single nanoparticle placed in the electric field, which, like polarization, penetrates inside this nanoparticle through the hot spots at the poles, see Section\,\ref{sec:Switching}. The polarization texture in a nanoparticle splits into two chiral topological states, the localized Hopfion, not transferring the flux between nanoparticles, and the flux-transferring helical vortex, traversing the nanoparticle and penetrating them via the hot spots. This helical vortex either pierces or envelopes the Hopfion, assuming a configuration that is not covered by Arnold theorem.   

Exploring the formation of the long-range disordered vortex-tube textures in ferroelectric heterostructures presents yet another possibility for the development of topological hydrodynamics. In fact, because of the unbounded planar space such textures may not necessarily follow Arnold theorem but present more elaborated topological states. A recent report \cite{Rijal2023} has established a connection between the topology of vortex tube patterns in ultra-thin films of PbZr$_{0.4}$Ti$_{0.6}$O$_3$ and the Klein bottle topology of the corresponding order parameter space but does not fully explain the local structure of the texture dislocations. It is also interesting to understand the evolution of such a system under a biased field studied in \cite{Nahas2020} from the hydrodynamic viewpoint.

}

\newcommand\teeonehalf{\mkern-2mu 1\mkern-1.5mu/2}
\begin{table}[t!]
 \caption{Space symmetry groups of chiral ferroelectrics in the polar (ferroelectrics) and non-polar (paraelectric) phases and transition temperatures. The symmetry groups of the chiral states are in bold.  The  data is collected from\marker{\,\cite{LandoldtOx,LandoldtNonOx,Long2020,Fabry2022}}}
{\footnotesize
\centering
\renewcommand{\arraystretch}{1.2}
\begin{tabular}{ c c l c l H } 
\hline \hline
  {} 
& {} 
& Polar phase
& $T_{c}$,$^\circ$C 
& Non-polar phase
& Type \\
\hline 
\parbox[t]{3mm}{\multirow{14}{*}{\rotatebox[origin=c]{90}{Polarization-induced  chirality}}}
& (NH$_{2}$CH$_{2}$COOH)$_{3}\cdot$ 
&  \fullbold{P2$_{1}$ (C$_{2}^{2}$)}
&  49 
&  P2$_{1}$/m (C$_{2h}^{2}$)
&  \textbf{TGS} \\
&  \multicolumn{1}{r|}{H$_{2}$SO$_{4}$}  &&&& \\
&  SrTeO$_{3}$
&  \fullbold{C2   \   (C$_{2}^{3}$)}
&  485 
&  C2/m (C$_{2h}^{3}$) 
&  {} \\
&  NH$_{4}$HSO$_{4}$ 
&  {Pc   \ \  (C$_{s}^{2}$)}
& -3 
&  P2$_{1}$/c  (C$_{2h}^{5}$) 
&  {} \\ 
&  {}
&  \fullbold{P1 \ (C$_{1}^{1}$)}
& -119 
&  {}
&  {} \\
&  NaH$_{3}$(SeO$_{3}$)$_{2}$ 
&  \fullbold{P1 \ (C$_{1}^{1}$)}
& -79 
&  P2$_{1}$/a (C$_{2h}^{5}$) 
&  one more PT to unknown polar phase at -162C p178\\
&  Ca$_{2}$B$_{6}$O$_{11}\cdot$5H$_{2}$O 
&  \fullbold{P2$_{1}$ (C$_{2}^{2}$)}
& -7 
&  P2$_{1}$/a (C$_{2h}^{5}$) 
&  {} \\
&  CH$_{2}$ClCOONH$_{4}$ 
&  \fullbold{C2 \ (C$_{2}^{3}$)}
& -150 
&  C2/c (C$_{2h}^{6}$) 
&  {} \\
& KLiSO$_{4}$ 
& \fullbold{P6$_{3}$ (C$_{6}^{6}$)}
& 708 
& Pmcn (D$_{2h}^{16}$) 
& ? \\
& Pb$_{5}$Ge$_{3}$O$_{11}$ 
& \fullbold{P3  \ (C$_{3}^{1}$)}
& 177 
& P$\overline{6}$ (C$_{3h}^{1}$) 
& \textbf{PGO}  \\
& RbNO$_{3}$ 
& \fullbold{P3$_{1}$  (C$_{2}^{3}$)} 
& 164 
&  \markertwo{Pm$\overline{3}$m (O$_{h}^{1}$)}
& 2 phase transitions at 219 and 291C p64 \\
& CsNO$_{3}$ 
& \fullbold{P3$_{1}$  (C$_{2}^{3}$)}
& 154 
& \markertwo{ Pm$\overline{3}$m  (O$_{h}^{1}$)}
& {} \\
& La$_{2}$Ti$_{2}$O$_{7}$ 
& \fullbold{P2$_{1}$ (C$_{2}^{2}$)}
& 1500 
& Fd$\overline{3}$m (O$_{h}^{7}$) 
& Fd3m was found from papers ? \\
& Nd$_{2}$Ti$_{2}$O$_{7}$ 
& \fullbold{P2$_{1}$ (C$_{2}^{2}$)} 
& $>$1500 
& Fd$\overline{3}$m \ \ (O$_{h}^{7}$) 
& Fd3m was found from papers ? \\
\\
\parbox[t]{3mm}{\multirow{14}{*}{\rotatebox[origin=c]{90}{Structural built-in chirality}}}
& NaKC$_{4}$H$_{4}$O$_{6}\cdot$4H$_{2}$O 
& \fullbold{P2$_{1}$ (C$_{2}^{2}$)}
& 24 
& \fullbold{P2$_{1}$2$_{1}$2 (D$_{2}^{3}$)}
& \textbf{Rochelle}  one more PT to D2 nonpolar P phase at -18C p253 \\
&  NaNH$_{4}$SeO$_{4}\cdot$2H$_{2}$O 
&  \fullbold{P2$_{1}$ (C$_{2}^{2}$)}
& -93 
& \fullbold{P2$_{1}$2$_{1}$2$_{1}$ (D$_{2}^{4}$)}
&  {} \\
&  RbH$_{3}$(SeO$_{3}$)$_{2}$ 
&  \fullbold{P2$_{1}$ (C$_{2}^{2}$)}
& -120 
&  \fullbold{P2$_{1}$2$_{1}$2$_{1}$ (D$_{2}^{4}$)}
&  {} \\
& Ca$_{2}$Sr(CH$_{3}$CH$_{2}$COO)$_{6}$ 
& \fullbold{P4$_{1}$ (C$_{4}^{2}$)}
& 10 
& \fullbold{P4$_{1}$2$_{1}$2 (D$_{4}^{4}$)} 
&  \textbf{DSP}  one more PT to unknown phase at -169C p217 \\
& BaAl$_{2}$O$_{3}$ 
& \fullbold{P6$_{3}$ (C$_{6}^{6}$)}
& 123 
& \fullbold{P6$_{3}$22 (D$_{6}^{6}$)}
& {} \\
&  Rb$_{2}$Cd$_{2}$(SO$_{4}$)$_{3}$ 
&  \fullbold{P2$_{1}$ (C$_{2}^{2}$)}
& -144 
&  \fullbold{P2$_{1}$3  (T$^{4}$)} 
&  {}\\
&  {}
&  \fullbold{P1 \ (C$_{1}^{1}$)}
& -170 
&  {}
&  {}  \\
&  Tl$_{2}$Cd$_{2}$(SO$_{4}$)$_{3}$ 
&  \fullbold{P2$_{1}$ (C$_{2}^{2}$)}
& -145 
&  \fullbold{P2$_{1}$3  (T$^{4}$)} 
&  {} \\
&  {}
&  \fullbold{P1 \ (C$_{1}^{1}$)}
& -153 
&  {}
& one more PT to D2 nonpolar phase at -181C p148 \\
&  CH$_{3}$NH$_{3}$Al(SO$_{4}$)$_{2}\cdot$
&  \fullbold{P2$_{1}$ (C$_{2}^{2}$)}
& -96 
&  \fullbold{P2$_{1}$3 (T$^{4}$)}
&  {} \\
& \multicolumn{1}{r|}{12H$_{2}$O} &&&& \\
& R-3-FP-MnCl$_{3}$ 
& \fullbold{P2$_{1}$ (C$_{2}^{2}$)}
& 90 
& \fullbold{C222$_{1}$ (D$_{2}^{5}$)} 
& \textbf{HOIP 1D} 1D (ref.Long2020) (R)-3-(fluoropyrrolidinium)MnCl3 \\
& (R-CMBA)$_{2}$PbI$_{4}$ 
& \fullbold{P1 \ (C$_{1}^{1}$)}
& 210 
& \fullbold{P422 (D$_{4}^{1}$)}
& \textbf{HOIP 2D} 2D (ref.Long2020) [R-1-(4-chloro-phenyl)ethylammonium]2PbI4 \\
& MDABCO-NH$_{4}$I$_{3}$ 
& \fullbold{R3 \ (C$_{3}^{4}$)}
& 175 
& \fullbold{P432  (O$^{1}$)} 
& \textbf{HOIP 3D} 3D (ref.Ye2018) MDABCO (N-methyl-N'-diazabicyclo[2.2.2]octonium)–ammonium triiodide \\ 
\\
\parbox[t]{3mm}{\multirow{4}{*}{\rotatebox[origin=c]{90}{Undefined}}}
& H(CH$_{3}$)$_{4}\cdot$ HgCl$_{3}$ 
& \fullbold{P2$_{1}$ (C$_{2}^{2}$)}
& {} 
& {?} 
& {} \\
& H(CH$_{3}$)$_{4}\cdot$ HgBr$_{3}$ 
& \fullbold{P2$_{1}$ (C$_{2}^{2}$)}
& {} 
& {?}
& {}  \\
& (NH$_{2}$ CH$_{2}$ COOH)$_{2} \cdot$ 
& \fullbold{P2$_{1}$ (C$_{2}^{2}$)}
& {} 
& {?} 
& {} \\
& \multicolumn{1}{r|}{MnCl$_{2}\cdot $2H$_{2}$O} &&&& \\
\hline \hline
\end{tabular}
}
\end{table}

\begin{figure*}
\begin{center}
\includegraphics [width=0.9\linewidth] {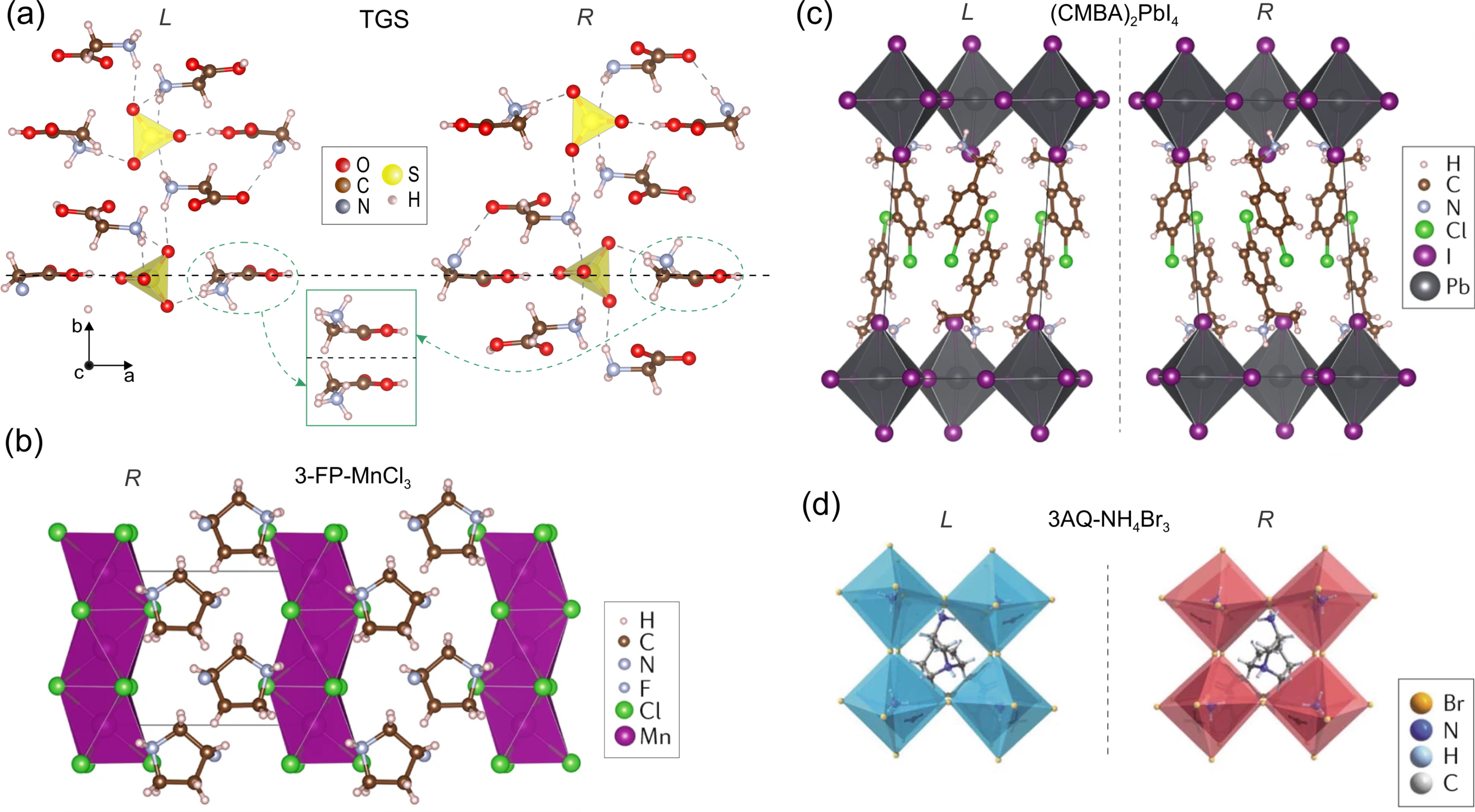}
\caption{ \marker{\textbf{Typical chiral ferroelectrics.}}
\textbf{(a)}, The left-handed (L) and the right-handed (R) enantiomers of the TGS, (NH$_2$CH$_2$COOH)$_3$H$_2$SO$_4$, in the ferroelectric phase. The dashed line shows the mirror plane, transforming the L-TGS to the R-TGS. Inset shows the L and R glycine groups, providing the handedness to the whole system. 
\textbf{(b)}, The R enantiomer of a quasi-1D chiral hybrid organic-inorganic perovskite HOIP 3-FP-MnCl$_3$ in the ferroelectric phase. 
\textbf{(c)}, The L and R enantiomers of a quasi-2D chiral HOIP (CMBA)$_2$PbI$_4$ in the ferroelectric phase. 
\textbf{(d)}, The  L and R enantiomers of a 3D chiral  HOIP 3AQ-NH$_4$-Br$_3$ in the ferroelectric phase.  
Panels (b)-(d) are from Ref.\,\cite{Long2020}. 
}
\label{FigMaterials}
\end{center}
\end{figure*}

\section{\label{sec:ExtTheor} Experiment and theory}

\subsection{\label{sec:Sources}Sources of chirality in ferroelectrics}

\subsubsection{\label{sec:Struct}Structural and polarization-induced chirality}

Remarkably, the very first, discovered in 1920, ferroelectric, the \textbf{Rochelle salt} (NaKC$_{4}$H$_{4}$O$_{6}\cdot$4H$_{2}$O) belongs in the enantiomorphs family. 
This fact did not call for any curious attention at that time. 
At present, the family of chiral ferroelectrics, whose symmetry is described by the Sohncke symmetry groups, see Sec.\,\ref{sec:symetry}, Fig.\,\ref{FigNature}(i), contains several dozens of materials, including quite known ferroelectrics like triglycine sulfate (\textbf{TGS}), NH$_{2}$CH$_{2}$COOH)$_{3}\cdot$H$_{2}$SO$_{4}$ (See Fig.\,\ref{FigMaterials}(a)), lead germanate (\textbf{PGO}) Pb$_5$Ge$_3$O$_{11}$, and dicalcium strontium propionate (\textbf{DSP})  Ca$_{2}$Sr(CH$_{3}$CH$_{2}$COO)$_{6}$. 
The most typical chiral ferroelectrics are listed in Table\,{IV}, where we indicated the symmetry group of the characteristic chiral ferroelectric phase, the symmetry group of the original high-temperature paraelectric phase, and the temperature of the phase transition between them. The  data is collected from the reference book\,\cite{LandoldtOx,LandoldtNonOx} and also from\marker{\,\cite{Long2020,Fabry2022}}. The symmetry groups of the chiral states are highlighted in bold. 


The past two decades have been marked by findings and intensive development of novel chiral ferroelectric materials\,\cite{Long2020} based on the family of the chiral hybrid organic-inorganic perovskites (\textbf{HOIP})\,\cite{Dong2019,Zheng2023,Manzi2023}. The HOIPs have gained explosive attention as one of the most promising platforms for optoelectronics and related technological developments. The HOIPs are characterized by incorporating chiral organic molecules into the perovskite materials, the perovskite corner-sharing anyon octahedra BX$_6$ with B being a metal or NH$_4$ group, and X=Cl, I, or Br, can form either quasi-1D chain-like, or quasi-2D layered, or 3D bulk structures. 
The first ferroelectrics based on the HOIPs have been synthesized in 2018\,\cite{Ye2018}, and are now constituting an advanced group of this family. 
Reported were the sets of ferroelectric chiral perovskites, ranging from the quasi-1D hybrid organic-inorganic manganese perovskites through quasi-2D hybrid organic-inorganic lead perovskites to 3D metal-free perovskites\,\cite{Dong2019} The typical representatives of the 1D, 2D, and 3D families of the HOIP ferroelectrics are shown in Fig.\,\ref{FigMaterials}(b-d) respectively. 

The chiral ferroelectrics are divided into two classes as illustrated in Table\,IV. 

In the ferroelectrics of the first class, which we refer to as ferroelectrics with polarization-induced chirality, the high-temperature non-polar paraelectric phase is achiral. The spontaneous emergence of the chirality takes place upon the transition to the low-temperature ferroelectric state, together with the spontaneous emergence of the polarization. The handedness of the system is related to the orientation of polarization and can be switched by reversing the polarization by the applied field, as we discuss in Sec.\,\ref{sec:Switching}.

 In the ferroelectrics of the second type, which we refer to as ferroelectrics with the structural built-in chirality, the high-temperature non-polar paraelectric phase is chiral, and the handedness of the ferroelectric state is predefined, and can not be switched.  Quite remarkably, this built-in structural chirality can induce novel unconventional modulated polarization states, which we discuss in section\,\ref{sec:Blue}. 

\medskip

\subsubsection{\label{sec:DW}Bloch domain walls}

\begin{figure}
\begin{center}
\includegraphics [width=1\linewidth] {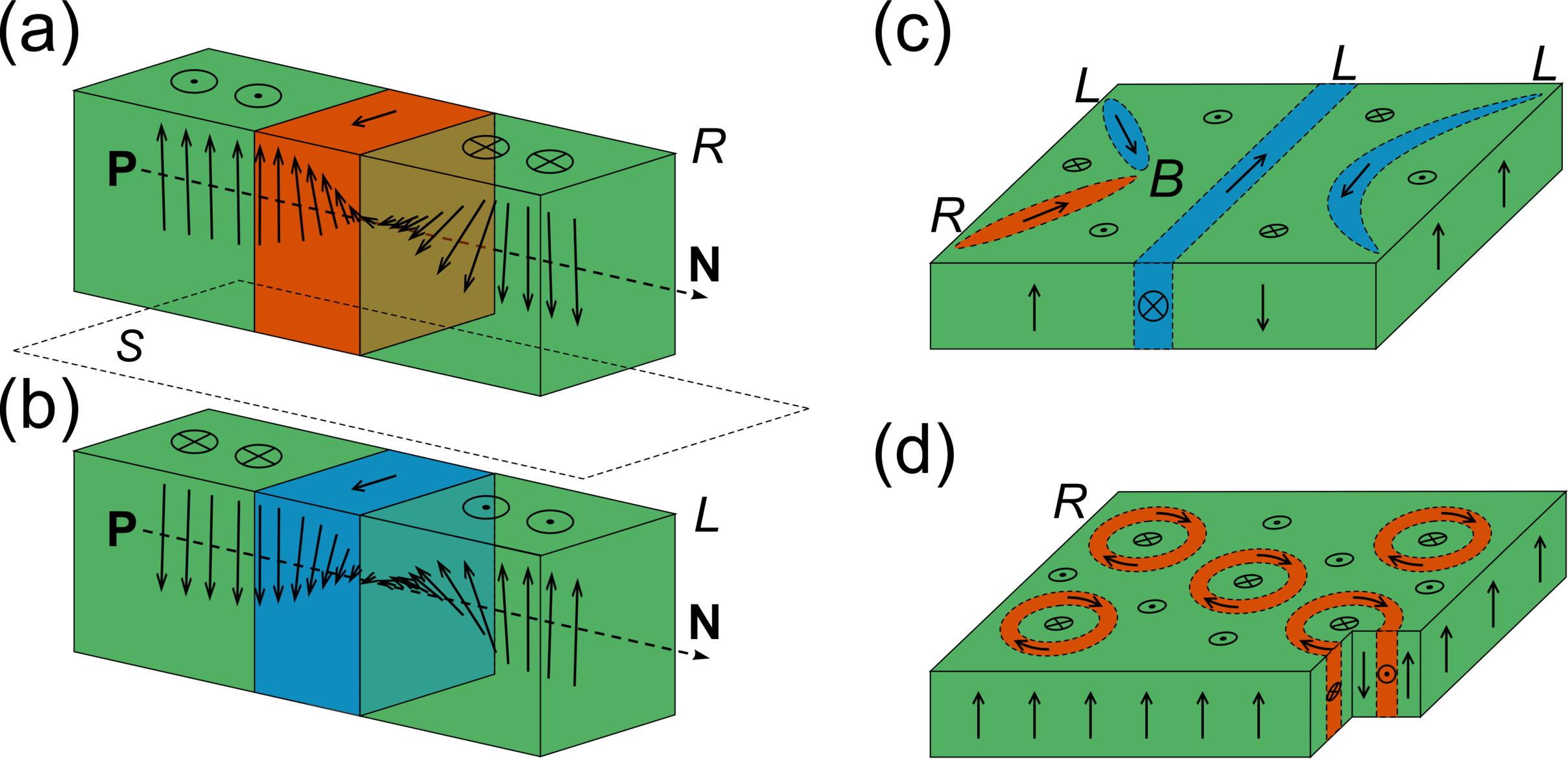}
\caption{ \textbf{Bloch domain walls.}  
\textbf{(a)}, The ferroelectric polarization inside the right-handed  Bloch domain wall (DW) rotates clockwise around the vector $\mathbf{N}$, normal to the wall, remaining directed along the wall plane when going across the thickness of the wall.
\textbf{(b)}, The same for the left-handed Bloch domain wall with the counterclockwise rotation. The left-handed DW is obtained from the right-handed DW by mirror imaging in the plane $S$. %
\textbf{(c)}, The chiral DWs can have straight or bent configurations. Junction of the DWs with different handedness, occurring at the Bloch lines (B), results in the change of the direction of the polarization vector rotation in a wall.
\textbf{(d)}, Looping of the chiral Bloch DWs into the rings results in the formation of the skyrmion tubes. 
}
\label{ChiralDW}
\end{center}
\end{figure}

One of the sources of handedness in ferroelectrics is the chiral polarization twist inside the so-called Bloch domain walls where the vector of polarization, changing its orientation from ``up" to ``down," rotates around the axis perpendicular to the domain wall plane instead of vanishing at the domain wall center, see Fig.\,\ref{ChiralDW}(a). 
The ferroelectric state with the Bloch wall is chiral since it cannot be superimposed with its mirrored image (Fig.\,\ref{ChiralDW}(b)) after any space rotation transformations. 
Accordingly, the symmetrical non-equivalent domain wall textures with the clock- and counterclockwise polarization vector rotation, formed by mirroring each other, correspond to the right-hand and left-hand chiral states. 

Multiple chiral domain walls may pierce the ferroelectric crystal providing an average chirality of the material. 
They can be either planar or form the chiral cylindrical domains, see Fig.\,\ref{ChiralDW}(c),(d). \marker{The small chiral cylindrical domains are the skyrmion tubes. They may be induced, for instance, by the electrical pulses applied to thin films\,\cite{Baudry2011} or nanodots\,\cite{Sene2011} or may emerge at the interface of the ferroelectric nanowire embedded into another ferroelectric media\,\cite{Nahas2015}}.    

Bloch domain walls have a predefined chirality only in ferroelectric materials possessing an intrinsic handedness, see Section\,\ref{sec:Struct}. In this case, their chirality is provided by the specific Dzyaloshinskii-Moriya interaction, as described in Section\,\ref{sec:Blue}. In the more common case of the achiral ferroelectric materials, the handedness of the Bloch domain walls is random, and special care should be taken to ensure their chiral coherence, see Section\,\ref{sec:Thermal}. 
Moreover, the chirality can change within the same domain wall by changing the direction of the polarization rotation, see Fig.\,\ref{ChiralDW}(c). At the loci where the chiralities of the opposite sign meet, a linear topological defect, the so-called Bloch line arises. It is the dynamics of the Bloch lines that eventually control the chirality switch in the Bloch domain wall. 

Bloch domain walls and Bloch lines are a common subject in magnetic systems\,\cite{Malozemoff2016}, but for a ferroelectric, 
they have been remaining intangible for quite a long time. Their existence in ferroelectrics was predicted in a number of theoretical works\,\cite{Tagantsev2001, Lee2009,Wojdel2014}. 
A nice pattern of Bloch domain walls, interspersing by Bloch lines, called as antiskyrmion, was predicted for the hexagonal domain in BaTiO$_3$ in the recent work\,\cite{Gonccalves2023}.  
The switchable Bloch lines were suggested for use as a functional element of memory devices\,\cite{Salje2014}.
The experimental evidence for the Bloch domain walls and Bloch lines in the periodically poled LiTaO$_3$ has been recently reported using the  second-harmonic generation depth-resolved microscopy\,\cite{Cherifi2017}. Also, the concomitant observation of antiferromagnetic and ferroelectric chiral textures at domain walls in the multiferroic material BiFeO$_3$ was reported in work \,\cite{Chauleau2020}.

\medskip
\subsubsection{\label{sec:Blue} Chiral modulated phases induced by Dzyaloshinskii-Moriya interaction}
\marker{

We consider here the source of chirality related to the specific anisotropic interaction of the atomic units that is explosively studied in magnetic systems but still remains elusive for ferroelectrics. 
In magnetism, such an interaction is known as the Dzyaloshinskii-Moriya interaction (DMI)\,\cite{Dzyaloshinsky1958, Moriya1960} and is provided by the antisymmetric exchange between the two neighbouring magnetic spins emerging in the non-centrosymmetric crystals. 
As a result, the magnetization forms different types of long-range nonuniform structures, including the skyrmions, merons, and domain walls, described above in Sections\,\ref{sec:TopApproach} and \,\ref{sec:DW}. 
In what follows we explore the description of the DMI-induced chiral phases in magnetism evolving it for ferroelectrics.

\begin{figure*}
\begin{center}
\includegraphics [width=0.8\linewidth] {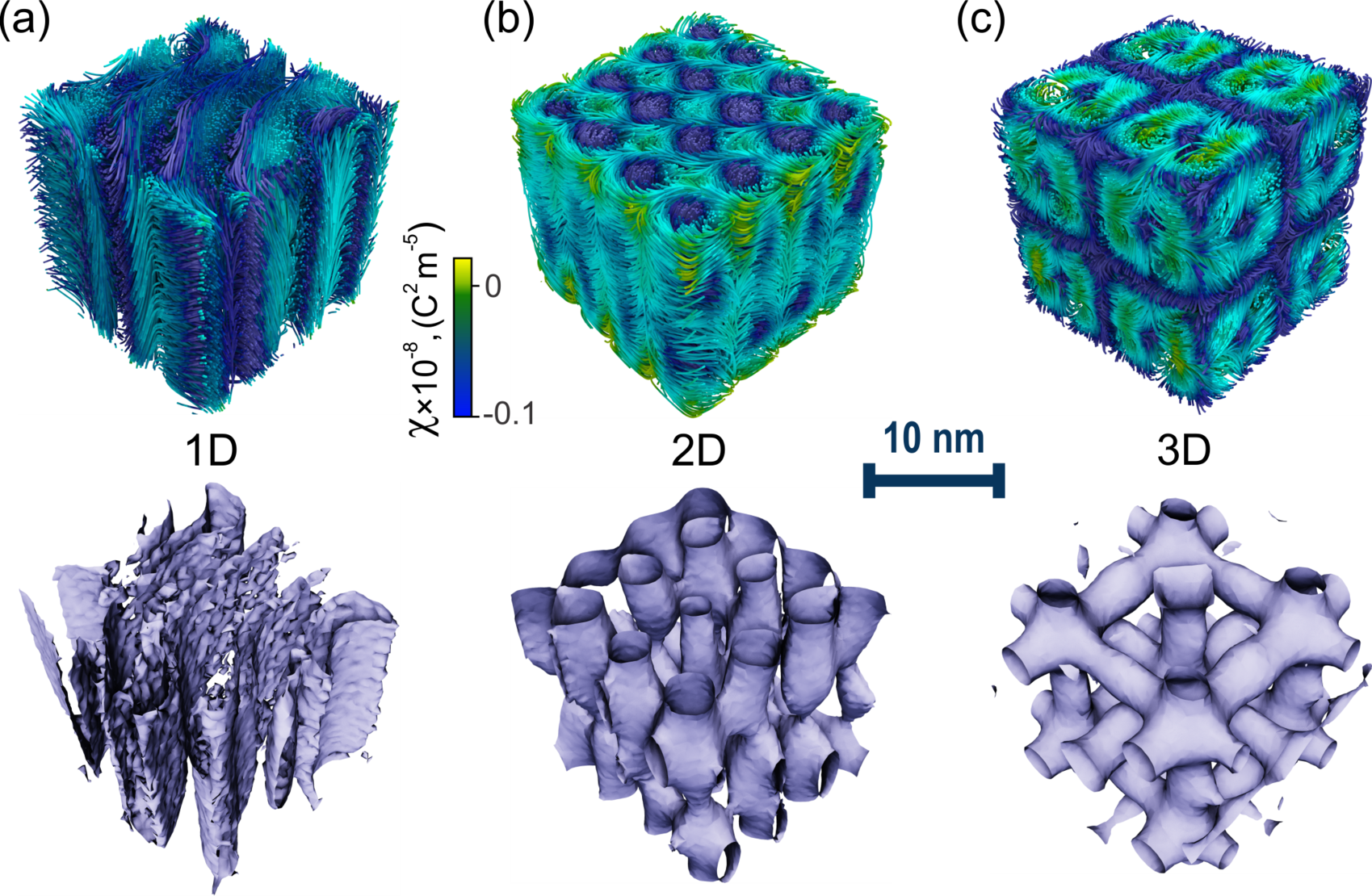}
\caption{ \textbf{The structure of the modulated phases in chiral cubic ferroelectrics.} 
\textbf{(a)}, \marker{ The 1D-modulated helicoidal phase. 
\textbf{(b)}, The double-periodic 2D-modulated phase of meron tubes.
\textbf{(c)}, The 3D-modulated phase.
The upper panels demonstrate the distribution of polarization streamlines and chirality as color maps. The lower panels visualize the spatial structure of the phases, presenting the iso-surfaces of the polarization magnitude $P=|\mathbf{P}|$, normalized on its maximum value $P_{m}$,  and taken at   $P/P_{m}=$0.8, 0.7  and 0.5 for 1D, 2D, and 3D phases respectively. The details of simulations are given in\,(\citeyear{SM}).}  
}
\label{FigBlue}
\end{center}
\end{figure*}

In the continuum limit, the contribution of the DMI to the energy of the system is characterized by the Lifshitz invariants (LI)\,\cite{Landau5} that are composed of the terms linear in spatial derivatives of magnetization, which is the order parameter of the magnetic system and of the bilinear in the order parameter terms. It is the linearity in the derivatives that promotes the instability towards the formation of the long-range modulated polarization textures, also referred to as incommensurate phases of type I\,\cite{BlincBook}. Importantly, because of the bilinearity in order parameter of the LIs, their transformation properties are similar in magnetics and ferroelectrics making these terms equally appropriate for both systems.  In particular, the LIs are possible only for crystal symmetries with no inversion element in both cases. The non-centrosymmetric point groups are listed in Table\,I, and all the enantiomorphic point groups, $C_n$, $D_n$, $T$, and $O$  belong to this family. 

There are three LIs,  
\begin{gather}
{L}^{(x)}=P_y\partial_x P_z-P_z\partial_x P_y, 
\quad
{L}^{(y)}=P_z\partial_y P_x-P_x\partial_y P_z, \nonumber
\\
{L}^{(z)}=P_x\partial_z P_y-P_y\partial_z P_x, 
\label{LI}
\end{gather}
relevant to ferroelectrics with the structural built-in chirality.
 These LIs induce chiral modulated structures, with modulation directed along $x$, $y$, and $z$ directions respectively. The reverse is also true, the LI-induced long-range nonuniform textures with distinct handednesses emerge only in the crystals with the structural build-in chirality. 
 This remarkably agrees with the maxim known as the Neumann-Curie principle: ``The symmetry of the physical phenomenon is at least as high as the crystallographic symmetry"\,\cite{Neumann1885} and ``The symmetries of the causes are to be found in the effects."\,\cite{Curie1894}  

At present, the possibility of the DMI-induced non-uniform states in ferroelectric materials is still discussed. 
Examples of the potential materials with built-in chirality, dicalcium strontium propionate Ca$_2$Sr(CH$_3$CH$_2$COO)$_6$ (group $D_4$) and quinuclidinol (group $D_6$), were suggested by\,\cite{Erb2020}. 
Another example is the ferroelectric RbH$_3$(SeO$_3$)$_2$ possessing the chiral point group $D_4$ in the paraelectric phase.
This compound has the LI-instability\,\cite{Levstik1985} resulting in the incommensurate phase discovered in\,\cite{Gladkii1972}. Recently the vortex domains, that can be the precursors of skyrmions, were discovered in 2D HOIP structures\,\cite{Zhang2020}.
Other potential materials are presented in Table\,IV. In these materials, the rotational degrees of freedom of the polarization vector compete with the crystal anisotropy, whereas the DMI term tends to induce long-range helical instability.

The atomistic simulation of the DMI-like interaction has recently been developed in the works\,\cite{Zhao2021,Chen2022b} where the anisotropic interaction of the off-center dipoles for the cubic perovskite materials was calculated. 
The calculated contourribution has mostly the local character, which unlikely results in the LIs inducing long-range chiral topological states, since the macroscopic symmetry of the system, $O_h$  (m$\overline 3$m), contouraining the inversion center does not belong to any non-centrosymmetric group listed in Table\,I.  Indeed, the simulations in\,\cite{Stengel2023} indicate that the uniform DMI-like forces sum up to zero in SrTiO$_3$, the compound which possesses the same symmetry $O_h$. 
Yet, by exploiting the developed methodology, one can generalize the atomistic DMI consideration to the non-centrosymmetric systems, in particular, to the ferroelectrics with built-in chirality and discover the long-range chiral modulated phases.

The symmetry analysis identifies a relationship between the possible combination of LIs and crystal point groups\,\cite{Bogdanov1989ftt,Ado2020}, identifying, therefore, the possible nonuniform textures. 
In cubic crystals with structural built-in chirality, an exciting realization of various modulated phases may arise.  
Those materials are, for example, the Rb$_{2}$Cd$_{2}$(SO$_{4}$)$_{3}$,  CH$_{3}$NH$_{3}$Al(SO$_{4}$)$_{2}\cdot$12H$_{2}$O, or similar compounds, possessing the T$^4$ (P2$_1$3) symmetry or the 3D chiral HOIPs-based ferroelectrics with the O$^1$ (P432) symmetry, see Table\,IV. 
 
The LI term, appropriate for $T$ and $O$ point groups, has the spherically symmetric form   
\begin{equation}
  T,\,O:\, F_{LI}=D({L}^{(x)}+{L}^{(y)}+{L}^{(z)})=D\,\mathbf{P}\cdot [\mathbf{\nabla}\times \mathbf{P}],
   \label{FDMI}
\end{equation}
where $D$ is the coupling coefficient.
Our original phase-field simulations in the exemplary cubic ferroelectrics with the LI (\ref{FDMI}), see\,(\citeyear{SM}) for detail, reveal chiral nonuniform phases,  characterized by the 1D, 2D, and 3D modulations of the polarization.
Upper panels in Fig.\,\ref{FigBlue} present the distribution of streamlines and chirality in these phases. 

In the 1D-modulated phase, see Fig.\,\ref{FigBlue}a, the polarization distribution has a layered \marker{helicoidal} structure in which the polarization streamlines are oriented parallel to each other inside the layers, but the layers do twist around a perpendicular axis. 
In the 2D-modulated phase, see Fig.\,\ref{FigBlue}b, the polarization streamlines form an array of twisted co-oriented cylinders, having a structure of alternating merons $M^+_{cw}$ and $M^-_{ccw}$.
In the 3D-modulated phase, see Fig.\,\ref{FigBlue}c, the polarization streamlines form a periodic cubic network of twisted cylinders. 

Lower panels  visualize the spatial structure of the phases, \marker{presenting the isosurfaces of the polarization magnitude}
They are the parallel planes in the case of $1D$-modulated phase, the columnar array in the case of $2D$-modulated phase, and the network of the intricate single-connected surfaces, known as a plumber's nightmare\,\cite{Huse1988}, in the case of the $3D$-modulated phase. 

Various structures may, therefore, develop in the bulk of the cubic crystals with structural build-in chirality. They resemble the cholesteric and blue phases in chiral nematic liquid crystals\,\cite{Bahr2001,Chandrasekhar2006,DeGennes1993}, DMI-induced non-uniform phases in magnetic systems\,\cite{Nagaosa2013}, and highly-knotted states in ferromagnetic liquids\,\cite{Ackerman2016}. 

 In crystals  possessing the lower symmetry groups the LI term (\ref{FDMI}) splits accordingly
 \begin{align}
   C_n,\,D_n\,(n>2): \, F_{LI}&=D_1({L}^{(x)}+{L}^{(y)})+D_2{L}^{(z)}) 
  \nonumber \\
   C_1,\,C_2,\,D_2:\qquad F_{LI}&=D_1{L}^{(x)}+D_2{L}^{(y)}+D_3{L}^{(z)},  
   \label{FDMIlow}
\end{align}
and the 1D, 2D, and 3D modulated structures, similar to those shown in Fig.\,\ref{FigBlue} also can emerge, but their structure will be essentially modified by the corresponding crystal anisotropy.  For instance, in the crystals with the uniaxial point symmetry groups $C_n$ and $D_n$ with $n>2$, the polarization can have the longitudinal conical precession around the spontaneous polarization component $\mathbf{P}_0$ directed along the crystal symmetry axis or transverse conical precession in the direction, perpendicular to the spontaneous polarization component.  At the same time, the double-periodic meron- and skyrmion-tube lattices induced by the superposition of modulations in the basal plane and aligned along the crystal symmetry axis can have the rectangular, triangular, square, or hexagonal symmetries of the elementary cells corresponding to $C_n$ and $D_n$ point groups with  $n=$2,3,4 and 6 respectively.
Notably, the point symmetry groups $C_n$ ($n\geq 1$) also accept the non-chiral LI of the type $P_x\partial_x P_z-P_z\partial_x P_x$ and similar\,\cite{Bogdanov1989ftt,Leonov2016}. These invariants may induce the Néel-type component in the emerging skyrmions, merons, and domain walls. We believe, however, that such admixture will be substantially dumped by depolarization forces emerging due to the bound charges, characteristic for the Néel-type structures, see Section\,\ref{sec:TopStates}.

It's worth mentioning that beyond the LIs there may exist more complex gradient coupling terms that exhibit nonlinearity either in the gradients or in the order parameter\,\cite{Ado2020} or involve coupling of the order parameter with other degrees of freedom, such as elasticity\,\cite{Stengel2023}.  These terms also might give rise to modulated phases. Yet, unlike the LI scenario, the non-uniform modulation emerges in a critical way, when the coupling constant exceeds some threshold value. Such modulated phases are referred to as type II incommensurate phases\,\cite{BlincBook}.

Other systems in which the modulated phases can be established are the systems with surface- and interface-induced DMI. The corresponding research which is the focus of attention in magnetism\,\cite{Hellman2017,Jiang2019} can be also relevant for studying thin films and heterostructures in ferroelectrics, and may lead to the discovery of the surface-induced skyrmions in ferroelectric heterostructures. 
In such systems, the surface of the crystal may result in the reduction of the crystal point symmetry group and, in particular, in the removal of the operation of inversion from it, inducing the specific LIs at the surface. 
Furthermore, when the surface-induced DMI disappears while moving into the bulk of the crystal, the surface-induced skyrmions should also disappear. Then, the skyrmion tube starting at the surface-induced skyrmion can soon end in the volume forming the structure, called bobber in magnetism\,\cite{Rybakov2015,Zheng2018}.

Multiferroics, the materials possessing both magnetic and ferroelectric properties, enable a mutual control between magnetism and electricity\,\cite{Cheong2007,Tokura2014,Fiebig2016,Wang2018,Spaldin2019}. 
The violation of the inversion symmetry and the emergence of chirality in magnetic systems induce the corresponding modifications in the symmetry group of ferroelectrics. 
The materials in which the ionic relaxation in the DMI-driven magnetic helicoidal states induces the polar lattice distortions resulting in the ferroelectric polarization are known as multiferroics of type II\,\cite{Khomskii2009}. 
The inverse effect, inducing the chiral polarization modulations by magnetism is yet unknown and is the promising direction of research.  

We have discussed the structurally chiral ferroelectrics and the effects related to their handedness. Now we turn to the topological chirality, the associated topological states, and other related phenomena arising due to the long-range electrostatic forces. Topological chirality is the general phenomenon inherent to all nanostructured ferroelectrics, independent of whether they are intrinsically chiral or not. Importantly, for the discussed above structurally chiral ferroelectrics one can expect the synergy of the intrinsic and topological chiralities. 
}

\subsubsection{\label{sec:Films}Thin films and superlattices}

\begin{figure*}
\begin{center}
\includegraphics [width=1\linewidth] {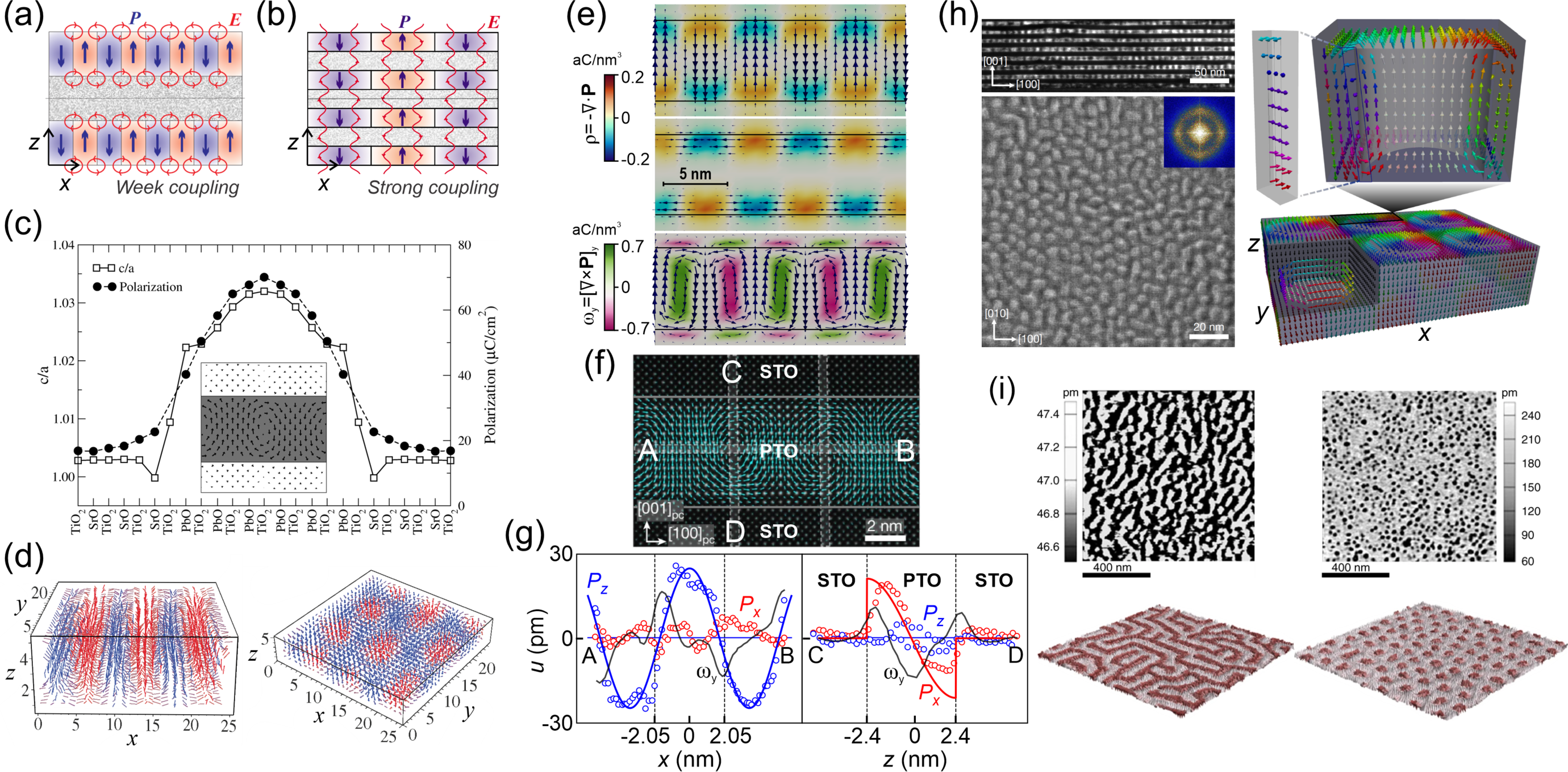}
\caption{ \textbf{Vortex matter in ferroelectric heterostructures.}  
\textbf{(a)}, The side view of the soft polarization domains \markertwo{ (blue arrows) and depolarization fields (red arrows) in ferroelectric layers in ferroelectric/paraelectric superlattice in weak-coupling regime. The smooth distribution of the spontaneous polarization is demonstrated by varying 
the color intensity. } From Ref.\,\cite{Stephanovich2003,Stephanovich2005}.
\textbf{(b)}, \markertwo{The view of the same domain structure as in\,\textbf(a) but in the case of the strong-coupling regime.}
\textbf{(c)}, \marker{The root-mean-squared polarization and average lattice parameter for each (001) layer of a PbTiO$_3$/SrTiO$_3$ superlattice. The inset shows the local dipole magnitudes and orientations.} From\,Ref.\,\cite{Zubko2012}.
\textbf{(d)}, Polarization patterns in strained films of PbZr$_{0.5}$Ti$_{0.5}$O$_3$ demonstrating the stripe vortices and vortex bubbles obtained by the first-principle simulations. From Ref.\,\cite{Kornev2004}.
\textbf{(e)}, A soft-domain distribution of spontaneous polarization in the vertical film plane (top) results in the emergence of the in-plane polarization component, compensating the emerging bound charges. \markertwo{The colour maps show the charge densities, $\rho$, produced by the respective polarization components}. The superposition of the induced polarization and spontaneous polarization yields the vortex polarization texture (bottom). 
\markertwo{The corresponding colour map shows the distribution of vorticity $\omega_y$}.
Adopted from Ref.\,\cite{Lukyanchuk2022} \marker{(Supplementary Information)}.
\marker{
\textbf{(f)}, The polarization vector map in a cross-sectional high-angle angular dark-field imaging STEM image of a PbTiO$_3$ layer embedded within a PbTiO$_3$/SrTiO$_3$ (PTO/STO) superlattice. From Ref.\,\cite{Yadav2019}.
\textbf{(g)}, The experimental values of polar atomic displacements $u$, quantified variations of the components of polarization $P_x$ and $P_z$, measured along the shown at panel \textbf{(f)} horizontal line A–B (left plot), and vertical line C-D (right plot), are presented by the red and blue circles. The bold red and blue lines show the theoretical curves corresponding to the equation\,(\ref{SoftPolar}) \markertwo{The black lines show the variation of the polarization vorticity $\omega_y$ (in arb. units) calculated from the experimental variation of polarization}. Adopted from Ref.\,\cite{Yadav2019}.
}
\textbf{(h)}, The experimentally observed (left) and simulated (right) skyrmion bubbles in SrTiO$_3$/PbTiO$_3$/SrTiO$_3$ trilayer heterostructures. From Ref.\,\cite{Das2019}.
\textbf{(i)}, The experimental (top) and simulated (bottom) dipolar configurations of PbZr$_{0.4}$Ti$_{0.6}$O$_3$ films reveal vortex labyrinth (left) and vortex bubble-skyrmion (right) configurations depending on the sample preparation. From Ref.\,\cite{Nahas2020}.
}
\label{FigVF}
\end{center}
\end{figure*}

The most studied nanostructured ferroelectric systems are the strained ferroelectric thin films and ferroelectric/paraelectric superlattices in which a wealth of topological states was discovered and explored during the past two decades. 
The general state-of-art is well described in recent reviews\,\cite{Zheng2017,Das2018,Ramesh2019,Tian2021,Tan2021,Chen2021,Das2020,Guo2022,Junquera2023}. 

\medskip
a)\textit{ Vortex matter} 

We now understand that the observed richness of nanostructures stems from the interplay between the ferroelectric order and the electrostatic depolarization forces. 
As a result, ferroelectrics split into domains\,\cite{Bratkovsky2000,Stephanovich2003,Stephanovich2005} that are similar to the Landau-Kittel domains in ferromagnetic samples\,\cite{Landau1935,Kittel1946}. 
The specific feature of ferroelectrics is that their domain-forming depolarization forces are much stronger than the demagnetization forces in their magnetic counterparts, see Section\,\ref{sec:Extensions}. The minimization of the associated depolarization energy results in reducing the polarization discontinuity near the sample boundary making the distribution of the spontaneous polarization component inside domains smooth. The corresponding polarization profile was introduced in Ref.\,\cite{DeGuerville2005, Lukyanchuk2009} and called the soft domain structure to highlight its difference from the Landau-Kittel magnetic domain profiles, which are normally flat.

\markertwo{Figures\,\ref{FigVF}(a,b) show the soft domain structure in the ferroelectric layer inside the ferroelectric/paraelectric superlattice. The smooth distribution of the spontaneous polarization components diminishing from the center of the domain to its boundaries is demonstrated by varying the color intensity. 
When the domains are well separated by the paraelectric layers, see Fig.\ref{FigVF}(a), the electrostatic coupling between domain structures in ferroelectric layers is weak. The residual fringing depolarization fields do interact vanishingly weak and are mostly localized within the narrow layer -- having the thickness of the order of the domain structure period -- near the ferroelectric/paraelectric interfaces. When the distance between the ferroelectric layers becomes less than the characteristic fringing field localization length, see Fig.\ref{FigVF}(b), the fringing fields emerging from the adjacent ferroelectric layers start to overlap leading to a strong coupling interaction between the vortex domains in neighboring layers. This regime marks the discussed in Section\,\ref{sec:Extensions}c transition from the planar 2D topology where domains in the ferroelectric layer are weakly coupled to the volumetric 3D topology of the entire system where they are coupled strongly.}

The experiment and simulations\,\cite{Zubko2012} do confirm the reduction of the polarization amplitude upon approaching the surface, see Fig.\,\ref{FigVF}(c).  
The total polarization texture in strained films of PbZr$_{0.5}$Ti$_{0.5}$O$_3$, is visualized by simulations as an array of alternately rotating polarization vortices or vortex bubbles lying in the plane of the ferroelectric layers\,\cite{Kornev2004}, see Fig.\,\ref{FigVF}(d), right and left panels respectively. 

\marker{The emergence of the vortex texture in the ferroelectric layers is easy to understand using the soft domain approach. The upper panel of  Fig.\,\ref{FigVF}(e), based on the simulations of the polarization distribution in PbTiO$_3$/SrTiO$_3$ superlattice\,\cite{Lukyanchuk2022}, (Supplementary Information), demonstrates the smooth distribution of the spontaneous out-of-plane component of polarization inside domains, obtained from the solution of the Ginzburg-Landau equations\,\cite{Stephanovich2003,Stephanovich2005}. 
\markertwo{However, such a distribution still produces the residual bulk bound charges visualized by the color map.  
To diminish the induced depolarization field, the in-plane component of the polarization, producing the compensating charges of the opposite sign emerges, see the middle panel. 
The superposition of both components gives rise to the almost divergence-free total polarization distribution in the form of an array of alternating CW and CCW vortices shown in the lower panel. The colour map of the panel illustrates the distribution of the vorticity $\omega_y=[\nabla \times \mathbf{P}]_y$ of the vortex structure. In the vortex core regions, this parameter alternatively takes positive and negative signs for the CC and CCW vortices respectively. However, in the vicinity of the paraelectric-dielectric interface the sign of $\omega_y$ changes to the opposite one for the same vortex, indicating that vorticity itself does not reflect the direction of the vortex rotation, as we discussed in Section\,\ref{sec:Quantification}}.

The basic calculated distribution of the polarization in the vortex array is 
\begin{equation}
\mathbf{P}_{v} =\mathbf{e}_z\, P_{3}\cos \frac{\pi z}{2a_{f}}\cos \frac{\pi x}{d} 
+ \mathbf{e}_x\, P_{1}\sin \frac{\pi z}{2a_{f}}\sin \frac{\pi x}{d}.
\label{SoftPolar}
\end{equation}%
Here the amplitude of the out-of-plane polarization component, $P_3$, is the maximal spontaneous polarization at the domain center, the amplitude of the in-plane polarization component, $P_{1}=P_{3}({d}/{2a_{f}})$, is taken from the condition $\mathrm{div}\mathbf{P}_v=0$. The parameter $d$ is the half-period of the domain structure and $2a_f$ is the film thickness.

The existence of the vortex arrays in ferroelectric PbTiO$_3$ layers embedded into the PbTiO$_3$/SrTiO$_3$ superlattice grown on the DyScO$_3$ substrate was confirmed experimentally in the seminal work\,\cite{Yadav2016}. Figure\,\ref{FigVF}(f) shows the atomic-scale mapping of the polar atomic displacements in such a system.   These atomic displacements reveal the presence of the long-range ordered arrays of the clock- and anticlockwise stripes of polarization vortices with nearly continuous polarization rotation\,\cite{Yadav2019}. 

The experimentally measured gradual variation of the polarization components along and across the vortex structure shown in Fig.\,\ref{FigVF}(g) well agrees with variation obtained from the expression\,(\ref{SoftPolar}) giving the soft domain profile. Equation\,(\ref{SoftPolar}) presents a convenient approximation of the polarization distribution in the observed vortex arrays without any fitting parameters. \markertwo{Figure\,\ref{FigVF}(g) also demonstrates the distribution of the vorticity of the polarization, $\omega_y$ along and across the vortex structure, obtained from the experimental polarization variation. In accordance with the simulation results, presented in the lower panel of Fig.\,\ref{FigVF}(g), $\omega_y$ changes the sign not only when going along the horizontal $x$-direction from vortex to vortex, but also within the same vortex when going from the vortex core to vortex periphery in the vertical $z$-direction.  }

Notably, at the same time, the internal field, $E_i\approx 2MV/cm$, was measured inside the ferroelectric layers of the same system\,\cite{Yadav2019}. 
Yet, this residual depolarization field, breaking up the condition $\mathrm{div}\mathbf{P}=0$ is very small, see Section\,\ref{sec:Extensions}, and can be accounted for by the less than $0.3\%$ disbalance of coefficients $P_3$ and $P_1$ in Eq,\,(\ref{SoftPolar}).  
Furthermore, there is also no visible escape of the polarization flux outside the layer in the experimental data presented in  Fig.\,\ref{FigVF}(f). This observation is in line with Eq.\,(\ref{SoftPolar}) and supports the topological approach in which the polarization field is confined within the ferroelectric film and is tangent to its surface, see Section\,\ref{sec:Foundations}.

In many simulations\,\cite{Zubko2012,Yadav2016,Pavlenko2022}, a part of the polarization flux does escape from the film forming the fringing field localized in the vicinity of the film surface, see Fig.\,\ref{FigVF}(e). The length of fringing field localization is typically of the order of the domain structure period.
However, as we discussed in Section\,\ref{sec:Extensions}, the relative magnitude of this effect is as small as $(T_c/2C)^{1/2}$$\simeq$0.03-0.05 or even less (here $C$ is a Curie constant and $T_c$ is a critical temperature of the ferroelectric phase transition). The rigorous treatment of this issue was given in Ref.\,\cite{Lukyanchuk2009}.

\markertwo{Simulations also reveal that in ferroelectric/paraelectric superlattices, the polarization streamlines may propagate from one ferroelectric layer to another, thus going through the entire system\,\cite{Aguado-Puente2012,Abid2021}. Meanwhile, a set of parallel antivortices appears in the middle of the paraelectric layer. 
This situation is realized when the distance between ferroelectric layers becomes very small and apparently corresponds to the discussed above strong coupling regime where the emerging paraelectric polarization tends to screen the residual depolarization fields traversing the paraelectric layers.    
The polarization flux leakage between ferroelectric nanospecies was also observed in the system of the compactified nanoparticles hosting polar topological states, see Section\,\ref{sec:Nanorods}.
}}

Another type of an emergent vortex matter is the ferroelectric bubbles envisioned theoretically\,\cite{Kornev2004} (Fig.\,\ref{FigVF}(d), right panel) and found experimentally in the same  PbTiO$_3$/SrTiO$_3$ superlattices but grown on the SrTiO$_3$ substrate\,\cite{Das2019}. It was shown that these bubble domains are vortices rolled up into a bagel, see Fig.\,\ref{FigVF}(h); in \,\cite{Das2019} they were referred to as skyrmions. 
\marker{Similar polar bubbles, revealing a zoo of topologies were recently found in BiFeO$_3$/SrTiO$_3$ superlattices\,\cite{Govinden2023superlat}}.

To reiterate, a salient feature of the ferroelectric topological states extended on the stripe and bubble domains is that the polarization vector at the domain boundary and at the layer surface is almost tangent to the boundary to avoid the appearance of the surface-bound charges which would have induced the huge depolarization fields. 

A plethora of even more sophisticated configurations of vortex matter, including labyrinthine patterns of stripe vortices transforming into disordered arrays of bubbles, was found in PbTiO$_3$/SrTiO$_3$ superlattices on the SrTiO$_3$ substrate\,\cite{Das2019} and also in the ultrathin films of PbZr$_{0.4}$Ti$_{0.6}$O$_3$\,\cite{Zhang2017,Nahas2020,Nahas2020b}, see Fig.\,\ref{FigVF}(i).  \marker{ The labyrinth in domain pattern (upper-left panel) was obtained after annealing the as-grown PbZr$_{0.4}$Ti$_{0.6}$O$_3$ films at 525\,K for 10\,min in air, and cooling it down to room temperature at a cooling rate of 10–15\,K/min. The bubble domains (upper-right panel) were then created by scanning the labyrinth domains using a scanning probe microscopy probe with an AC amplitude of 500\,mV. The low-left and low-right panels show the dipolar configurations obtained from simulations within corresponding labyrinthine and bubble phases.} 

Formation of the complex micro- and nanostructured patterns of domains having different shapes is a phenomenon well known for various physical systems\,\cite{Seul1995}, in particular, for the magnetic\,\cite{Hubert2008}, and superconducting ones\,\cite{Huebener2001}. 
In all cases, a variety of patterned structures arises including stripes, meanders (labyrinth-like patterns), and either ordered or disordered droplet (bubble) arrays. 
A particular realization results from the competition between the different types of structure-forming forces, which often have a non-local long-range character. 
\markertwo{For instance, recent studies brought in phase diagrams revealing different polar patterns under different temperatures, strains, and fields and highlights how elastic and electric forces trigger transitions between the described above vortex, labyrinth, and bubble topological states  \cite{Dai2023,Yuan2023}.}
It is the subtle balance between these forces that results in the observed richness of the polarization topological patterns in ferroelectric heterostructures upon even slight variations of the temperature, geometrical parameters, and elastic strains.  Further analysis of these patterns in ferroelectrics reveals much more fine substructures of the polarization field, which we discuss below.

\begin{figure*}
\centering
\includegraphics [width=0.93\linewidth] {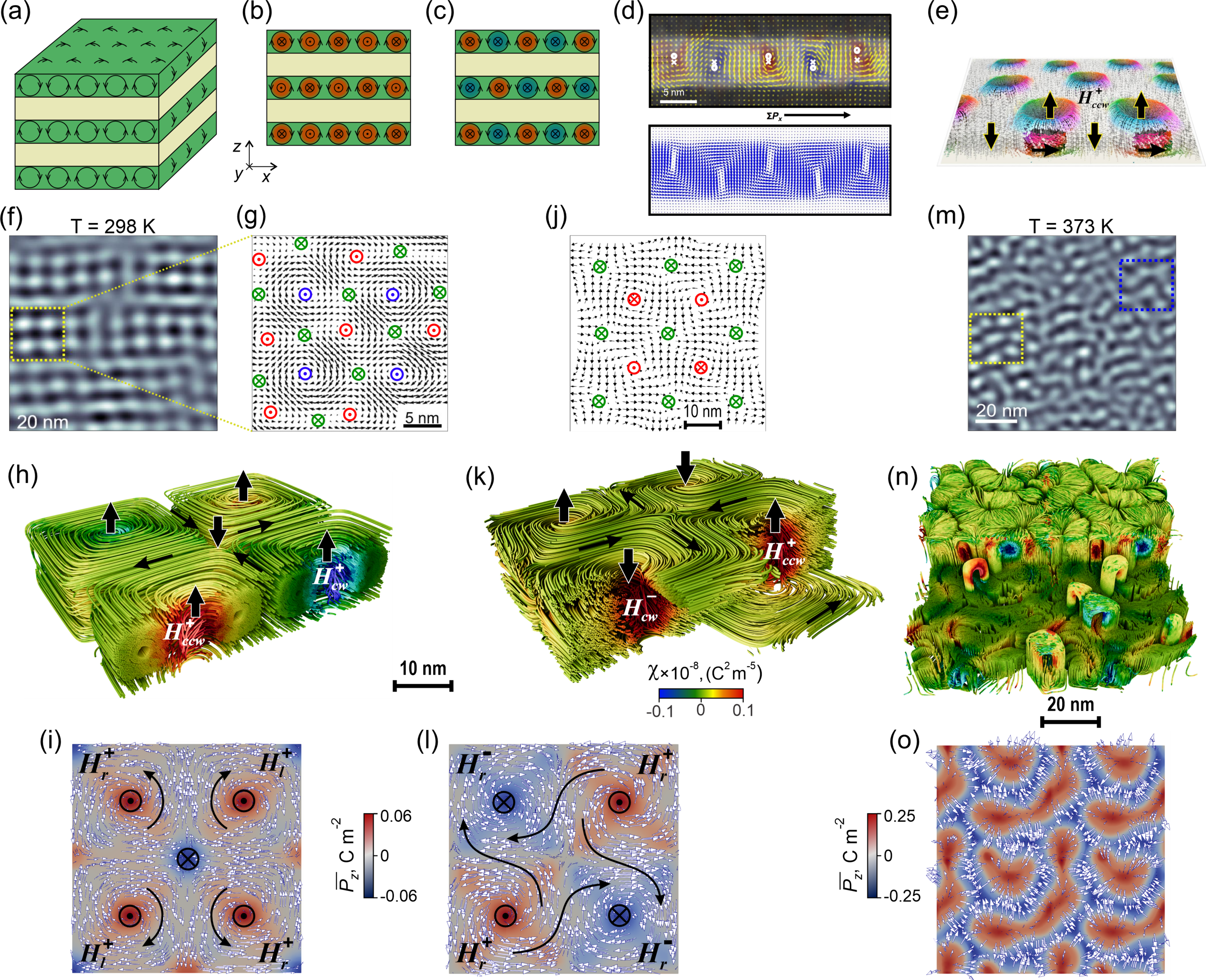}
\caption{ \textbf{Emerging of chirality in ferroelectric heterostructures.} 
\textbf{(a)},\,A periodic array of stripe in-plane vortices with alternatively, clockwise (CW) and counterclockwise (CCW), rotating polarization emerging in heterostructures; the ferroelectric layers with vortices are shown in turquoise and the paraelectric layers in light yellow.
\textbf{(b)},\,The vertical cut of the heterostructure in which vortices have the component along their axis, hence acquiring chirality. The case where the axial component and rotating component jointly alternatively change their direction from vortex to vortex is shown. In such a configuration, all the vortices have the same (right) handedness shown in red-brown.   
\textbf{(c)},\, The same as panel \textbf{(b)} but for vortices with the same direction of the in-plane component. The handedness, in this case, alternatively varies from vortex to vortex, which is indicated by color variations from red-brown (right) to blue (left). 
\textbf{(d)},\,Experimental (top) and simulation (bottom) studies of the heterostructure having vortices with the periodically varying chirality that emerges in the DyScO$_3$ substrate-deposited SrTiO$_3$/PbTiO$_3$/SrTiO$_3$ trilayer. This corresponds to the configuration of the panel \textbf{(c)}. Here, however, the vortices are alternatively displaced up and down, which provides the definite net chirality to the vortex array. From Ref.\,\cite{Behera2022}.
\textbf{(e)},\,The periodic polarization structure of the dilute array of vortex bubbles having the structure of chiral $H_{cw}^{-}$ Hopfions. The picture is adopted from \cite{Das2021} where this structure was experimentally discovered in SrTiO$_3$ substrate-deposited SrTiO$_3$/PbTiO$_3$/SrTiO$_3$ superlattice and referred to as a skyrmion array. 
\textbf{(f)},\,The regular array of vortex bubbles with ascending polarization flux in their cores in the strain-free SrTiO$_3$/PbTiO$_3$ superlattice observed at $T=373$\,K and interpreted in\,\cite{Shao2023} as a meron lattice.
\textbf{(g)},\,Schematic presentation of the polarization vector field distribution, in the bubble array shown in panel \textbf{(f)}. The bubble cores have an alternative, left and right handednesses, shown as the blue and red circles. Green circles correspond to the chiral-neutral points of descending polarization flux. Adopted from Ref.\,\cite{Shao2023}.
\marker{
\textbf{(h)},\,Simulated reconstructed volume distribution of the polarization streamlines for the structure, see\,(\citeyear{SM}), shown in panels  \textbf{(f)} and \textbf{(g)}, indicated by the color map. 
\textbf{(i)},\, \markertwo{The thickness average} of the shown in the panel \textbf{(h)} vector field, demonstrates the similarity with the observation presented in the panels \textbf{(f)} and \textbf{(g)}.
\textbf{(j)},\,Another possible arrangement of the chiral vortex bubbles alternatively ascending and descending polarization fluxes in their cores. All the bubbles possess the same chirality.
\textbf{(k)},\,Simulated reconstructed volume distribution of the polarization streamlines and chirality for the structure shown in panel \textbf{(j)}, see SM. The chiral bubbles have the structure of coupled  $H^-_{cw}$ and $H^+_{ccw}$ Hopfions possessing the same (right) handedness.  
\textbf{(l)},\,\markertwo{The thickness average}  of the shown in the panel \textbf{(k)} vector field.  
\textbf{(m)},\,The labyrinth-like disordered array of vortices (blue square) and vortex bubbles (yellow square) observed in the strain-free SrTiO$_3$/PbTiO$_3$ superlattice at $T=298$\,K. Adopted from Ref.\,\cite{Shao2023}. 
\textbf{(n)},\,The corner crosscut of the ferroelectric film with the simulated reconstructed polarization vortex labyrinth structure, see\,(\citeyear{SM}). The spatial distribution of the polarization streamlines demonstrates an emergence of the disordered array of Hopfions, possessing different chiralities. To better illustrate the 3D Hopfion structure, the streamlines forming the Hopfions were not cut and left to continue into space.
\textbf{(o)},\,\markertwo{The thickness average}  of the shown in the panel \textbf{(n)} vector field, demonstrates the similarity with the observation presented in panel\,\textbf{(m)}. 
}
}
\label{FigHF}
\end{figure*}

\medskip

\textit{b) Chirality and topology} 

Now we analyze the observations of vortex matter in ferroelectric heterostructures and the emergence of the chirality there from the topological point of view, building upon the topological description of the confined ferroelectrics, discussed in Sec.\,\ref{sec:Foundations}.
The particularity of the ferroelectric polarization patterns is that they have an intricate swirled texture resulting from the requirement of vanishing of the bound charges.  This internal-scale swirl structure defines the properties of the observed global configurations in compliance with the general topological impact. According to Arnold's theorem, the region of the confined and divergenceless polarization field can be partitioned into cells, each hosting either of two possible topological states, vortex or Hopfion. In ferroelectric planar layers, these cells correspond to the structural units of the polarization patterns, periodic or labyrinth-like stripes, or separate bubbles. The cells hosting vortices are achiral, whereas the cells hosting Hopfions are chiral.

The vortex vector field itself is achiral. Hence the arrays composed of alternately rotating vortices, emerging in ferroelectric heterostructures, Fig.\,\ref{FigHF}(a), do not bring the handedness to the system by themselves. 
However, chirality will appear provided that the vector field in the core of the vortex acquired the axial component to avoid the energy-costly singularity, resulting in the helical rotation of polarization\,\cite{Louis2012}.

 Packing the alternately-rotating vortices with the alternating axial components and having the same handedness
 into the vortex array results in the appearance of the macroscopic chirality of the system, see Fig.\,\ref{FigHF}(b). The chirality of such types of vortex arrays has indeed been discovered in the DyScO$_3$-grown PbTiO$_3$/SrTiO$_3$ superlattice by measuring the circular dichroism of the resonant soft X-ray diffraction pattern\,\cite{Shafer2018} and by the direct STEM imaging \,\cite{Chen2022}, see Sec.\,\ref{sec:Methods} for details.
 
A different type of packing alternately-rotating vortices having the same direction of the axial component,  hence  alternately-changing  handedness,
was reported in the SrTiO$_3$/PbTiO$_3$/SrTiO$_3$ trilayer grown on the DyScO$_3$ substrate\,\cite{Behera2022,Susarla2023}, see Fig.\,\ref{FigHF}(c).
They use the optical second-harmonic generation–based circular dichroism validated by the electron microscopy examination, see Sec.\,\ref{sec:Methods} for details.
Because of the mirror symmetry between the CW and CCW rotating vortices compensating their chiralities, the total chirality in such a system, in general, should be zero. 
However, in this particular system, the sequential alternating up-and-down shifts with respect to the middle of the ferroelectric layer of the left- and right-handed vortices, see Fig.\,\ref{FigHF}(d), result in breaking the mirror symmetry and emerging the weak chirality of the vortex array.

\markertwo{
An acquisition of chirality by the vortex array is accounted for within the soft domain approach as an admixture of the additional vortex component to solution $\mathbf{P}_{v}$, given by Eq.\,(\ref{SoftPolar}). For instance, for the shown in Fig.\,\ref{FigHF}d  chiral array with up-down shifted vortices, the polarization structure has the form,
\begin{equation}
\mathbf{P}_{chi}=\mathbf{P}_{v}+\mathbf{e}_x\,P_{a}, 
\label{Pchi}
\end{equation} 
where the first term corresponds to the achiral vortex structure, given by Eq.\,(\ref{SoftPolar}),
and the second term is the small admixture of the uniform in-plane $a$-phase with magnitude $P_a$. The coexistence of the phases with the differently oriented polarization is an expected result for the low-strained films\,\cite{Pertsev1998PRL}. 
}

Rigorously speaking, the chiral in-plane vortex stripes represent the Hopfion configuration. Indeed, the polarization flux, produced by the axial polarization component of the vortex, does not terminate at the edge of the heterostructure to avoid the formation of the bound charges and, thus, turns back, creating polarization counterflux at the vortex margins. However, it would be difficult to experimentally disclose the Hopfion nature of such a structure because the edges where the vortices terminate are macroscopically remote, and various mechanisms screen the bound charges. 

Now, we turn to bubbles.
The important discovery of\,\cite{Das2019,Das2021} was that the polarization vector field of the vortex rolled up into a bagel acquires an axial component in the same way as it happens in the stripe chiral vortex domains, see Fig.\,\ref{FigHF}(e).
This component rotates counterclockwise with respect to the torus' circular axis 
providing, therefore, the chirality of the bubble, which was confirmed by X-ray diffraction circular dichroism, see Sec.\,\ref{sec:Xray}. 
It was observed that the descending polarization flux in the space separating the bubble domains is directed oppositely to the ascending flux in the centers of the bubbles and that the bubble polarization vector makes a $180^\circ$ rotation when going from the center of the bubble to its periphery. 
This occurs to avoid the collision of the in-plane field components of the neighbor bubbles.
Therefore, the texture of the polarization vector in the planar cross-section of the chiral bubble domain resembles the structure of a skyrmion, comprising N\'{e}el skyrmions at the top and at the bottom and Bloch skyrmion in the middle of the ferroelectric layer. 
That is why the observed array of bubbles was there referred to as the lattice of skyrmions.

\marker{This observation may look contradictory. The Néel skyrmions host substantial depolarization charges because of the non-zero divergence of the polarization. This creates a strong depolarization field which is detrimental to the ferroelectric state, see Section\,\ref{sec:Extensions}.  At the same time, the intrinsic electric field observed inside the bubble is of the order of 2\,MV/cm\,\cite{Das2021} which is hundreds of times smaller than the expected depolarization field from the bound charges, see Section\,\ref{sec:Extensions}. 

This paradox is resolved by going from the `planar' view to the consideration of the entire volume structure of the bubble. It appears that unlike in the conventionally defined 2D magnetic skyrmions, there is a leakage of the polarization flux into the third direction of the bubble which compensates the depolarization charges accumulated by the N\'{e}el-type skyrmion. To make this even more transparent, we describe the polarization texture inside the bubble using the streamline presentation. The constraint imposed by the vanishing of the bound charges-induced depolarization energy is equivalent to the continuity of the streamlines within the confined volume, ensuring the divergenceless of the depolarization field. This indicates that the emerging streamlines configuration carries the fingerprint of the Hopfion. Furthermore, the accompanying entanglement and knotting of the streamlines cause the chirality of the bubble, see Section\,\ref{sec:Hopfions}. 
}

Indeed, the detailed visualization of polarization bubbles\,\cite{Das2021} shown in Fig.\,\ref{FigHF}(e) clearly demonstrates that the polarization structure of the array of bubbles is topologically equivalent to that in an array of the right-handed Hopfions, $H^+_{ccw}$, in which the polarization field lines helically wind over the periodically distributed tori. In general, there exist four energetically equivalent Hopfions, left-handed $H_{cw}^+$ and $H_{ccw}^-$ and right-handed $H_{ccw}^+$ and $H_{cw}^-$. The packing of the Hopfions of the same type in a lattice provides a coherent chirality to the entire system. 

Analysis of the recent experiments reveals that the situation is more involved since a bubble array can comprise Hopfions of different types. A particular type of emerging Hopfions forming a lattice depends on the relationship between the competing ferroelectric, elastic, and gradient energies.   
In such structures, the overall polarization texture is governed by the matching of the polarization vector fields of the neighboring Hopfions.

In particular, arrays of vortex bubbles having an alternative, CW and CCW in-plane rotations of polarization vector in a checkerboard order, were reported in the ultrathin PbTiO$_3$ films under the tensile strain caused by the SmScO$_3$ substrate\,\cite{Wang2020} and also in free-standing membranes of the \marker{PbTiO$_3$/SrTiO$_3$} superlattices at $T=373$\,K\,\cite{Shao2023}, see Fig.\,\ref{FigHF}(f,g). They observed that similar to bubbles studied 
in\,\cite{Das2019}, the polarization vector field has on average an ascending direction through the film thickness
in the centers of bubbles. However, it does not make the complete half-rotation from ``up" to ``down" but preserves an in-plane direction at the bubble's periphery, enabling the perfect matching of the oppositely rotating polarization in-plane components of touching adjacent bubbles. In these papers, the bubbles with the ``half-rotating" polarization were referred to as merons due to having a structure similar to two-dimensional magnetic merons\,\cite{Yu2018}, see Fig.\,\ref{FigMeron}.

\marker{
Notably, the polarization in the part of the cross-sectional area, denoted by crosses in Fig.\,\ref{FigHF}(g) must have the descending orientation opposite to that at the centers of the bubbles, forming the dual antimeron array.
This ensures that the total polarization flux through the plane is zero. This guarantees that despite the existence of the surface bound charges, having different signs for the ascending and descending polarization fluxes, the total charge at the surface remains zero which reduces the depolarization energy.
 
 The volume reconstruction of the meron structures (see\,(\citeyear{SM}) for details) shows, see Fig.\,\ref{FigHF}(h), that in accordance with Arnold theorem, the polarization forms the lattice of the alternating $H^+_{ccw}$ and $H^+_{cw}$ Hopfions having the characteristic polarization vector winding over the tori seen at the crosssection of the structure. 
 Unlike the meron lattice, the Hopfions do not create surface bound charges, hence unfavorable depolarization energy contribution at all. This explains the transformation of merons into Hopfions.  
 Because the emerging Hopfions carry the inverse, right and left, handednesses, the net chirality of the system vanishes. 
 Figure\,\ref{FigHF}(i) demonstrates \markertwo{the image of the vector field resulting from the averaging of the reconstructed 3D vector over the film thickness}. It acquires the structure of the meron-antimeron array and appears similar to the experimental observation shown in panels\,\textbf{(f)} and\,\textbf{(g)}.}

The coherent chirality of the system arises when packing the Hopfion pairs possessing the same handedness in a lattice. For instance, for packing the right-handed $H^+_{ccw}$ and $H^-_{cw}$ Hopfions, sketched in Fig.\,\ref{FigHF}(j). 
 A remarkable feature of such a structure is that the upcoming part of the polarization flux in the core of an $H^+_{ccw}$ Hopfion returns down not only within the Hopfion itself but distributes over the adjacent $H^-_{cw}$ Hopfions via becoming a part of their downcoming flux. Then this flux closes via the joining back to the upcoming flux of the neighboring $H^+_{ccw}$ Hopfions. This results in the Hopfion superstructure united through the interconnection 
 of their polarization field lines, \marker{as shown in Fig.\,\ref{FigHF}(k) for the simulated reconstructed 3D view and in Fig.\,\ref{FigHF}(l) 
\markertwo{for the thickness-averaged image}, see\,(\citeyear{SM}) for details of simulation}. This structure manifesting a macroscopic chirality has not yet been observed, hence it offers an attractive direction for future research. 

 \marker{ Figure\,\ref{FigHF}(m) presents another type of arrangement, a labyrinthine-type disorder of vortices (represented by the blue square) and vortex bubbles (represented by the yellow square), observed within the same strain-free SrTiO$_3$/PbTiO$_3$ superlattice but at temperature $298$\,K\,\cite{Shao2023}}. A similar labyrinthine structure was shown in the work\,\cite{Nahas2020,Govinden2023bubble} to reveal various topological states, concave and convex disclinations or meron-antimeron pair, bimerons, skyrmions, and the rare instances of target skyrmions and dislocations. 
 The presented data offer a surface view of the field distribution. However, \marker{the 3D reconstruction of the polarization streamlines spatial distribution} at the point of transition from stripe vortices to bubbles, where labyrinths form, shows that the field configuration -- in line with Arnold theorem -- is topologically equivalent to the composition of randomly oriented Hopfion cells, possessing different chirality, 
 \marker{see Fig.\,\ref{FigHF}(n) for the simulated 3D view and Fig.\,\ref{FigHF}(o) \markertwo{for the thickness-averaged image}. The details of simulations are given in\,(\citeyear{SM})}. Remarkably, the randomly distributed chirality in such a glassy state can be trained into a desirable coherent state using the field-, thermal-, and photo-induced manipulation methods described in Sec.\,\ref{sec:Manipulation}. 

Now we understand the nature of the global transition from the vortex stripes to the bubble droplet states governed by varying the parameters of the system. According to the above analysis, this transition presents the rotation of the Hopfion array from the in-plane to the out-of-plane direction. 
Furthermore, our analysis demonstrates that the trove of the numerous experimentally observed patterns acquiring various empirical names \marker{might be} just a greasepaint of the ensembles of differently oriented Hopfion and vortex configurations presented by their side \markertwo{or thickness averaged} views. The elicitation of the individual Hopfions emerging due to Arnold theorem and composing these ensembles is a challenging but appealing task.

\subsubsection{\label{sec:nanopart}Nanoparticles and nanodots}

\begin{figure*}
\begin{center}
\includegraphics [width=0.75\linewidth] {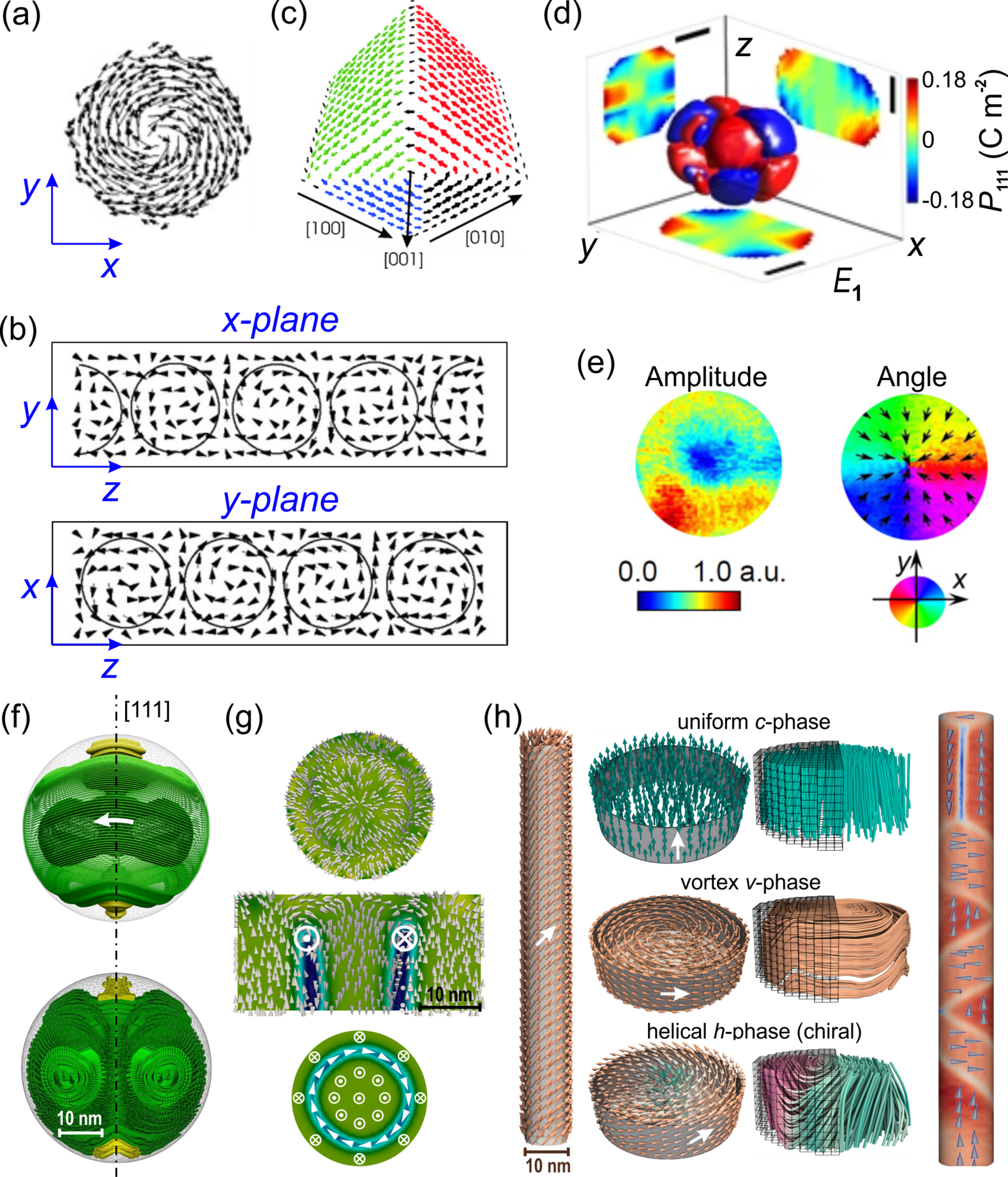}
\caption{ \textbf{Chirality in ferroelectric nanoparticles and nanorods.}  
\textbf{(a)}, Ab-initio simulation of local dipoles distribution in disc-shape nanoparticle of PbZr$_{0.5}$Ti$_{0.5}$O$_3$.  
\textbf{(b)}, The same for rectangular nanorod of PbZr$_{0.5}$Ti$_{0.5}$O$_3$. (\textbf{(a)} and \textbf{(b)}, are from Ref.\,\cite{Naumov2004}). 
\textbf{(c)}, Ab-initio simulation of a dipole vortex in a cubic nanoparticle of PbZr$_{0.4}$Ti$_{0.6}$O$_3$. From Ref.\,\cite{Prosandeev2007nd}. 
\textbf{(d)}, Experimental observation of a three-dimensional nanodomain in the BaTiO$_3$ spherical nanoparticle. Color map shows $x$, $y$, and $z$ components of polarization. From Ref.\,\cite{Karpov2017}.
\textbf{(e)}, Experimental PFM analysis of the vector field distribution in cylindrical nanodot of BiFeO$_3$, top view. Color maps show the amplitude and phase of the polarization vector. From Ref.\,\cite{Li2017}.
\textbf{(f)}, Phase-field simulation of streamlines in the ferroelectric nanoparticle of PbZr$_{0.6}$Ti$_{0.4}$O$_3$. Three-dimensional (top) and cross-section (bottom) views reveal the existence of the topological states, Hopfion (green) and two vortices (yellow). From Ref.\,\cite{Lukyanchuk2020}.
\textbf{(g)}, Half-Hopfion topological state in the cylindrical nanodot of PbTiO$_3$; the top view (top), vertical cross-section (middle), and bottom view (bottom). This state was simulated in\,\cite{Tikhonov2020} and referred to as a skyrmion. 
\textbf{(h)}, Topological strain-driven states simulated in a cylindrical nanorod of PbTiO$_3$. The middle panel shows the uniform c-phase, where the polarization is directed along the nanorod axis, vortex $v$-phase hosting the polarization vortex tube, and helical $h$-phase with helically-rotating polarization. The $c$ and $v$ phases are achiral, while the $h$ phase is chiral. The total view of the nanorod hosting $h$-phase is shown on the left. The terminal part of the nanorod demonstrating the domain segregation of the $c$-phase due to depolarization effects is shown on the right. From Refs.\,\cite{Pavlenko2022, Kondovych2023}. 
}
\end{center}
\label{FigNanopart}
\end{figure*}

Ferroelectric free-standing nanoparticles and substrate-deposited nanodots appeared as one of the first systems providing a material ground for the study of polarization behavior in confined systems. As has been discussed above, a straightforwardly observable implication of the presence of topological states carrying the chirality is a nonuniform character of the distribution of the polarization field.  
Figure\,\ref{FigNanopart} illustrates the development of the studies of the nanoparticles and nanodots.
That the ferroelectric state in the disk-shaped nanoparticles and rectangular nanorods of  PbZr$_{0.5}$Ti$_{0.5}$O$_3$ indeed hosts the vortex state was first demonstrated in simulations by\,\cite{Naumov2004}, see Fig.\,\ref{FigNanopart}(a),(b).
Yet, the observed size effect in thermodynamic properties of ferroelectric nanoparticles \,\cite{Zhao2004,Erdem2006} was interpreted in terms of a uniform state\,\cite{Glinchuk2013}. However, the subsequent development brought in a bunch of simulations demonstrating the inherently inhomogeneous vortex states in cubic nanoparticles of PbZr$_{0.4}$Ti$_{0.6}$O$_3$ \,\cite{Prosandeev2007nd} (see Fig.\,\ref{FigNanopart}(c)), and  PbTiO$_3$\,\cite{Stachiotti2011},  spherical nanoparticles of PbTiO$_3$ and BaTiO$_3$\,\cite{Mangeri2017}, cylindrical nanoparticles of PbTiO$_3$\,\cite{DiRino2020,Kondovych2023}, and domain states in 
ferroelectric nanodiscs of TGS\,\cite{Martelli2015} and PbTiO$_3$\,\cite{DiRino2020}. Notably, these described inhomogeneous states did not possess handedness.  

Finally, the inhomogeneous states in ferroelectric nanoparticles and nanodots were experimentally observed in 2017\,\cite{Karpov2017,Li2017,Li2023}. In work \,\cite{Karpov2017}, the presence of the polarization vortex state, revealing the toroidal moment distribution, was demonstrated for the BaTiO$_3$ nanoparticles using the Bragg coherent diffractive imaging, see Fig.\,\ref{FigNanopart}(d). Unfortunately, as was discussed in Sec.\,\ref{sec:Quantification}, the presence of the toroidal moment alone does not allow for the detection of the chirality of the system. 

The various types of spontaneous topological domain structures, including center-convergent domains, center-divergent domains, and double-center domains were discovered in the array of epitaxial ferroelectric nanodots of BiFeO$_3$ by piezoresponse force microscopy (PFM)\,\cite{Li2017}. Although the observed surface distribution of the polarization, see Fig.\,\ref{FigNanopart}(e),  allows for an assumption that the polarization structure in bulk can have an intricate vortex structure, the identification of chiral properties of the nanoparticles is again left off the board. 
In summary, the listed research set in quite a productive study direction. 
In this Section, we discuss the ways to explore it. 

To move on and reveal the onset of chirality in ferroelectric nanoparticles and nanodots, we build on the consideration of the emergent polarization topological states under confinement, vortices, and Hopfions, combining the topological approach with their energy consideration.  The energy advantage of the vortex state is that it allows for the polarization distribution in the plane of anisotropy, reducing thus the anisotropic contribution. 
At the same time, the polarization suppressing at the vortex-core line is unfavorable from the Ginzburg-Landau energy viewpoint. 
In contrast, the polarization singularity region in Hopfions is reduced to two points at the poles, making thus Hopfion more energetically favorable. Concurrently, the out-of-anisotropy plane escape of polarization increases the Hopfion anisotropic energy. 
Therefore, to achieve the chiral Hopfions, one has to reduce the anisotropy energy. This can be done, in particular, either by modification of the chemical composition or by application of the appropriately designed strain. 

The first way was explored in ferroelectric spherical nanoparticles by selecting the PbZr$_{1-x}$Ti$_x$O$_3$ compound with $x=0.4$\,\cite{Lukyanchuk2020}. 
At this concentration, corresponding to the vicinity of the so-called morphotropic point, the polarization vector is only slightly pinned by the crystal anisotropy, rotating from the [001] crystallographic orientation (corresponding to the tetragonal phase stable at higher $x$) to the  [111] crystallographic orientation (corresponding to the rhombohedral phase stable at lower $x$). The volume of the nanoparticle having the radius 25\,nm was shown to split into three cells, 
hosting different topological states in compliance with the Arnold theorem, see Fig.\,\ref{FigNanopart}(f). The major part of the volume was occupied by the chiral Hopfion, whose structure is similar to that of the model case discussed in Sec.\,\ref{sec:Foundations}, but deformed due to the residual crystal anisotropy. Two small cells in the vicinity of the poles are hosting the tiny achiral vortices.  
For spherical nanoparticles of  PbZr$_{0.6}$Ti$_{0.4}$O$_3$ larger than 30\,nm the polarization lines maintaining the Hopfion/vortex structure become more deformed. Then the new cells with the newly-born little Hopfions, chiral Hopfioninos, bud off from their parent Hopfion. 
At the same time, the spherical nanoparticles built from the pure free-standing PbTiO$_3$ host only the achiral vortices\,\cite{Mangeri2017,DiRino2020}, which is a consequence of the relatively strong anisotropy of this compound. \marker{The extreme example of a Hopfion is the achiral anapole,  discussed in Section,\ref{sec:Hopfions}. This configuration, termed toron, was predicted for BiFeO$_3$/BaTiO$_3$ core-shell nanoparticles in\,\cite{Lich2023}.}

A plethora of formations, including different vortices and antivortices, hedgehogs and antihedgehogs, and a few skyrmions, 
were simulated in conical nanoparticles of \marker{BaTiO$_3$} embedded into SrTiO$_3$ matrix\,\cite{Prosandeev2019} and referred there as a topological eclecton. 
This observation is in line with the general concept of confinement-induced topological states, which can vary depending on the nanoparticle's geometry. Possibly the observed cocktail of topological states can be decomposed into the vortex and Hopfions elementary cells upon analyzing the streamlines configuration.  

The strain-engineering method was employed to reveal the chirality in the disc-shape nanodots of PbTiO$_3$\,\cite{Tikhonov2020}, deposited on the conducting substrate. The chiral bubbles were shown to emerge under the appropriately selected compressive misfit strain $u_m\simeq-0.002$, nanodot diameter 40\,nm, and thickness 20\,nm, see Fig.\,\ref{FigNanopart}(g). Although, according to the characteristic polarization distribution in the down cross-section, they were called ``skyrmions," they, in fact, have the 3D structure of the half-Hopfion in the whole volume of the nanodot. This half-Hopfion transforms into the complete chiral Hopfion by the mirroring because of the mirror boundary condition at the conducting substrate plane. 

The obtained results call for the extension of the described above experiments on sensing polarization distribution in ferroelectric nanoparticles and nanodots. For instance, the fingerprints of topological states observed in ferroelectric nanoparticles\,\cite{Karpov2017} can correspond to the in-particle chiral Hopfion states, whereas the center-convergent and center-divergent polarization domains observed at the upper surface of the ferroelectric nanodot in\,\cite{Li2017} can be a manifestation of the half-Hopfion residing in the bulk of the nanodot and having the similar polarization texture at the top of the nanodot. Moreover, the described chiral topological states can be operated by the electric field with switching handedness. We will discuss the corresponding protocols in Sec.\,\ref{sec:Switching}.

\begin{figure*}
\begin{center}
\includegraphics [width=0.9\linewidth] {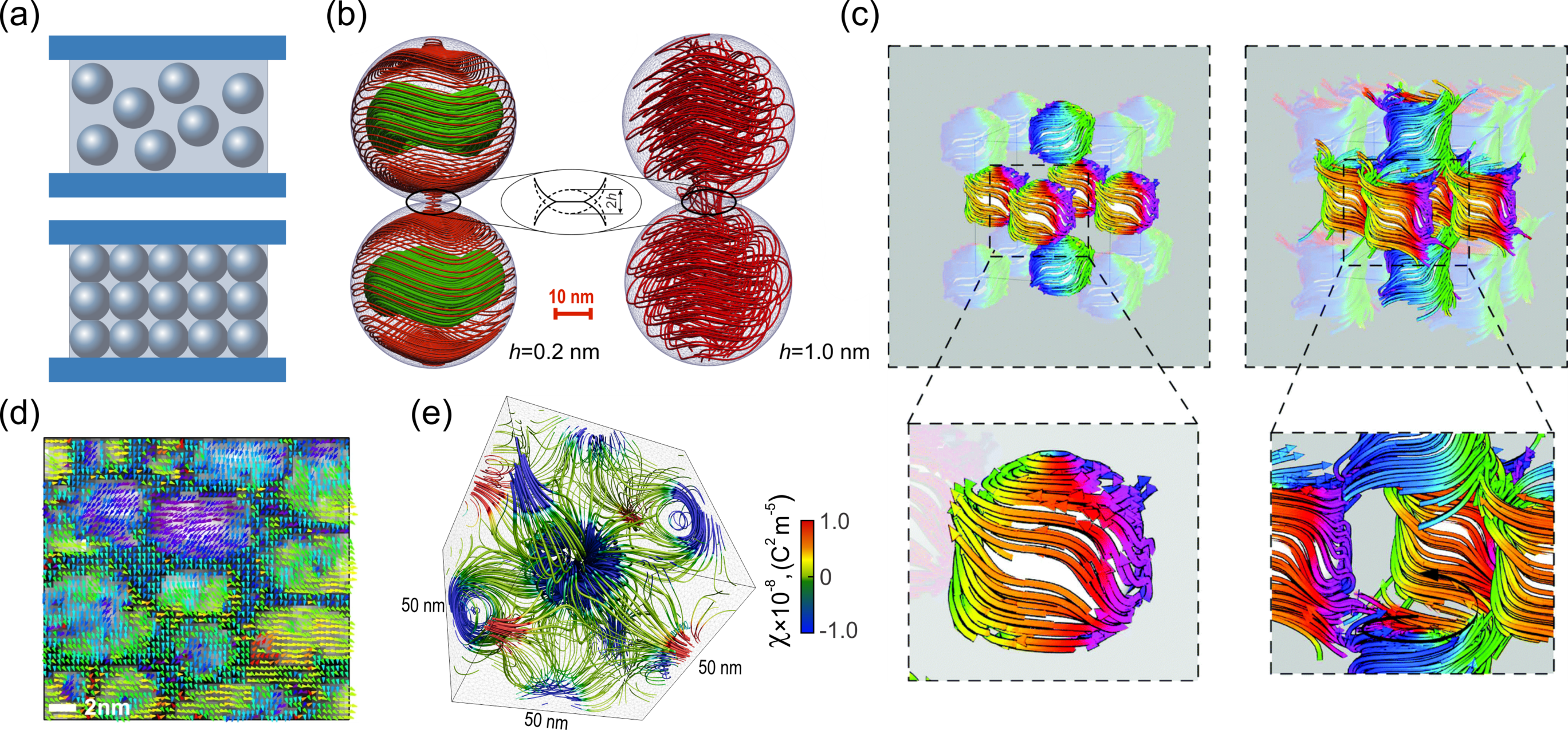}
\caption{ \textbf{Chiral structures in composites and disordered ferroelectrics}  
\textbf{(a)}, Dilute composite of isolated nanoparticles in a capacitor and sintered composite of contacting nanoparticles.
\textbf{(b)}, (Left part) Field lines flow between moderately densified nanoparticles of PbZr$_{0.6}$Ti$_{0.4}$O$_3$ with radius $R=25$ nm, contacted along the $[111]$ direction. The densification degree is characterized by the thickness of the connecting neck $2h=0.4$ nm, see inset. The delocalized helical lines flowing between the particles are embraced by the localized lines forming the toroidal Hopfion states. (Right part) In a composite of strongly densified nanoparticles of the same size and material but with the thickness of the connecting neck  $2h=2$ nm, all the field lines form delocalized helices flowing across the entire composite \,\cite{Lukyanchuk2020}). 
\textbf{(c)}, The polarization distributions in the ferroelectric nanostructures formed in the face-centered cubic nanoporous SrTiO$_3$ \,\cite{Masuda2019}. 
\textbf{(d)}, Atomically resolved contrast-inverted STEM ABF image of bulk Bi$_{0.5}$Na$_{0.5}$TiO$_3$ sample with ferroelectric bubbles. The $\delta_{\rm Ti-O}$ displacement vector maps overlaid on their corresponding polarization intensity; the displacement vectors are indicated as colored arrows according to their rotation angles, and the polarization intensity is indicated by bright-dark contrast\,\cite{Yin2021}.
\textbf{(e)}, Simulation of the polarization and chirality distribution in \marker{PbZr$_{0.6}$Ti$_{0.4}$O$_{3}$} ferroelectrics where the space non-uniform Gaussian distribution of critical temperature mimics structural disorder.
}
\label{FigComposites}
\end{center}
\end{figure*}

\subsubsection{\label{sec:Nanorods} Nanorods and nanowires}

Nanorods (or nanowires), long cylindrical nanoparticles with tens nanometers cross-sections, represent another family of nanostructured ferroelectrics\,\cite{Rorvik2011}. Synthesized already in the 90's\,\cite{Cheng1992}, they still remain underexplored as compared to other nanostructured ferroelectrics, especially regarding the aspects of polarization distribution.
Theoretical considerations indicate that there are two possible principal directions of polarization promoted by the axial symmetry of the nanorods, namely, either along or perpendicular to their axis. Similarly to magnetism, we refer to the first case as an easy-axis symmetry and to the second one as an easy plane symmetry.  The realization of either possibility depends on the polarization anisotropic energy induced by the external strain\,\cite{Pavlenko2022} or surface tension\,\cite{Glinchuk2013}, as illustrated in Fig.\,\ref{FigNanopart}(h), \cite{Pavlenko2022}.
In the first case, polarization is uniformly elongated along the whole length of the nanorod\,\cite{Glinchuk2013}, forming the uniform $c$-phase, see\,\cite{Pavlenko2022}. In the second case, it hits the nanorod walls creating the energy-costing bound charges and the associated with them depolarization fields. 
To avoid the formation of the bound charges, polarization may swirl into the vortex, coaxial with the nanorod axis\,\cite{Lahoche2008}, forming the vortex $v$-phase. Both phases, $c$ and $v$, are achiral. However, at certain tensions, the intermediate state between the $c$- and $v$-phases in which the polarization is inclined with respect to the nanorod axis is also possible. 
Similar to the vortex case, to avoid the emergence of bound charges at the boundary, the polarization rotates around the nanotube axis, forming the helical chiral $h-$phase\,\cite{Pavlenko2022}. 

At this stage, the question is whether the more complex chiral topological states similar to what does exist in thin films and nanoparticles can emerge for nanorods. 
Actually, already in the first simulation of the polarization in nanostructured ferroelectrics\,\cite{Naumov2004}, the system of coupled vortices elongated along the axis $z$ of the nanorod with the square cross-section $x$-$y$ was discovered. Figure\,\ref{FigNanopart}(b) demonstrates the observed structure. The edges of the vortices on the $x$ plane go through the centers of the vortices in the $y$ plane; two sets of vortices are therefore interconnected like links in a chain. Such a structure does not possess a mirror plane symmetry and has a streamline networking similar to the array of the interconnected Hopfions in thin films, Fig.\,\ref{FigHF}(j). Hence it does have a definite handedness. 

In works\,\cite{Prosandeev2013} and\,\cite{Nahas2015},
chiral skyrmion-like tubes were revealed in nanocomposites made out of the periodic squared arrays of BaTiO$_3$ nanowires embedded into matrices formed by Ba$_{0.15}$Sr$_{0.85}$TiO$_3$ solid solution with $x=0.15$ and $x=0$, respectively.  The chirality in these systems emerges due to the Bloch-like change of the vertical polarization component, from positive to negative, when passing from the nanowire region to the embedding media. \marker{Even more elaborated structures in which the core of the helical $h$-phase helically rotates itself around the nanorod axis,  were discovered recently in nanorods, composed from alternating PbTiO$_3$ and PbZr$_{0.05}$Ti$_{0.95}$O$_3$ cylinders\,\cite{Lich2022}.}
An essential feature in such structures is the possibility of manipulating chirality by the electric field, as we discuss in Sec.\,\ref{sec:Switching}. 

Note, finally, the chiral states can arise at the ends of nanorods hosting the uniform $c$-phase. The emergence of the depolarization fields 
at the terminal points at the tops and bottoms of the nanorods leads to the swirling of the polarization 
at the terminal regions and the formation of the chiral domains, shown in  Fig.\,\ref{FigNanopart}(h) (right panel) \cite{Pavlenko2022}. 
The effect of the emergence of specific terminal chiral polarization domains (referred to as flexons) can also be associated with the intervening flexoelectric effect\,\cite{Morozovska2021}.

The described chirality researches indicate the perspectives opening in further exploring of nanorods.

\subsubsection{\label{sec:Composites} Composites and disordered ferroelectrics}

Nanocomposite ferroelectrics in which the ferroelectric nanoparticles are embedded into the matrix of other materials, present high interest for the applications. The influence of the confinement on polarization distribution and on the emergence of the topological states of polarization depends on the electrostatic cross-communication between the nanoparticles, hence the density of their packing, see Fig.\,\ref{FigComposites}(a). The left panel of Fig.\,\ref{FigComposites}(b) shows the configuration of streamlines in adjusting nanoparticles of PbZr$_{0.4}$Ti$_{0.4}$O$_3$ that forms under the conditions of moderate densification of nanoparticles. Only a fraction of polarization helical streamlines pass through the interfacial aperture. Another part of the lines forms the internal Hopfions. In the case of densely packed nanoparticles, all the field lines form helices flowing through the area of the contact\,\cite{Lukyanchuk2020}, as shown in the right panel of Fig.\,\ref{FigComposites}(b).
A similar configuration of inter-communication between nanoparticles by virtue of the helically-screwed streamlines, piercing the spots of the contact between nanoparticles was studied in work\,\cite{Masuda2019} for the strained nanoparticles of SrTiO$_3$, see Fig.\,\ref{FigComposites}(c). The remarkable feature of such systems is that depending on the strength of electrostatic interaction between nanoparticles, the chirality emerging in them can be switched either on the individual or on the collective level by external fields, as suggested in Sec.\,\ref{sec:Manipulation}. This offers a platform for various optoelectronic and computing devices, see Sec.\,\ref{sec:Perspectives}. 

An interesting situation arises in the bulk of disordered ferroelectric materials where the structural or composition disorder is coupled with the polar degrees of freedom. Then, in order to avoid the formation of depolarization charges, the nonuniform polarization field swirls into the bubble configurations which according to Arnold theorem should be either vortices or Hopfions. Such self-confined bubble-like features with multiple polar topologies were recently observed in bulk Bi$_{0.5}$Na$_{0.5}$TiO$_3$ relaxor-ferroelectric having chemically-driven disorder morphology\,\cite{Yin2021}, see Fig.\,\ref{FigComposites}(d). Figure\,\ref{FigComposites}(e) presents the simulation of polarization distribution in \marker{PbZr$_{0.6}$Ti$_{0.4}$O$_{3}$} ferroelectrics, in which the space non-uniform Gaussian distribution of critical temperature mimics the structural disorder, and shows the highly entangled structure of the streamlines.  In accordance with Arnold theorem, the streamlines tangle into toroidal knots which are the chiral Hopfions. Understanding the topology of disordered ferroelectrics as a composition of the Hopfion topological states is important not only because it holds high potential for applications but since it posits a deep fundamental appeal. This concept offers ways of targeted manipulation of space-disordered chirality by operating individual Hopfions, thus enabling the groundbreaking technology of neuromorphic nano-optoelectronics. On the other hand, it sheds light on the seminal problem of the physics of relaxor-ferroelectrics.


\subsection{\label{sec:Sensing}Chirality sensing}

\begin{figure*}
\begin{center}
\includegraphics [width=0.8\linewidth] {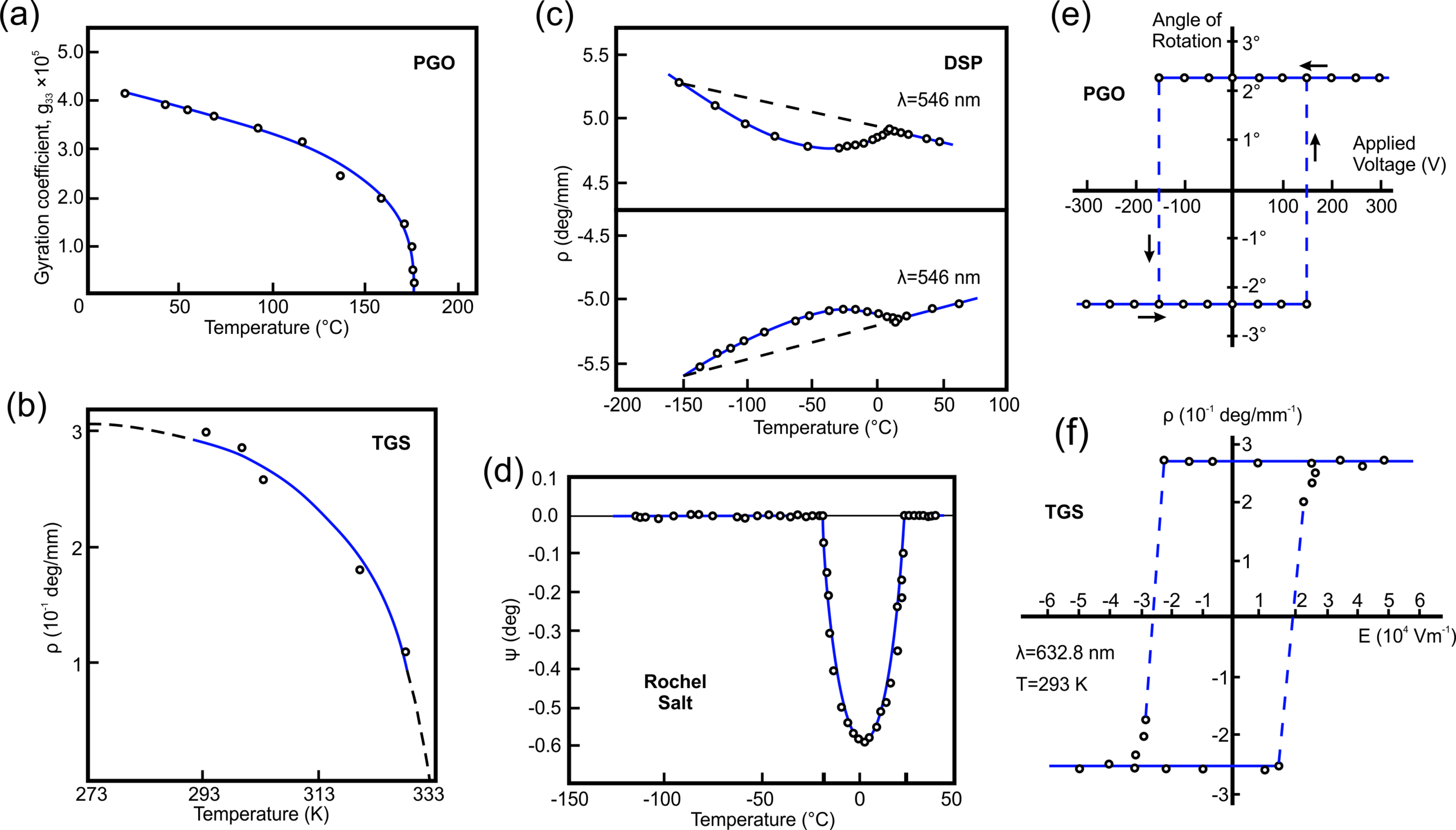}
\caption{ \textbf{Optical activity of bulk ferroelectrics.}  
\textbf{(a)}, Temperature dependence of gyration coefficient g$_{33}$ of PGO. From Ref.\,\cite{Iwasaki1971}.
\textbf{(b)}, Temperature dependence of optical rotatory power $\rho$ of TGS. From Ref.\cite{Koralewski1983}.
\textbf{(c)}, Temperature dependence of optical rotatory power $\rho$ of the DSP. From Ref.\,\cite{Kobayashi1971}.
\textbf{(d)}, Temperature dependence of the rotation angle $\Psi$ of the optical indicatrix of the Rochelle salt\,\cite{Kobayashi1991}.
\textbf{(e)}, The DC field hysteresis of the angle of rotation of the plane of polarization of the PGO. From Ref.\,\cite{Iwasaki1971}.
\textbf{(f)}, Optical rotary power hysteresis loop for the TGS crystal. From Ref.\,\cite{Koralewski1983}.
}
\label{FigOptAct}
\end{center}
\end{figure*}

\subsubsection{\label{sec:Methods}Overview of the methods}

Now we describe the progress made in the detection and measurement of chirality in ferroelectric materials as well as the development of the advanced methods used for other emergent chiral nanosystems. We emphasize the opportunities for employing \marker{these} methods for nanostructured ferroelectrics. 

We focus first on optical methods, which are a golden foundational base for studying chiral crystals and molecular systems.
After a brief introduction to the electrodynamics of chiral materials, we address the state of the art of traditional ways of investigating of optical activity of bulk ferroelectrics.
Notably, while the research of bulk materials has achieved remarkable maturity, the study of the optical activity in the nanostructures ferroelectrics, in particular, possessing the topological chirality, is still in its infancy. The emerging challenge is related to both, the unusual topological structuring of the polarization in nanosystems and their small size, which is much smaller than the light wavelength, resulting in a very weak outcoming signal. Thus, we put a special emphasis on the innovative linear and nonlinear optics methods that have recently emerged in other areas of nanophysics. 

Next, we address another approach utilizing the chirality of the polarized X-ray radiation that was successfully implemented for revealing the structure of topological states in various ferroelectric systems.  Its apparent advantage is that the typical X-ray wavelength, 2-3 nm, is smaller than the size of the polarization topological defects which allows for good precision.  

Finally, we mention another, still underexplored in ferroelectrics, methods of chirality testing, NMR, NQR, and ESR, and also chemical sensing.

\subsubsection{\label{sec:LinOpt}Linear optics}

The chiral states of the light, possessing either the right or left circular polarization of the beam, interact differently with the chiral matter. This property called the material optical activity, manifests itself as a circular birefringence, the rotation of the polarization plane of the linearly polarized light when passing a chiral medium, and also as circular dichroism which is the different attenuation of left- and right-handed circularly polarized light. 

\textit{a) Bulk materials.} 
The optical activity of the anisotropic media tested by the light beam with the wave vector $\mathbf{k}$ and frequency $\omega$ is determined by the non-local electromagnetic response described by the non-local dielectric permittivity tensor\,\cite{Landau8}, 
\begin{equation}
\varepsilon_{ik}(\omega, \mathbf{k})=\varepsilon_{ik}^0(\omega)+i \frac{c}{\omega}e_{ikm}g_{ml}k_l. 
\end{equation}
Here $c$ is the speed of light,  $\varepsilon_{ik}^0(\omega)$ is the local achiral part of the dielectric tensor,  $e_{ikm}$ is the anti-symmetric index, and $g_{ml}$ is the so-called gyration tensor, which actually determines the chiral properties of the media. 
The optical rotary power, $\rho$, quantifies the polarization rotation angle per mm of the light propagation. It is expressed through the gyration tensor as $\rho=(\pi/\lambda n_0) g_{ij} s_i s_j$, where $\lambda$, is the wavelength of light, $n_0$ is the ordinary refraction index, and $s_i$ is a unit vector in the direction of the light propagation. \marker{Recently, the accurate and computationally efficient first-principles methodology for the calculation of natural optical activity was suggested\,\cite{Zabalo2023}. This methodology, in combination with experimental methods, can be especially useful to understand the optical activity in ferroelectrics.}

The circular birefringence was detected and studied in the most common chiral ferroelectrics, see Fig.\,\ref{FigOptAct}.
In ferroelectrics with the polarization-induced chirality, see Table\,IV in Sec.\,\ref{sec:Struct}, the emergence of the optical activity in the polar phase has been demonstrated by measuring the temperature dependence of the $g_{11}$ component of the gyration tensor in the PGO material\,\cite{Iwasaki1971,Iwasaki1972}, (Fig.\,\ref{FigOptAct}(a)) and optical rotary power $\rho$ in the  TGS material\,\cite{Koralewski1983} (Fig.\,\ref{FigOptAct}(b)). Also, in ferroelectrics with structural built-in chirality, the specific ferroelectricity-induced features of optical activity were revealed in the temperature dependencies of the optical rotary power in DSP crystals\,\cite{Kobayashi1971,Sawada1977} (Fig.\,\ref{FigOptAct}(c)), and in the rotation angle of the optical indicatrix in the Rochelle salt\,\cite{Kobayashi1991} (Fig.\,\ref{FigOptAct}(d)).
The hysteresis-type switchability of the optical rotary power associated with the polarization switching by the electric field was also demonstrated for the ferroelectrics with polarization-induced chirality,  PGO\,\cite{Iwasaki1971} (Fig.\,\ref{FigOptAct}(e)), and TGS\, \cite{Koralewski1983} (Fig.\,\ref{FigOptAct}(f))). 
The studies were carried out mostly on single-domain crystals. 
The consideration of the optical activity for the multi-domain systems requires special analysis \cite{Shuvalov1964}.

The other approach comprises measurements of vibrational optical rotation and vibrational circular dichroism\cite{Nafie2011,Collins2017}, reflecting the differential response of a medium to the left- and right-handed circularly of the polarized light at wavelengths corresponding to vibrational transitions.
These effects are seen by use of the Raman and infrared measurements catching the small differences in the intensity of the vibrational scattering on chiral substances for the right- and left-circularly polarized light. The drawback of these techniques is the very weak response which is by orders of magnitude smaller than the background non-polarized scattering. This may be a possible reason why they are not yet widely used for ferroelectric materials.

\markertwo{
\textit{b) Thin films and superlattices.}
The emergence of controllable ferroelectric chirality in thin ferroelectric films and heterostructures promises unprecedented ways to explore the optical activity of these materials. The chiroptical effects in layered structures are mostly studied in artificially created 2D plasmonic structures in which the nanoscale metallic patterns possessing distinct handedness are created by the electrochemical deposition or drawn on a resistive layer using an electron beam lithography\,\cite{Hossain2014}. Notably, a substantial optical rotation was observed in the 2D arrays of the chiral gold nanostructures\,\cite{Kuwata-Gonokami2005}. Although these arrays were much thinner than the plasmon resonance wavelength, they exhibited chiroptical effects comparable in magnitude to similar effects in chiral molecular materials that are several orders of magnitude thicker. In this context, the spatially-extended chiral vortex arrays and polar chiral bubbles emergent in ferroelectric thin films can become even more operationable, due to their manipulability by electric fields.  

Despite their advantages, the light-matter interaction in planar chiral nanostructures is rather weak due to their significantly reduced material thickness. Furthermore, while chirality is inherently a three-dimensional property, the three-dimensionality of planar nanostructures is limited. This problem can be resolved by stacking multiple layers of chiral nanostructures\,\cite{Hossain2014} which increases the light beam optical pass, thereby making the chirality of the material more efficient\,\cite{Chen2022Chir}. Indeed, the optical activity of mutually twisted dielectric layers patterned by the sub-wavelength gold plasmonic arrays of cross-shaped\,\cite{Decker2009} and U-shaped\,\cite{Liu2009} islands, and Moiré textures\,\cite{Wu2017}, was demonstrated to enhance the chiroptical properties of these materials. Thus,
the discussed in Section\,\ref{sec:Films} ferroelectric superlattices, hosting the chiral polarization topological states, represent a promising direction for chiroptic development. They will pave the way towards the design of the artificial 3D materials with optical properties significantly surpassing those of naturally-active ferroelectric crystals.}

\textit{c) Nanostructures.} The study of the optical activity of ferroelectric nanostructures where geometry plays an essential role in scattering requires a more elaborate consideration. We survey the nanoparticles of radius $R$, smaller than the wavelength $\lambda$ of the incident light.

In this case, Rayleigh scattering is a predominant mechanism of light scattering and special techniques for circular differential Rayleigh scattering are to be developed. The circular differential Rayleigh scattering, $\Delta C_{scat}$ and absorption cross sections $\Delta C_{abs}$ per unit area for the nanoparticle, defined as a difference between the scattering cross sections of the left- and right-hand polarized light and describing a circular birefringence and  circular dichroism of the nanoparticle are given by\,\cite{Yoo2015}
\begin{eqnarray}
\frac{\Delta C_{scat}}{R^{2}} &=&\frac{32\pi }{3}\frac{\left( \varepsilon
_{1}^{\prime }-\varepsilon \right) ^{2}+\varepsilon _{1}^{\prime \prime 2}}{%
\left( \varepsilon _{1}^{\prime }+2\varepsilon \right) ^{2}+\varepsilon
_{1}^{\prime \prime 2}}n_0^{3}\left( \frac{2\pi R}{\lambda }\right) ^{4}\kappa\,, 
\label{Cscat}
\\
\frac{\Delta C_{abs}}{R^{2}} &=&16\pi \frac{\varepsilon _{1}^{\prime \prime
}\varepsilon }{\left( \varepsilon _{1}^{\prime }+2\varepsilon \right)
^{2}+\varepsilon _{1}^{\prime \prime 2}}\frac{2\pi R}{\lambda }\kappa\,.
\label{Cabc}
\end{eqnarray}
Here $\varepsilon _{1}=\varepsilon
_{1}^{\prime }+i\varepsilon _{1}^{\prime \prime }$ is the complex permittivity of the particle, $\kappa$ is the chirality parameter which is related with gyration tensor as $g_{ml}=2 \kappa \delta_{ml}$, we assume here that material is isotropic, and $\varepsilon$ is the permittivity of the media in which the nanoparticle is embedded. 


Similar to the non-polarized Rayleigh scattering, the cross-section of the circular differential
Rayleigh scattering decreases with the size of the particle as $(R/\lambda)^4$ for circular birefringence and as $(R/\lambda)$ for circular dichroism. The additional reduction due to small $\kappa$  makes the employment of the circular differential Rayleigh scattering for chirality sensing of small nanoparticles challenging. 

To enhance the interaction between the circularly polarized light and chiral nanoparticles for more than an order of magnitude, the local surface plasmon resonance at the resonance wavelength, $\lambda_r$, defined by the condition 
$\varepsilon _{1}^{\prime }(\lambda_r)=-2\varepsilon$ in Eqs.\,(\ref{Cabc}),(\ref{Cscat}), can be used.
 The plasmonic amplification was employed for finding and investigating chirality in noble metals as well as in semiconductor nanostructures\cite{Kong2020,Mun2020,John2021}.

Another attractive methodology to measure the circular dichroism on individual nano-objects was suggested in the work\,\cite{Vinegrad2018}.
It was demonstrated that electrically controlled switching between discrete circular polarized states substantially reduces detrimental linear optical
activity effects allowing to perform diffraction limited measurements with of the order of $10^6$ extinction in sensitivity. 

The surface-enhanced Raman optical activity offers one more way for amplifying chiroptical effects in nanostructures. Its amplifying effect is provided by the surface plasmons localized on a neighboring noble metal surface interacting with the incident light\cite{Pour2011,DasM2021raman}. 

To conclude here, we mention the circularly polarized luminescence, which manifests the difference in the emission intensity for the right and left circularly polarized light that have been extensively used for the investigation of
various molecular systems 
and which has recently attracted considerable attention as a beneficial way of studying the chiral nanoparticles\,\cite{Zhang2022}.

\subsubsection{\label{sec:NonLinOpt}Nonlinear optics}
Sensing the low-intensity chiral signals from nanostructured materials with a characteristic scale less than the wavelength of light may require more advanced tools, which are provided by nonlinear optics. 

The second-harmonic generation with the typical resolution of hundreds of nanometers meets the challenge and offers powerful means to detect and investigate local properties of the polar topological states even though their size is far below the resolution limit of the optical methods, see \cite{Bonacina2020,Cherifi2021}.  
The second-harmonic generation confocal microscopy combined with local polarimetry was used to reveal the chiral Bloch-like configuration of DWs in LiTaO$_3$ crystal 
\cite{Cherifi2017}. A similar method was used by \cite{Behera2022} for studying the chirality of vortex domains in the \marker{PbTiO$_3$/SrTiO$_3$} superlattices. Enantioselective sensing by collective circular dichroism\,\cite{Kim2022enantio} is another emergent method recently used for quantitative determination and in situ monitoring of nanoscale chirality.

We now touch upon the prospective method of the enantioselective hyper-Rayleigh scattering that was successfully used to test the structural chirality of the inorganic nanomaterials \cite{Liu2021}. In this method, the nonlinear hyper-Rayleigh scattering optical activity 
measures the differential Rayleigh scattering of a nano-object irradiated with the left- and right-circular polarized ultrafast laser pulses, promising the hyper-sensitive detection of chirality.

\subsubsection{\label{sec:Xray}X-ray scattering}

The first detections of polarization chirality in nanostructured ferroelectrics were undertaken by \cite{Shafer2018} and \cite{Das2019,Das2023} for domain structures in the PbTiO$_3$/SrTiO$_3$ superlattices. They used the method of the resonant soft X-ray diffraction-based circular dichroism (RSXD-CD), suggested previously by \cite{Hill1995, Lovesey2005}. 
There, the soft X-ray resonant scattering was produced by the synchrotron radiation at the titanium L absorption edge. The observed circular dichroism of the X-ray diffraction peaks was attributed to a chiral rotation of the electric polarization within domains. 
The development of the interpretation of the experiment was proposed by \cite{Lovesey2018}, who has indicated that, while the electrical polarization in the electron–dipole approximation does not directly affects the X-rays scattering, yet, the latter may sidewards feel the chiral polar arrangements. 
At the same time, the observed chirality may be caused by the chiral array of charge quadrupole moments that form in these heterostructures.

The subsequent investigation of the chirality of the ferroelectric domains in the films of BiFeO$_3$ films  \cite{Chauleau2020, Fusil2022}, took  into account the quadrupole part of the scattering amplitude and confirmed the existence of chirality in ferroelectric domains. However, recent studies showed that the interpretation of the RSXD-CD data can be by far more complex in the case of the realistic geometries of thin films and superlattices because in these systems the loss of the single-crystal symmetry occurs \cite{Kim2022}.  The emergence of cutting-edge microscopies utilizing coherent X-ray techniques\,\cite{Carbone2023} heralds an era of deeper insights into yet-studied chiral nano-ferroelectrics and paves the way for transformative breakthroughs in nanoscale chirality research.

\subsubsection{\label{sec:OtherMethods}Other methods}
Beyond the well-recommended chiral radiation scattering methods, other techniques for chirality sensing at the nanoscale are either already employed or suggested for other types of nanosystems. They can be further successfully used for nanostructured ferroelectrics.
The chiral plasmonic activity of nanoparticles can be measured by the single-particle circular differential scattering \cite{Liu2021}. 
PFM\cite{Li2017}. \marker{Recently the novel methodology of transferring orbital angular momentum to an electron beam to reveal toroidal and chiral order in nanostructured ferroelectrics was suggested\,\cite{Nguyen2023}}.
Also, one can apply the NMR, NQR, ESR spectroscopic, and STEM imaging for the local-environment chirality testing \cite{Lazzeretti2017,Wenzel2018,Chen2022}. Finally, one can adopt the chemical surface functionalization by chiral molecules allowing one to discriminate between the left- and right-handed chiral nanoparticles\,\cite{Kumar2016}.

\subsection{\label{sec:Manipulation}Chirality manipulation.}

\begin{figure*}
\begin{center}
\includegraphics [width=1\linewidth] {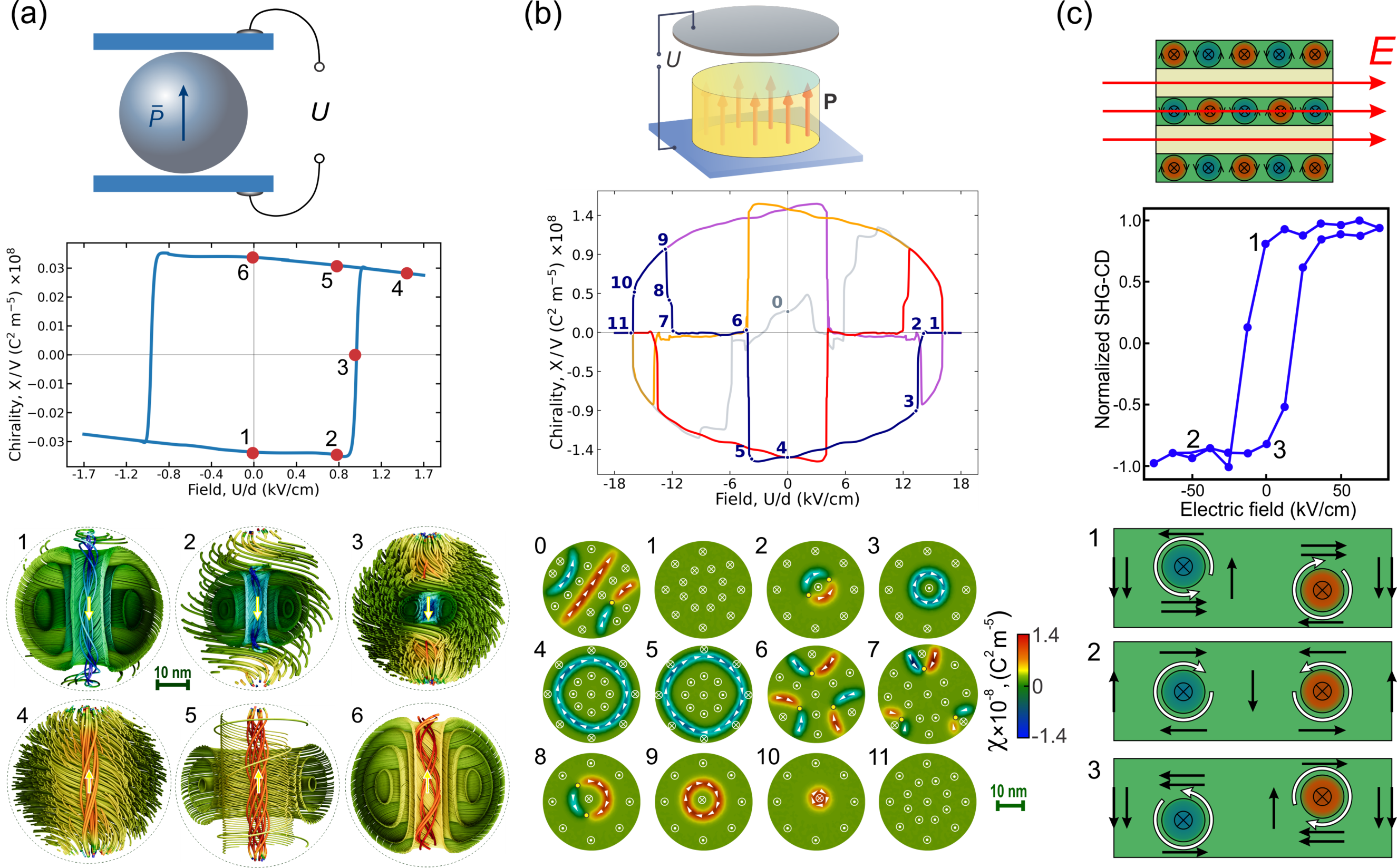}
\caption{
\textbf{Chirality switching by the electric field.} The upper panels illustrate the geometry of the switching device. Middle panels show switching hysteresis loops. The bottom panels demonstrate polarization distribution (arrows) and chirality (color map) of the initial states, intermediate states through which the switching occurs, and the final chiral states.   
\textbf{(a)}, Simulation of the chirality switching in spherical nanoparticles hosting Hopfions. Light yellow arrows indicate the direction of the average polarization flux in the Hopfion core.
\textbf{(b)}, Simulation of chirality switching in cylindrical nanodots of PbTiO$_3$, hosting half-Hopfions. The bottom panels demonstrate the view of the polarization distribution (white arrows, crosses, and dots) in the pathway states from the bottom of the nanodot.  From Ref.\,\cite{Tikhonov2020}.
\textbf{(c)},  Experimental study of the chirality dependent second harmonic generation circular dichroism (SHG-CD) switching in PbTiO$_3$/SrTiO$_3$ superlattices hosting helical vortices deposited on DyScO$_3$ substrate. White arrows indicate the direction of polarization rotation in the vortices. Black arrows indicate the preferable polarization directions outside vortices. Black crosses inside the vortices indicate the component of the polarization orientation in the core. From Ref.\,\cite{Behera2022}.  
}
\label{FigSwitch}
\end{center}
\end{figure*} 

\subsubsection{\label{sec:Switching}Field-induced switching}


Chiral ferroelectric materials, in which the structural polar ordering is directly coupled to the electric field, appear to be the unique choice of materials for exploring field-tunable chirality.  They offer a remarkable operational platform for controlled optical activity manipulation by the electric field.
Notably, yet underexplored kinds of ferroelectrics are the structurally chiral ferroelectrics in which the chirality emerges together with the polarization via the spontaneous symmetry breaking from the non-chiral paraelectric state, referred to in Sec.\,\ref{sec:Struct} as a ferroelectric with polarization-induced chirality. The handedness of such materials is related to the orientation of polarization and is switchable by reversing the polarization by the applied field. The field-induced switchability of the sign of the optical rotary power associated with the handedness was demonstrated for the typical representatives of this class, the PGO\,\cite{Iwasaki1971} and TGS\,\cite{Koralewski1983}, see Fig.\,\ref{FigOptAct}(e),(f).

Here we focus mostly on the possibility of chirality switching in topological states of ferroelectrics. Note, first, that when the polarization topological state possesses definite handedness, the chirality can not be inversed just by the field-induced reversion of polarity in each point of the structure since the structure of the streamlines will remain the same. 
Therefore, the efficient changing handedness of the system requires reconstruction of the topological polarization ordering, occurring via passing through a series of intermediate metastable states via the pathway, guided by an electric field. 
  
Several theoretical studies addressed the problem of switching chirality by the electric field.
In particular, the work\,\cite{Prosandeev2013} showed that the optical activity of the chiral skyrmion-like tubes emerging in the array of nanowires of ferroelectric BaTiO$_3$ embedded in a solid solution Ba$_{0.15}$Sr$_{0.85}$TiO$_3$ can indeed be switched by the electrical field.
The calculations were further optimized in\,\cite{Walter2016}. The possibility of switching the chirality of polarization helical state arising in the helically-screwed nanowire by electric field was reported in\,\cite{Shimada2016}. Switching of the helical coreless polarization domain, having the structure of the chiral skyrmion, was also studied in\,\cite{Baudry2014}.  
Combining the application of the electric field with the special design of the geometry of the nanostructure or with the applied strain field was suggested as a  tool for manipulating chirality and vorticity\,\cite{VanLich2017,Yuan2018switch,Yuan2018}. \marker{ The dynamic motion of polar skyrmions in oxide heterostructures, that in particular can host chiral states, with integrated external thermal, electrical, and mechanical stimuli was recently suggested in\,\cite{Hu2023}}.

The switching of chirality in nanoparticles is of special interest because of its high technological potential. In particular, it offers remarkable perspectives for tunable chiral optoelectronics.  
Figure\,\ref{FigSwitch}(a) shows how the polarizing-depolarizing cycle for the Hopfion-hosting nanoparticle inverses its chirality.
For illustrative purposes, we simulated the isotropic case using the phase-field method, in which the strain-renormalized coefficients in the Ginzburg-Landau functional are close to those in  PbZr$_{0.6}$Ti$_{0.4}$O$_3$ but possesses the spherical symmetry. 
The switching pathway using the full Ginzburg-Landau functional for PbZr$_{0.6}$Ti$_{0.4}$O$_3$ is described in\,\cite{Lukyanchuk2020}. 

We consider a nanoparticle with the left-handed Hopfion $H_{ccw}^-$ (state 1). As we have already mentioned, the combination of the directed-down CCW helix-like central flux and directed-up CCW peripheral helix-like fluxes constitutes a Hopfion's structure. The total chirality is defined by the dominating contribution of the central part. Application of the electric field in the up direction i.e., against the polarization direction at the core, favors the peripheral helix and suppresses the central helix. 
\marker{With the increasing absolute value of the field, the topological phase transition occurs, at which the new topological states abruptly emerge inside the nanoparticles.}
The outer part of the upcoming helical streamlines unlink in the nanoparticle pole regions from the down-streaming central flux and pierces the nanoparticle, enveloping, thus, the shrinking Hopfion (states 2,3). Another topological phase transition occurs when the Hopfion completely disappears, and the up-directed helical stream occupies all the space of the nanoparticle (state 4).  Then, the helix continues the unwinding, leading finally to a complete poling of the nanoparticle. 

Notably, the chirality of the final upcoming helix is a succession of the chirality of the peripheral upcoming flux of the Hopfion, hence is opposite to the initial total Hopfion chirality. This chirality conserves on the way back to the zero-field state, where another sequence of topological phase transitions occurs. Upon decreasing the absolute value of the applied field, the new Hopfion appears in the nanoparticle. However, unlike in the field-increasing case, this Hopfion emerges at the periphery of the nanoparticle and embraces the upcoming polarization helix (state 5). Upon decreasing the field, it propagates to the nanoparticle center and, finally, occupies the full nanoparticle when the field vanishes (final state 6). The right-handed chirality and polarity of the newly-formed Hopfion, $H_{ccw}^+$ (state 6), dominated by its central part of the CCW-swirled flux, is inherited from the upcoming helix that remained from the high field configuration. To summarize here, the described process of the re-polarization of the ferroelectric nanoparticle results in deterministic chirality switching.  Importantly, the close-packing of the ferroelectric nanoparticles in the dielectric matrix enables the collective effects, resulting in even more exciting opportunities.

Experimentally, the possibility of the electrical control of chirality in the ferroelectric BaTiO$_3$ nanoparticles was announced in the already mentioned work\,\cite{Karpov2017}. There, the evolution of the toroidal moment under the action of the field was studied. However, as we mentioned above, measurements of the toroidal moment cannot be employed for chirality detection.  

Manipulation by the chirality in ferroelectric nanodots of PbTiO$_3$ was studied in the work\,\cite{Tikhonov2020}. It was shown that the cleverly designed protocol of the electrical field variation enables the field-induced switching between the degenerate topological states carrying the different polarities and chiralities Fig\,\ref{FigSwitch}(b). 
These states are the left-handed $H^+_{cw}$ and $H^-_{ccw}$, and right-handed $H^+_{ccw}$ and $H^-_{cw}$ half-Hopfions, referred to in\,\cite{Tikhonov2020} as skyrmions because of the particular skyrmion-like distribution of polarization at the nanodot-substrate interface, see also Sec.\,\ref{sec:nanopart}. The polarity and handedness of the first Hopfion emerging under the field- or zero-field cooling via the spontaneous symmetry loss is undetermined. This is, for example, the left-handed half-Hopfion $H_{ccw}^-$ (state 5) in Fig\,\ref{FigSwitch}(b). Then, the cycle of the consecutive topological transformations through states 6-8 is driven by the re-matching of the streamlines in the volume of the nanodot. Similarly to the described above case of the Hopfion chirality switching in ferroelectric nanoparticles, it results in the changing of the handedness of the nanodot, leading to the right-handed half-Hopfion  $H_{ccw}^+$ (state 9).  
%
%
We expect that in the ferroelectric nanodots the technology of manipulation by the topological states which is already in place,\,\cite{Li2017,Li2023} can enable the switching of chirality by the electric field.

Recently, more reports on the experimental realization of switching of topological chirality in ferroelectric/paraelectric superlattices with vortices have emerged. 
The approach of\,\cite{Behera2022} demonstrated in Fig.\,\ref{FigSwitch}(c) was based on switching the average chirality of the chiral vortex array in PbTiO$_3$ layers in PbTiO$_3$/SrTiO$_3$ superlattices arising when the sequential vortices with opposite chirality shift up and down with respect to the middle of the layer (see Sec.\,\ref{sec:Films}). 
\marker{As follows from Eq.\,(\ref{Pchi}), this switch implies the reversing direction of the polarization in the uniform in-plane $a$-phase, admixed to the vortex phase. Accordingly, to provide this switch, the application of the field, parallel to the film surface is required.}
The applied in-plane electric field switches the direction of the average in-plane polarization, and, therefore, the coupled to it vortex displacement and associated with its chirality. 
Also, the work\,\cite{Chen2022} demonstrated the possibility of the vortex chirality manipulation in a similar system through its complete poling of the superlattice to the monodomain c-phase by the electric filed, perpendicular to the superlattice layers with the destruction of the vortex state. 
It was shown that after the system relaxation following the removal of the electric fields back to the vortex state, the rotation of certain vortices and their handedness reverses. 
Although the last approach offers a good potential strategy for manipulating vortex chirality, more development is necessary to make the process more controllable.

\subsubsection{\label{sec:Thermal}Thermal and photo-induced manipulation}

\begin{figure*}
\begin{center}
\includegraphics [width=1\linewidth] {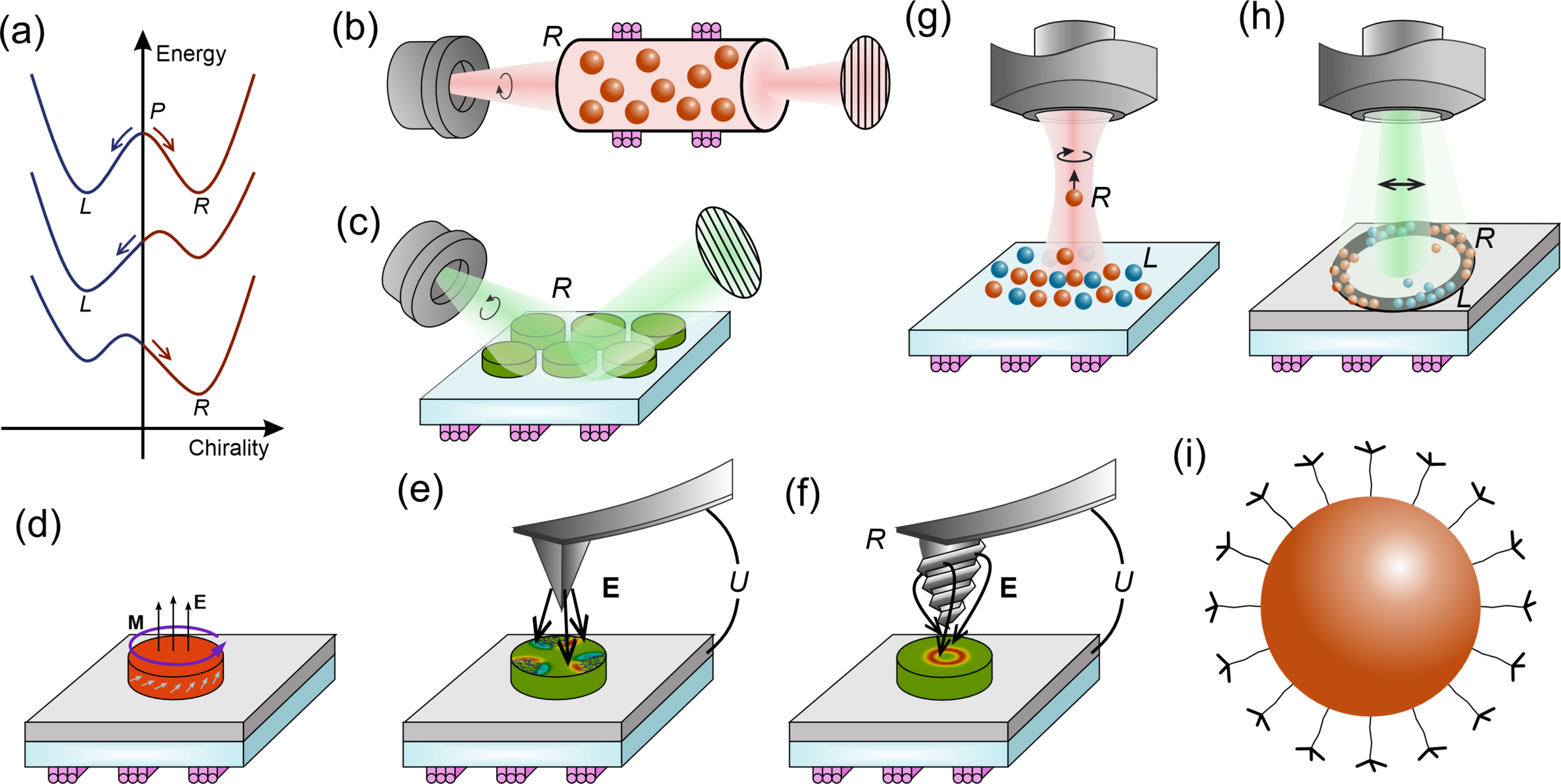}
\caption{ \textbf{Thermaly Methods of chirality manipulation.}  
\textbf{(a)}, The energy landscape enables the thermally stimulated methods of formation of the desirable chirality. 
\textbf{(b)}, The process of the formation of the desirable chirality in a ferroelectric nanoparticle-based composite material using thermal stimulation with the polarized laser beam as an external stimulus. 
\textbf{(c)}, The same for the ferroelectric nanodot array. 
\markertwo{\textbf{(d)} Torsion (M) - induced chirality is operating by the field (E) and heating. }
\textbf{(e)} Writing and operating chiral topological states in ferroelectric nanodots, using the AFM/PFM tip and heating.
\textbf{(f)} The same for the specially designed chiral AFM/PFM tip.
\textbf{(g)}, The enantioselective manipulating by the right-handed (red) and left-handed (blue) nanoparticles using the optical tweezer beam.
\textbf{(h)}, The enantioselective localization of the right-handed (red) and left-handed (blue) nanoparticles by the plasmonic traps, stimulated by the polarized optical tweezer beam.
\textbf{(i)}, The functionalization of the surface of the ferroelectric nanoparticle by chiral molecules serves to establish the desirable chirality of polarization texture in a nanoparticle.
}
\label{FigManip}
\end{center}
\end{figure*}

\marker{Figure\,\ref{FigManip} presents the ways of switching between the energy-equivalent ferroelectric states with opposite chiralities by the external chiral stimuli, in particular, by the chiral electromagnetic fields. The possibility of the chirality switching of the Hopfions confined within the nanoparticles and nanodots by employing the circularly polarized laser beam was suggested in\,\cite{Lukyanchuk2020,Tikhonov2020}. Another remarkable opportunity of chirality switching was presented in a recent work\,\cite{Gao2023} that has shown that the twisted light penetrating into the ultrathin ferroelectric film of PbZr$_{0.4}$Ti$_{0.6}$O$_3$ sandwiched between two electrodes can excite and drive the skyrmions emerging in the film and switch their handedness by passing through an intermediate achiral Néel skyrmion state.

However, the direct chirality switching by the circularly polarized light requires an overpassing of the energy barriers between the states with opposite chiralities that may require some substantial power of the incident beam.
In what follows, we present the developed in\,\cite{Lukyanchuk2021pat} thermally stimulated methods for switching and manipulating chirality in nanostructured ferroelectrics that are capable of overcoming such a requirement. 
}
To that end, we note that in accord with the fundamental principles of spontaneous symmetry breaking, the transition under the cooling from the achiral paraelectric (P) state occurs to one of the energy-equivalent ferroelectric states possessing the right- (R) or left- (L) hand chiralities.  
Figure\,\ref{FigManip}(a) presents the corresponding energy profile for the spontaneous chiral symmetry breaking. In the suggested  thermally stimulated methods of
formation of the desirable chirality, the final state is achieved  by small chiral perturbations having the required handedness, for instance, by chiral built-in disorder, chiral electric field, chiral mechanical forces, and structural or chemical stimuli that trigger a preferable transition pathway to the enantiomer of either the left- (the middle curve) or the right-hand (the lower curve) type.  

Figure\,\ref{FigManip}(b) shows the desirable chirality formation and switching in the transparent ferroelectric nanocomposite material in \marker{which} ferroelectric nanoparticles are embedded. Achieving the desirable chirality is realized by the incident beam of the circularly polarized radiation produced by the laser or by another radiation source.  Proper thermal cycling can be achieved by either the built-in heater (shown in violet) or by electromagnetic radiation, in particular by the same laser beam that triggers chirality. 
A cleverly designed thermally-cycling protocol in a proper combination with the circularly polarized external radiation flux or pulses triggers the desirable chirality and also stimulates the switching between the different chiral states while passing through the intermediate high-temperature achiral paraelectric state. 
To control the degree of chirality of the nanocomposite, the optical activity of the nanocomposite is measured by the external polarimeter.

Figure\,\ref{FigManip}(c) shows a similar  method for the desirable chirality formation and switching in the ferroelectric nanodot array deposited on the substrate. The method exploits the circularly polarized beam incident on the surface carrying nanodots. The control of the chirality is realized by measurements of the polarization rotation of the reflected radiation.

The methods discussed above can be applied separately or in combination. \markertwo{For instance, the torsion-induced chirality\,\cite{Yuan2018} can be manipulated by the field and temperature, see (Fig.\,\ref{FigManip}(d))}. The thermal activation of the handedness can be used in combination with manipulation by AFM/PFM tips (Fig.\,\ref{FigManip}(e)), in particular, with the specially-designed tips, producing electric field with the predefined chirality\,\cite{Zhao2017} (Fig.\,\ref{FigManip}(f)). These methods can be used for local manipulation with single nanoparticles, nanodots, films, and heterostructures, as well as for the global manipulation of the ensembles and arrays of nanoparticles and nanodots, and polarization textures in bulk materials. The thermal and photo-induced manipulation can be realized using ultra-short picosecond laser pulses or other thermally-assisted ways of chirality changing. 
Importantly, the corresponding technologies for the photo-induced and thermal domain creation and switching do already exist\,\cite{Stoica2019,Kimel2020,Zhang2023}.

\subsubsection{\label{sec:Tweezers}Tweezers}

The very efficient tool for operating the ferroelectrics is the optical tweezers and plasmonic optical tweezers\,\cite{Polimeno2018,Bradac2018,Crozier2019,Zhang2021,Ren2021}, which has already been used extensively for enantioselective manipulation with metallic nanoparticles. The optical tweezers and plasmonic optical tweezers enable us not only to sort the nanoparticles according to their handedness but also to deliver them at the required location. On top of that, the polarized laser beam of the tweezer can be used for the photo-induced switching and testing of the nanoparticles' chirality. 
The application of the optical tweezers to ferroelectrics has recently been suggested\,\cite{Lukyanchuk2021pat} but not yet sufficiently employed. 

\textit{Optical Tweezers} are the tools utilizing a tightly focused laser beam to trap particles, cells, viruses, and design nanostructures\,\cite{Polimeno2018,Bradac2018}. In a typical optical tweezer design, shown in Fig.\,\ref{FigManip}(g), the focused laser beam passes through a microscope lens to focus at the chosen spot near the surface of the sample. This spot acts as a trap capable to hold and manipulate a small dielectric object at the desired place.

The acting on the nanoparticle optical force $\mathbf{F}$ 
is calculated as a flux of the  Maxwell stress tensor $ \hat{\mathbf{T}}$ through the surface $S$, surrounding the particle\,\cite{Patti2019,Mun2020,Ye2017}: 
\begin{equation}
 \mathbf{F} 
 =\oint_S  \left\langle \hat{\mathbf{T}}\right\rangle d\mathbf{s}\,.
\label{MaxwellF}
\end{equation}
Here  $d\mathbf{s}$ is a surface element, and brackets stand for the
time averaging. The  Maxwell stress tensor 
is expressed through the electric, $\mathbf{E}$, and magnetic, $\mathbf{B}$, fields that include the incident
and scattered components \,\cite{Polimeno2018,Mun2020,Pfeifer2007},
\begin{equation}
T_{ij}=\varepsilon _{0}\left( E_{i}E_{j}+\frac{c^{2}}{n_0^2}B_{i}B_{j}-\frac{1}{2}\left(
E^{2}+\frac{c^{2}}{n_0^2}B^{2}\right) \delta _{ij}\right) .
\label{MaxwellT}
\end{equation}

Besides the straightforward transfer of the linear momentum that realizes the optical forces, the circularly polarized light is capable to transfer also the angular momentum producing the light-induced rotation of the irradiated particles. 
Within the same approach, the  optical torque 
$\mathbf{\Gamma }$ exerted on a particle is 
\begin{equation}
 \mathbf{\Gamma } =-\oint_S \left\langle 
\hat{\mathbf{T}}\times \mathbf{r}\right\rangle d\mathbf{s}.
\label{MaxwellG}
\end{equation}

A promising tweezer technique is the use of circularly-polarized light to trap chiral particles distinctively 
depending on their handedness. The basis of the principle is that the right-handed enantiomer experiences different optical force and torque than the left-handed enantiomer for a given input optical handedness. For instance, in the plane-wave decomposition model for a focused laser beam\,\cite{Patti2019,Ali2020} the optical forces and torques  are decomposed as 
\begin{equation}
\mathbf{F}=\mathbf{F}_{0}\left( \kappa \right) \pm \mathbf{F}_{\sigma }\left( \kappa
\right) ,\qquad \mathbf{\Gamma} =\mathbf{\Gamma }_{0}\left( \kappa \right) \pm 
\mathbf{\Gamma }_{\sigma }\left( \kappa \right).
\label{ChiralForce}
\end{equation}
Here the sign $\pm$ corresponds to the sign of the circular light polarization $\sigma=\pm 1$. Unfortunately, both light polarization-independent and light polarization-dependent contributions, denoted by the subscripts $0$ and $\sigma$ respectively,  depend on the particle's chirality parameter $\kappa $ in a
non-trivial way. The full expression for the optical force and torque is quite cumbersome and cannot be easily decomposed with respect to particle chirality parameter 
$\kappa $. Nevertheless, the chirality-dependent optical forces acting on small particles, such as enantioselective optical pressure,  torque, and optical of a chiral particle in a tightly focused circularly polarized laser beam can be calculated and used for the optical separation of enantiomers\,\cite{Canaguier2013,Patti2019,Ali2020,Li2020b}. The electromagnetic laser-induced binding force acting between illuminated by the polarized beam particles can be also strongly enantioselective\,\cite{Forbes2015}, allowing for the handedness-dependent manipulation and sorting of 
chiral particles.

However, conventional optical tweezers based on traditional optical microscopes are subject to the diffraction limit, making precise trapping and manipulating the nano-size particles challenging. 

\textit{Plasmonic Optical Tweezers} are capable to break the diffraction barrier\,\cite{Crozier2019,Zhang2021}. They have become commonly used for high-precision trapping and manipulation by micro- and nanometer-sized species in various areas of science and technology.  The plasmonic optical tweezers technique, shown Fig.\,\ref{FigManip}(h), is based on the excitation of surface plasmons at a  metal/dielectric interface by the evanescent light of the optical tweezer.  Plasmons localized at the specifically designed hotspots trap the targeted nanoparticles. Various methods have been suggested to deliver the nanoparticles into the hotspot area using the optical, electrodynamic, hydrodynamic, and other forces\,\cite{Ren2021}.

The plasmonic optical tweezers can be very useful in manipulating  the ferroelectric nanoparticles, in particular, those possessing definite chirality. Excitation of chiral surface plasmons by the circular-polarized laser beam induces the opposite transverse forces on the chiral specimens with opposite 
handedness\,\cite{Zhao2016}, providing an exciting opportunity for their sorting and other enantioselective applications.

\subsubsection{\label{sec:OtherManip}Other methods}

Apart from the switching of the chirality of nanostructural ferroelectrics by external electrical and strain fields and temperature, other methods of manipulation are possible. The handedness of nanoparticles can be controlled, for example, by the appropriate boundary conditions\,\cite{Govinden2021}, and by geometrical shaping of the sample\,\cite{Xiong2022}. 
Chemically assisted switching by the functionalization of the nanoparticle surface by chiral molecules of the definite handedness can enable another, the chemical way of chirality manipulation\,\cite{Zafar2020}, see Fig.\,\ref{FigManip}(i).

\section{\label{sec:Perspectives}Perspectives.}

\subsection{\label{sec:perspI}Optoelectronics and plasmonics.}
The discussed in the Review topological chirality of nanostructured ferroelectrics is a novel field creating a matchless opportunity for advancing optoelectronics and plasmonics. 
The unique emerging functionality is provided by the possibility of switching chirality by the electric field and other stimuli. This aligns with the occurring revolutionary development of the physics and applications of chiral plasmonic metallic nanoparticles\,\cite{Govorov2011,Hentschel2017,Kong2020,Valev2013,Neubrech2020,Wu2022} that usually host the built-in chirality. Combining switchable ferroelectric materials with plasmonic structures creates new opportunities for controlling light at the nanoscale. Implementing the ferroelectric nanostructures for plasmonic switches, modulators, and filters that can be actively controlled provides a high degree of flexibility in plasmonic device design.
This offers new opportunities for developing materials and metamaterials with switchable optical activity and plasmonic platforms and for enhancing chiroptical signals and developing chiral plasmonics for communication technologies.

Ferroelectric chiral nanoparticles can also be used to enhance the nonlinear optoelectronic properties of  materials. Combining chiral nanoparticles with plasmonic materials enables inducing strong second-harmonic generation and other nonlinear effects. This property can become useful in a wide range of applications, including chirality-dependent frequency conversion, harmonic generation, and nonlinear microscopy. 
A possible application lies in the development of chiral plasmonic devices, which can manipulate light with high sensitivity and specificity. 
The field-tunable circular dichroism of ferroelectric nanoparticles can be exploited for developing sensitive chiral field-effect transistors, sensors for chiral molecules\cite{Torsi2013}, and switchable chiral light-emitting transistor\,\cite{Zhang2014}.

Notably, the dynamical behavior of ferroelectric topological states has a vibrational character with the resonance frequency lying in the sub-Terahertz frequency band, 0.1--1\,THz.
These oscillations were simulated, calculated, and measured in\,\cite{Zhang2011,Pakhomov2013,Gui2014,Li2021}, and, as indicated in\,\cite{Lukyanchuk2018,Lukyanchuk2021}, they may be 
related to oscillations of the surface-bound charges being, therefore, nothing but the plasmonic resonance. This opens new perspectives for optoplasmonics at the sub-Terahertz frequencies.
Furthermore, the field-switchable handedness of the vibrating topological states makes the foreseen THz plasmonic applications tunable and chirality-sensitive.

The discussed in the Review topological chirality of nanostructured ferroelectrics is a novel field creating a matchless opportunity for advancing optoelectronics and plasmonics. 
The unique emerging functionality is provided by the possibility of switching chirality by the electric field and other stimuli. This aligns with the occurring revolutionary development of the physics and applications of chiral plasmonic metallic nanoparticles\,\cite{Govorov2011,Hentschel2017,Kong2020,Valev2013,Neubrech2020,Wu2022} that usually host the built-in chirality. Combining switchable ferroelectric materials with plasmonic structures creates new opportunities for controlling light at the nanoscale. Implementing the ferroelectric nanostructures for plasmonic switches, modulators, and filters that can be actively controlled provides a high degree of flexibility in plasmonic device design.
This offers new opportunities for developing materials and metamaterials with switchable optical activity and plasmonic platforms and for enhancing chiroptical signals and developing chiral plasmonics for communication technologies.

Ferroelectric chiral nanoparticles can also be used to enhance the nonlinear optoelectronic properties of materials. Combining chiral nanoparticles with plasmonic materials enables inducing strong second-harmonic generation and other nonlinear effects. This property can become useful in a wide range of applications, including chirality-dependent frequency conversion, harmonic generation, and nonlinear microscopy. 
A possible application lies in the development of chiral plasmonic devices, which can manipulate light with high sensitivity and specificity. 
The field-tunable circular dichroism of ferroelectric nanoparticles can be exploited for developing sensitive chiral field-effect transistors, sensors for chiral molecules\cite{Torsi2013}, and switchable chiral light-emitting transistor\,\cite{Zhang2014}.

\markertwo{The emergence of spatially extensive periodic chiral vortex stripes and polar chiral bubble arrays within ferroelectric thin films and superlattices introduces a novel capability for electric field-driven operational manipulation. The advanced engineering of such structures markedly improves interactions between light and matter, thereby offering the possibility of extraordinary control over chiroptical characteristics. Such progress lays the foundation for the creation of synthetic 3D materials, whose optical qualities may greatly exceed those found in natural optically active crystals. This development opens new horizons in fields such as advanced sensing technologies, telecommunications, and the realm of optically operating information processing devices.}

Notably, the dynamical behavior of ferroelectric topological states hosted \markertwo{by heterostructures}, nanoparticles, and nanorods has a vibrational character with the resonance frequency lying in the sub-Terahertz frequency band, 0.1--1\,THz.
These oscillations were simulated, calculated, and measured in\,\cite{Zhang2011,Pakhomov2013,Gui2014,Li2021}, and, as indicated in\,\cite{Lukyanchuk2018,Lukyanchuk2021}, they may be 
related to oscillations of the surface-bound charges being, therefore, nothing but the plasmonic resonance. This opens new perspectives for optoplasmonics at the sub-Terahertz frequencies.
Furthermore, the field-switchable handedness of the vibrating topological states makes the foreseen THz plasmonic applications tunable and chirality-sensitive.

\subsection{\label{sec:perspII}Nanoelectronics and computing.}

Ferroelectrics are traditionally considered a promising material for the construction of computing units, in particular, memory elements\,\cite{ScottBook}.  By applying the electric field, the polarization of ferroelectrics can be reversed allowing for binary data storage. In multistate ferroelectrics, for instance, in multiaxial cubic bulk ferroelectrics and strained films, the logic of the memory elements can  be extended to the multi-level state\,\cite{Baudry2017,Boni2017,Xu2019,Chen2022,Razumnaya2023},  realizing thus the non-von-Neumann computing circuits. The emergence of the topological states enables even more prospects for nanoelectronics,\,\cite{Catalan2012,Seidel2019}. In particular, multi-domain, multi-vortex, and skyrmion states offer quite a widened scope for implementing multi-level elements\,\cite{Martelli2015,Lukyanchuk2019,Omari2014}. Furthermore, ferroelectric domain wall-based synaptic and neuronal devices were suggested as a platform for neuromorphic computing\,\cite{Wang2022b,Jiang2019}. 
Ferroelectrics are traditionally considered a promising material for the construction of computing units, in particular, memory elements\,\cite{ScottBook}.  By applying the electric field, the polarization of ferroelectrics can be reversed allowing for binary data storage. In multistate ferroelectrics, for instance, in multiaxial cubic bulk ferroelectrics and strained films, the logic of the memory elements can  be extended to the multi-level state\,\cite{Baudry2017,Boni2017,Xu2019,Chen2022,Razumnaya2023},  realizing thus the non-von-Neumann computing circuits. The emergence of the topological states enables even more prospects for nanoelectronics,\,\cite{Catalan2012,Seidel2019}. In particular, multi-domain, multi-vortex, and skyrmion states offer quite a widened scope for implementing multi-level elements\,\cite{Martelli2015,Lukyanchuk2019,Omari2014}. Furthermore, ferroelectric domain wall-based synaptic and neuronal devices were suggested as a platform for neuromorphic computing\,\cite{Wang2022b,Jiang2019}. 

The switchable chirality of nanostructured ferroelectrics holds the potential for being utilized as an extra dimension for information storage. By using chiral polarization topological states as storage elements, the memory cells can be designed to store information in the form of enantiomers, which could potentially increase the storage density of the memory that can be readable and writable by the circularly polarized light. 
Another approach is to use chiral THz-plasmonic ferroelectric nanoparticles to enhance the performance of the existing RAM technologies. Ferroelectric nanoparticles hosting chiral plasmonic oscillations can be used as the THz antennas and emitters to control the  propagation of the THz radiation waves between communicating computing chips, allowing for a faster and more efficient data transfer. The discrimination between the left and right-handed polarization rotation of the propagating waves can be used as one more degree of freedom for the densification of the transferring data flux.  

 The tunable chirality  has potential applications in quantum communications by improving the efficiency of the photon sources used for transmitting quantum information\,\cite{Long2020}. The originally proposed photon sources were supposed to emit unpolarized photons since using the polarized photon states was viewed as reducing their efficiency. Modern protocols propose to use quantum phase variations for encrypting information\,\cite{Kirsanov2021}. Ferroelectric nanodots and nanoparticles, hosting chiral topological states offer a remarkable platform for high-efficiency devices capable to produce photons with circular polarization. These devices realizing modern quantum communications protocols will enable quantum communication technologies, in particular, quantum internet\,\cite{Singh2021}. 
 

\subsection{\label{sec:perspIII}Chemistry and  bio-medical applications.}
 Many biological molecules, such as DNA and proteins, have a distinct chirality. Accordingly, certain chiral biomolecules may appear as the cause of neurogenerative disorders such as Parkinson’s, Alzheimer’s, and Huntington’s when entering the human body if their original handedness is altered\,\cite{Lininger2022,Mitra2022}. This makes it crucial for the fields like pharmacology, toxicology, and pharmacodynamics to be capable to detect, separate, and synthesize substances according to their chirality\,\cite{Brooks2011,Pinto2020}. However, the synthesis of chiral molecules can present various challenges due to the difficulty of exercising an efficient stereochemical control, in particular, the ability to selectively produce one enantiomer over the other. The tunable chirality of nanostructured ferroelectrics can be used to overcome these challenges by allowing for selective synthesis of a single enantiomer, control of stereochemistry, scalability, cost-effectiveness, and high outcome. Furthermore, by tuning the chirality of the reaction environment, it is possible to achieve stereochemical control and improve the efficiency of chiral molecule synthesis. 
 Nanoparticles with tunable chirality also possess the potential for functioning as catalysts for targeted chemical reactions involving chiral molecules. This property can be harnessed for a range of practical applications such as drug development and other related fields.

Chiral plasmonic nanoparticles can serve as a useful tool in biosensing applications by modifying their surface with biomolecules to selectively detect certain analytes, such as viruses and bacteria. This is a powerful sensing technique that utilizes the optical properties of chiral plasmonic nanostructures to detect and analyze biomolecules with high sensitivity and specificity and has potential applications in various fields of biomedical research and diagnostics\,\cite{Xia2011}. However, the handedness of the currently used plasmonic metamaterials is defined by the shape of the constitutive metallic nanoparticles and is predetermined by their synthesis\,\cite{Mitra2022,Paiva-Marques2020,Mejia2018}. Ferroelectric nanoparticles with tunable chirality can become particularly useful for chiral plasmonic biosensing because they can be designed with precise and adjustable chiral properties, which can enhance the sensitivity and selectivity of the biosensing system. By tuning the chirality of the nanoparticles, it is possible to optimize their interactions with specific biomolecules and improve the accuracy of the biosensing measurements. Additionally, they can be functionalized with a variety of biomolecules and polymers\,\cite{Zhang2023bioelectronics}  to selectively target analytes with high specificity.

\section{\label{sec:ACKNOWLEDGEMENTS}ACKNOWLEDGEMENTS}
 The authors acknowledge Terra Quantum AG for the support of the work.
 I.L. and  Y.T. acknowledge the support of the European Union H2020-MSCA-ITN-MANIC action, (GA 861153), and of the European Union H2020-MSCA-RISE-MELON action (GA 872631), I.L. and S.K. acknowledge the support of the EU H2020-MSCA-RISE-MILEAGE action (GA 734931), A.R acknowledges the support of the European Union HORIZON-WIDERA-2022-TALENTS–ERA Fellowship-FerroChiral (GA 101090285), S.K. acknowledges the support from the Alexander von Humboldt Foundation.



\bibliographystyle{apalike_custom}
\bibliography{ChiRevREVMODPHYS}

\begin{thebibliography}{}

\bibitem[Abid et~al., 2021]{Abid2021}
Abid, A.~Y., Sun, Y., Hou, X., Tan, C., Zhong, X., Zhu, R., Chen, H., Qu, K., Li, Y., Wu, M., et~al. (2021).
\newblock Creating polar antivortex in {PbTiO$_3$/SrTiO$_3$} superlattice.
\newblock {\em Nat. Commun.}, 12(1):2054.

\bibitem[Ackerman and Smalyukh, 2016]{Ackerman2016}
Ackerman, P.~J. and Smalyukh, I.~I. (2016).
\newblock Static three-dimensional topological solitons in fluid chiral ferromagnets and colloids.
\newblock {\em Nat. Mater.}, 16(4):426--432.

\bibitem[Ado et~al., 2020]{Ado2020}
Ado, I., Qaiumzadeh, A., Brataas, A., and Titov, M. (2020).
\newblock Chiral ferromagnetism beyond {L}ifshitz invariants.
\newblock {\em Phys. Rev. B}, 101(16):161403.

\bibitem[Aguado-Puente and Junquera, 2008]{Aguado-Puente2008}
Aguado-Puente, P. and Junquera, J. (2008).
\newblock Ferromagneticlike closure domains in ferroelectric ultrathin films: {F}irst-principles simulations.
\newblock {\em Phys. Rev. Lett.}, 100(17):177601.

\bibitem[Aguado-Puente and Junquera, 2012]{Aguado-Puente2012}
Aguado-Puente, P. and Junquera, J. (2012).
\newblock Structural and energetic properties of domains in {PbTiO$_3$/SrTiO$_3$} superlattices from first principles.
\newblock {\em Phys. Rev. B}, 85(18):184105.

\bibitem[Ali et~al., 2020]{Ali2020}
Ali, R., Pinheiro, F.~A., Dutra, R.~S., Rosa, F.~S., and Neto, P. A.~M. (2020).
\newblock Enantioselective manipulation of single chiral nanoparticles using optical tweezers.
\newblock {\em Nanoscale}, 12(8):5031--5037.

\bibitem[Arnold and Khesin, 2021]{Arnold2021}
Arnold, V.~I. and Khesin, B.~A. (2021).
\newblock {\em Topological methods in hydrodynamics}, volume 125.
\newblock Springer Science \& Business Media.

\bibitem[Bahr and Kitzerow, 2001]{Bahr2001}
Bahr, C. and Kitzerow, H.-S. (2001).
\newblock {\em Chirality in liquid crystals}.
\newblock Springer.

\bibitem[Baudry et~al., 2017]{Baudry2017}
Baudry, L., Lukyanchuk, I., and Vinokur, V.~M. (2017).
\newblock Ferroelectric symmetry-protected multibit memory cell.
\newblock {\em Sci. Rep.}, 7(1):1--7.

\bibitem[Baudry et~al., 2011]{Baudry2011}
Baudry, L., Sen{\'e}, A., Luk'Yanchuk, I.~A., and Lahoche, L. (2011).
\newblock Vortex state in thin films of multicomponent ferroelectrics.
\newblock {\em Thin Solid Films}, 519(17):5808--5810.

\bibitem[Baudry et~al., 2014]{Baudry2014}
Baudry, L., Sen{\'e}, A., Luk'yanchuk, I.~A., Lahoche, L., and Scott, J.~F. (2014).
\newblock Polarization vortex domains induced by switching electric field in ferroelectric films with circular electrodes.
\newblock {\em Phys. Rev. B}, 90(2):024102.

\bibitem[Behera et~al., 2022]{Behera2022}
Behera, P., May, M.~A., G{\'o}mez-Ortiz, F., Susarla, S., Das, S., Nelson, C.~T., Caretta, L., Hsu, S.-L., McCarter, M.~R., Savitzky, B.~H., et~al. (2022).
\newblock Electric field control of chirality.
\newblock {\em Sci. Adv.}, 8(1):eabj8030.

\bibitem[Blinc and Levanyuk, 1986]{BlincBook}
Blinc, R. and Levanyuk, A. (1986).
\newblock Incommensurate phases in dielectrics. {1. Fundamentals, 2. Materials}.

\bibitem[Bogdanov et~al., 1989]{Bogdanov1989ftt}
Bogdanov, A., Kudinov, M., and Yablonskii, D. (1989).
\newblock To the theory of magnetic vortices in easy axis ferromagnets.
\newblock {\em Sov. Phys. Solid State}, 31(10):1707.

\bibitem[Bonacina et~al., 2020]{Bonacina2020}
Bonacina, L., Brevet, P.-F., Finazzi, M., and Celebrano, M. (2020).
\newblock Harmonic generation at the nanoscale.
\newblock {\em J. Appl. Phys.}, 127(23):230901.

\bibitem[Boni et~al., 2017]{Boni2017}
Boni, G.~A., Filip, L.~D., Chirila, C., Pasuk, I., Negrea, R., Pintilie, I., and Pintilie, L. (2017).
\newblock Multiple polarization states in symmetric ferroelectric heterostructures for multi-bit non-volatile memories.
\newblock {\em Nanoscale}, 9(48):19271--19278.

\bibitem[Bradac, 2018]{Bradac2018}
Bradac, C. (2018).
\newblock Nanoscale optical trapping: {A} review.
\newblock {\em Adv. Opt. Mater.}, 6(12):1800005.

\bibitem[Bratkovsky and Levanyuk, 2000]{Bratkovsky2000}
Bratkovsky, A. and Levanyuk, A. (2000).
\newblock Abrupt appearance of the domain pattern and fatigue of thin ferroelectric films.
\newblock {\em Phys. Rev. Lett.}, 84(14):3177--3180.

\bibitem[Bratkovsky and Levanyuk, 2009]{Bratkovsky2009}
Bratkovsky, A. and Levanyuk, A. (2009).
\newblock Continuous theory of ferroelectric states in ultrathin films with real electrodes.
\newblock {\em J. Comput. Theor. Nanosci.}, 6(3):465--489.

\bibitem[Brooks et~al., 2011]{Brooks2011}
Brooks, W.~H., Guida, W.~C., and Daniel, K.~G. (2011).
\newblock The significance of chirality in drug design and development.
\newblock {\em Curr. Top. Med. Chem.}, 11(7):760--770.

\bibitem[Canaguier-Durand et~al., 2013]{Canaguier2013}
Canaguier-Durand, A., Hutchison, J.~A., Genet, C., and Ebbesen, T.~W. (2013).
\newblock Mechanical separation of chiral dipoles by chiral light.
\newblock {\em New J. Phys.}, 15(12):123037.

\bibitem[Carbone and Bikondoa, 2023]{Carbone2023}
Carbone, D. and Bikondoa, O. (2023).
\newblock Focused and coherent {X}-ray beams for advanced microscopies.
\newblock {\em Nucl. Instrum. Methods Phys. Res. B: Beam Interact. Mater. At.}, 539:127--135.

\bibitem[Catalan et~al., 2012]{Catalan2012}
Catalan, G., Seidel, J., Ramesh, R., and Scott, J.~F. (2012).
\newblock Domain wall nanoelectronics.
\newblock {\em Rev. Mod. Phys.}, 84(1):119.

\bibitem[Chandrasekhar, 2006]{Chandrasekhar2006}
Chandrasekhar, S. (2006).
\newblock {\em Chirality in liquid crystals}.
\newblock Springer Science \& Business Media.

\bibitem[Chauleau et~al., 2020]{Chauleau2020}
Chauleau, J.-Y., Chirac, T., Fusil, S., Garcia, V., Akhtar, W., Tranchida, J., Thibaudeau, P., Gross, I., Blouzon, C., Finco, A., et~al. (2020).
\newblock Electric and antiferromagnetic chiral textures at multiferroic domain walls.
\newblock {\em Nat. Mater.}, 19(4):386--390.

\bibitem[Chen et~al., 2022a]{Chen2022}
Chen, P., Tan, C., Jiang, Z., Gao, P., Sun, Y., Wang, L., Li, X., Zhu, R., Liao, L., Hou, X., et~al. (2022a).
\newblock Electrically driven motion, destruction, and chirality change of polar vortices in oxide superlattices.
\newblock {\em Sci. China: Phys. Mech. Astron.}, 65(3):1--8.

\bibitem[Chen et~al., 2022b]{Chen2022b}
Chen, P., Zhao, H.~J., Prosandeev, S., Artyukhin, S., and Bellaiche, L. (2022b).
\newblock Microscopic origin of the electric {D}zyaloshinskii-{M}oriya interaction.
\newblock {\em Phys. Rev. B}, 106(22):224101.

\bibitem[Chen et~al., 2021]{Chen2021}
Chen, S., Yuan, S., Hou, Z., Tang, Y., Zhang, J., Wang, T., Li, K., Zhao, W., Liu, X., Chen, L., et~al. (2021).
\newblock Recent progress on topological structures in ferroic thin films and heterostructures.
\newblock {\em Adv. Mater.}, 33(6):2000857.

\bibitem[Chen et~al., 2022c]{Chen2022Chir}
Chen, Y., Du, W., Zhang, Q., {\'A}valos-Ovando, O., Wu, J., Xu, Q.-H., Liu, N., Okamoto, H., Govorov, A.~O., Xiong, Q., et~al. (2022c).
\newblock Multidimensional nanoscopic chiroptics.
\newblock {\em Nat. Rev. Phys.}, 4(2):113--124.

\bibitem[Cheng et~al., 1992]{Cheng1992}
Cheng, H., Ma, J., Zhao, Z., Qiang, D., Li, Y., and Yao, X. (1992).
\newblock Hydrothermal synthesis of acicular lead titanate fine powders.
\newblock {\em J. Am. Chem. Soc.}, 75(5):1123--1128.

\bibitem[Cheong and Mostovoy, 2007]{Cheong2007}
Cheong, S.-W. and Mostovoy, M. (2007).
\newblock Multiferroics: {A} magnetic twist for ferroelectricity.
\newblock {\em Nat. Mater.}, 6(1):13--20.

\bibitem[Cherifi-Hertel et~al., 2017]{Cherifi2017}
Cherifi-Hertel, S., Bulou, H., Hertel, R., Taupier, G., Dorkenoo, K. D.~H., Andreas, C., Guyonnet, J., Gaponenko, I., Gallo, K., and Paruch, P. (2017).
\newblock Non-{I}sing and chiral ferroelectric domain walls revealed by nonlinear optical microscopy.
\newblock {\em Nat. Commun.}, 8:15768.

\bibitem[Cherifi-Hertel et~al., 2021]{Cherifi2021}
Cherifi-Hertel, S., Voulot, C., Acevedo-Salas, U., Zhang, Y., Cr{\'e}gut, O., Dorkenoo, K.~D., and Hertel, R. (2021).
\newblock Shedding light on non-{I}sing polar domain walls: {I}nsight from second harmonic generation microscopy and polarimetry analysis.
\newblock {\em J. Appl. Phys.}, 129(8):081101.

\bibitem[Collins et~al., 2017]{Collins2017}
Collins, J.~T., Kuppe, C., Hooper, D.~C., Sibilia, C., Centini, M., and Valev, V.~K. (2017).
\newblock Chirality and chiroptical effects in metal nanostructures: {F}undamentals and current trends.
\newblock {\em Adv. Opt. Mater.}, 5(16):1700182.

\bibitem[Crozier, 2019]{Crozier2019}
Crozier, K.~B. (2019).
\newblock Quo vadis, plasmonic optical tweezers?
\newblock {\em Light Sci. Appl.}, 8(1):1--6.

\bibitem[Curie, 1894]{Curie1894}
Curie, P. (1894).
\newblock Sur la sym{\'e}trie dans les ph{\'e}nom{\`e}nes physiques, sym{\'e}trie d'un champ {\'e}lectrique et d'un champ magn{\'e}tique.
\newblock {\em J. Phys. Theor. Appl.}, 3(1):393--415.

\bibitem[Dai et~al., 2023]{Dai2023}
Dai, C., Hong, Z., Das, S., Tang, Y.-L., Martin, L.~W., Ramesh, R., and Chen, L.-Q. (2023).
\newblock Strain effects on stability of topological ferroelectric polar configurations in {(PbTiO$_3$)$_n$/(SrTiO$_3$)$_n$} superlattices.
\newblock {\em Appl. Phys. Lett.}, 123(5).

\bibitem[Das et~al., 2021a]{DasM2021raman}
Das, M., Gangopadhyay, D., {\v{S}}ebest{\'\i}k, J., Habartov{\'a}, L., Michal, P., Kapit{\'a}n, J., and Bou{\v{r}}, P. (2021a).
\newblock Chiral detection by induced surface-enhanced {R}aman optical activity.
\newblock {\em Chem. Commun.}, 57(52):6388--6391.

\bibitem[Das et~al., 2018]{Das2018}
Das, S., Ghosh, A., McCarter, M.~R., Hsu, S.-L., Tang, Y.-L., Damodaran, A.~R., Ramesh, R., and Martin, L.~W. (2018).
\newblock Perspective: {E}mergent topologies in oxide superlattices.
\newblock {\em APL Mater.}, 6(10):100901.

\bibitem[Das et~al., 2020]{Das2020}
Das, S., Hong, Z., McCarter, M., Shafer, P., Shao, Y.-T., Muller, D., Martin, L., and Ramesh, R. (2020).
\newblock A new era in ferroelectrics.
\newblock {\em APL Mater.}, 8(12):120902.

\bibitem[Das et~al., 2021b]{Das2021}
Das, S., Hong, Z., Stoica, V., Gon{\c{c}}alves, M., Shao, Y.-T., Parsonnet, E., Marksz, E.~J., Saremi, S., McCarter, M., Reynoso, A., et~al. (2021b).
\newblock Local negative permittivity and topological phase transition in polar skyrmions.
\newblock {\em Nat. Mater.}, 20(2):194--201.

\bibitem[Das et~al., 2023]{Das2023}
Das, S., McCarter, M.~R., G{\'o}mez-Ortiz, F., Tang, Y.-L., Hong, Z., Ghosh, A., Shafer, P., Garc{\'\i}a-Fern{\'a}ndez, P., Junquera, J., Martin, L.~W., et~al. (2023).
\newblock Pure chiral polar vortex phase in {PbTiO$_3$/SrTiO$_3$} superlattices with tunable circular dichroism.
\newblock {\em Nano Letters}.

\bibitem[Das et~al., 2019]{Das2019}
Das, S., Tang, Y., Hong, Z., Gon{\c{c}}alves, M., McCarter, M., Klewe, C., Nguyen, K., G{\'o}mez-Ortiz, F., Shafer, P., Arenholz, E., et~al. (2019).
\newblock Observation of room-temperature polar skyrmions.
\newblock {\em Nature}, 568(7752):368--372.

\bibitem[De~Gennes and Prost, 1993]{DeGennes1993}
De~Gennes, P.-G. and Prost, J. (1993).
\newblock {\em The physics of liquid crystals}.
\newblock Oxford {U}niversity {P}ress.

\bibitem[De~Guerville et~al., 2005]{DeGuerville2005}
De~Guerville, F., Luk’yanchuk, I., Lahoche, L., and El~Marssi, M. (2005).
\newblock Modeling of ferroelectric domains in thin films and superlattices.
\newblock {\em Mater. Sci. Eng. B.}, 120(1-3):16--20.

\bibitem[Decker et~al., 2009]{Decker2009}
Decker, M., Ruther, M., Kriegler, C., Zhou, J., Soukoulis, C., Linden, S., and Wegener, M. (2009).
\newblock Strong optical activity from twisted-cross photonic metamaterials.
\newblock {\em Opt. Lett.}, 34(16):2501--2503.

\bibitem[di~Gregorio et~al., 2020]{Gregorio2020}
di~Gregorio, M.~C., Shimon, L.~J., Brumfeld, V., Houben, L., Lahav, M., and van~der Boom, M.~E. (2020).
\newblock Emergence of chirality and structural complexity in single crystals at the molecular and morphological levels.
\newblock {\em Nat. Commun.}, 11(1):1--9.

\bibitem[Di~Rino et~al., 2020]{DiRino2020}
Di~Rino, F., Sepliarsky, M., and Stachiotti, M. (2020).
\newblock Topology of the polarization field in {PbTiO$_3$} nanoparticles of different shapes by atomic-level simulations.
\newblock {\em J. Appl. Phys.}, 127(14):144101.

\bibitem[Dong et~al., 2019]{Dong2019}
Dong, Y., Zhang, Y., Li, X., Feng, Y., Zhang, H., and Xu, J. (2019).
\newblock Chiral perovskites: {P}romising materials toward next-generation optoelectronics.
\newblock {\em Small}, 15(39):1902237.

\bibitem[Dzyaloshinsky, 1958]{Dzyaloshinsky1958}
Dzyaloshinsky, I. (1958).
\newblock A thermodynamic theory of “weak” ferromagnetism of antiferromagnetics.
\newblock {\em J. Phys. Chem. Solids.}, 4(4):241--255.

\bibitem[Erb and Hlinka, 2020]{Erb2020}
Erb, K. and Hlinka, J. (2020).
\newblock Vector, bidirector, and bloch skyrmion phases induced by structural crystallographic symmetry breaking.
\newblock {\em Phys. Rev. B}, 102(2):024110.

\bibitem[Erdem et~al., 2006]{Erdem2006}
Erdem, E., Semmelhack, H.-C., B{\"o}ttcher, R., Rumpf, H., Banys, J., Matthes, A., Gl{\"a}sel, H.-J., Hirsch, D., and Hartmann, E. (2006).
\newblock Study of the tetragonal-to-cubic phase transition in {PbTiO$_3$} nanopowders.
\newblock {\em J. Phys. Condens. Matter}, 18(15):3861.

\bibitem[F{\'a}bry et~al., 2022]{Fabry2022}
F{\'a}bry, J., Ku{\v{c}}er{\'a}kov{\'a}, M., Du{\v{s}}ek, M., Buixaderas, E., and Hlinka, J. (2022).
\newblock Structure of the high-temperature phase of caesium nitrate--the importance of high-resolution data.
\newblock {\em Acta Cryst. B}, 78(2).

\bibitem[Fernandez et~al., 2022]{Fernandez2022}
Fernandez, A., Acharya, M., Lee, H.-G., Schimpf, J., Jiang, Y., Lou, D., Tian, Z., and Martin, L.~W. (2022).
\newblock Thin-film ferroelectrics.
\newblock {\em Adv. Mater.}, 34(30):2108841.

\bibitem[Fert et~al., 2017]{Fert2017}
Fert, A., Reyren, N., and Cros, V. (2017).
\newblock Magnetic skyrmions: {A}dvances in physics and potential applications.
\newblock {\em Nat. Rev. Mater.}, 2(7):1--15.

\bibitem[Fiebig et~al., 2016]{Fiebig2016}
Fiebig, M., Lottermoser, T., Meier, D., and Trassin, M. (2016).
\newblock The evolution of multiferroics.
\newblock {\em Nat. Rev. Mater.}, 1(8):1--14.

\bibitem[Finn and Antonsen~Jr, 1985]{Finn1985}
Finn, J.~M. and Antonsen~Jr, T.~M. (1985).
\newblock Magnetic helicity: what is it and what is it good for?
\newblock {\em Comments Plasma Phys. Controlled Fusion}, 9:111--126.

\bibitem[Forbes and Andrews, 2015]{Forbes2015}
Forbes, K.~A. and Andrews, D.~L. (2015).
\newblock Chiral discrimination in optical binding.
\newblock {\em Phys. Rev. A}, 91(5):053824.

\bibitem[Fusil et~al., 2022]{Fusil2022}
Fusil, S., Chauleau, J.-Y., Li, X., Fischer, J., Dufour, P., L{\'e}veill{\'e}, C., Carr{\'e}t{\'e}ro, C., Jaouen, N., Viret, M., Gloter, A., et~al. (2022).
\newblock Polar chirality in {BiFeO$_3$} emerging from a peculiar domain wall sequence.
\newblock {\em Adv. Electr. Mat.}, page 2101155.

\bibitem[Gao et~al., 2023]{Gao2023}
Gao, L., Prokhorenko, S., Nahas, Y., and Bellaiche, L. (2023).
\newblock Dynamical control of topology in ferroelectric skyrmions via twisted light.
\newblock {\em arXiv preprint arXiv:2302.01402}.

\bibitem[Gladkii and Sidnenko, 1972]{Gladkii1972}
Gladkii, V. and Sidnenko, E. (1972).
\newblock Double dielectric hysteresis loop of {KD$_2$PO$_4$} crystals.

\bibitem[Glinchuk et~al., 2013]{Glinchuk2013}
Glinchuk, M., Ragulya, A., and Stephanovich, V. (2013).
\newblock {\em Nanoferroics}.
\newblock Springer.

\bibitem[G{\"o}bel et~al., 2021]{Gobel2021}
G{\"o}bel, B., Mertig, I., and Tretiakov, O.~A. (2021).
\newblock Beyond skyrmions: {R}eview and perspectives of alternative magnetic quasiparticles.
\newblock {\em Phys. Rep.}, 895:1--28.

\bibitem[Gon{\c{c}}alves et~al., 2023]{Gonccalves2023}
Gon{\c{c}}alves, M.~A., Pa{\'s}ciak, M., and Hlinka, J. (2023).
\newblock Antiskyrmionic ferroelectric medium.
\newblock {\em arXiv preprint arXiv:2303.07389}.

\bibitem[Govinden et~al., 2023a]{Govinden2023rev}
Govinden, V., Prokhorenko, S., Zhang, Q., Rijal, S., Nahas, Y., Bellaiche, L., and Valanoor, N. (2023a).
\newblock Spherical ferroelectric solitons.
\newblock {\em Nat. Mater.}, 22(5):553--561.

\bibitem[Govinden et~al., 2023b]{Govinden2023bubble}
Govinden, V., Rijal, S., Zhang, Q., Nahas, Y., Bellaiche, L., Valanoor, N., and Prokhorenko, S. (2023b).
\newblock Stability of ferroelectric bubble domains.
\newblock {\em Phys. Rev. Mater.}, 7(1):L011401.

\bibitem[Govinden et~al., 2021]{Govinden2021}
Govinden, V., Rijal, S., Zhang, Q., Sando, D., Prokhorenko, S., Nahas, Y., Bellaiche, L., and Valanoor, N. (2021).
\newblock Controlling topological defect transitions in nanoscale lead zirconate titanate heterostructures.
\newblock {\em Phys. Rev. Mater.}, 5(12):124205.

\bibitem[Govinden et~al., 2023c]{Govinden2023superlat}
Govinden, V., Tong, P., Guo, X., Zhang, Q., Mantri, S., Seyfouri, M.~M., Prokhorenko, S., Nahas, Y., Wu, Y., Bellaiche, L., et~al. (2023c).
\newblock Ferroelectric solitons crafted in epitaxial bismuth ferrite superlattices.
\newblock {\em Nat. Commun.}, 14(1):4178.

\bibitem[Govorov et~al., 2011]{Govorov2011}
Govorov, A.~O., Gun'ko, Y.~K., Slocik, J.~M., G{\'e}rard, V.~A., Fan, Z., and Naik, R.~R. (2011).
\newblock Chiral nanoparticle assemblies: circular dichroism, plasmonic interactions, and exciton effects.
\newblock {\em J. Mater. Chem.}, 21(42):16806--16818.

\bibitem[Gr{\"u}nebohm et~al., 2021]{Grunebohm2021}
Gr{\"u}nebohm, A., Marathe, M., Khachaturyan, R., Schiedung, R., Lupascu, D.~C., and Shvartsman, V.~V. (2021).
\newblock Interplay of domain structure and phase transitions: {T}heory, experiment and functionality.
\newblock {\em J. Phys. Condens. Matter}, 34(7):073002.

\bibitem[Guan et~al., 2020]{Guan2020}
Guan, Z., Hu, H., Shen, X., Xiang, P., Zhong, N., Chu, J., and Duan, C. (2020).
\newblock Recent progress in two-dimensional ferroelectric materials.
\newblock {\em Adv. Electron. Mater.}, 6(1):1900818.

\bibitem[Gui and Bellaiche, 2014]{Gui2014}
Gui, Z. and Bellaiche, L. (2014).
\newblock Terahertz dynamics of ferroelectric vortices from first principles.
\newblock {\em Phys. Rev. B}, 89(6):064303.

\bibitem[Guo et~al., 2022]{Guo2022}
Guo, X., Zhou, L., Roul, B., Wu, Y., Huang, Y., Das, S., and Hong, Z. (2022).
\newblock Theoretical understanding of polar topological phase transitions in functional oxide heterostructures: {A} review.
\newblock {\em Small Methods}, page 2200486.

\bibitem[Halasyamani and Poeppelmeier, 1998]{Halasyamani1998}
Halasyamani, P.~S. and Poeppelmeier, K.~R. (1998).
\newblock Noncentrosymmetric oxides.
\newblock {\em Chem. Mater.}, 10(10):2753--2769.

\bibitem[Harnagea and Pignolet, 2004]{Harnagea2004}
Harnagea, C. and Pignolet, A. (2004).
\newblock Challenges in the analysis of the local piezoelectric response.
\newblock In {\em Nanoscale Characterisation of Ferroelectric Materials: Scanning Probe Microscopy Approach}, pages 45--85. Springer.

\bibitem[Hegstrom and Kondepudi, 1990]{Hegstrom1990}
Hegstrom, R.~A. and Kondepudi, D.~K. (1990).
\newblock The handedness of the universe.
\newblock {\em Sci. Am.}, 262(1):108--115.

\bibitem[Hellman et~al., 2017]{Hellman2017}
Hellman, F., Hoffmann, A., Tserkovnyak, Y., Beach, G.~S., Fullerton, E.~E., Leighton, C., MacDonald, A.~H., Ralph, D.~C., Arena, D.~A., D{\"u}rr, H.~A., et~al. (2017).
\newblock Interface-induced phenomena in magnetism.
\newblock {\em Rev. Mod. Phys.}, 89(2):025006.

\bibitem[Hentschel et~al., 2017]{Hentschel2017}
Hentschel, M., Sch{\"a}ferling, M., Duan, X., Giessen, H., and Liu, N. (2017).
\newblock Chiral plasmonics.
\newblock {\em Sci. Adv.}, 3(5):e1602735.

\bibitem[Hill and McMorrow, 1996]{Hill1995}
Hill, J.~P. and McMorrow, D.~F. (1996).
\newblock {Resonant exchange scattering: {P}olarization dependence and correlation function}.
\newblock {\em Acta Crystallogr. A}, 52(2):236--244.

\bibitem[Hopf, 1931]{Hopf1931}
Hopf, H. (1931).
\newblock Über die abbildungen der dreidimensionalen sphäre auf die kugelfläche.
\newblock {\em Math. Ann.}, 104:637--665.

\bibitem[Hossain and Gu, 2014]{Hossain2014}
Hossain, M.~M. and Gu, M. (2014).
\newblock Fabrication methods of 3{D} periodic metallic nano/microstructures for photonics applications.
\newblock {\em Laser Photonics Rev.}, 8(2):233--249.

\bibitem[Hu et~al., 2023]{Hu2023}
Hu, L., Wu, Y., Huang, Y., Tian, H., and Hong, Z. (2023).
\newblock Dynamic motion of polar skyrmions in oxide heterostructures.
\newblock {\em arXiv preprint arXiv:2308.08219}.

\bibitem[Hubert and Sch{\"a}fer, 2008]{Hubert2008}
Hubert, A. and Sch{\"a}fer, R. (2008).
\newblock {\em Magnetic domains: the analysis of magnetic microstructures}.
\newblock Springer Science \& Business Media.

\bibitem[Huebener, 2001]{Huebener2001}
Huebener, R. (2001).
\newblock {\em Magnetic flux structures in superconductors: extended reprint of a classic text}, volume~6.
\newblock Springer Science \& Business Media.

\bibitem[Huse and Leibler, 1988]{Huse1988}
Huse, D.~A. and Leibler, S. (1988).
\newblock Phase behaviour of an ensemble of nonintersecting random fluid films.
\newblock {\em Journal de Physique}, 49(4):605--621.

\bibitem[Iwasaki et~al., 1972]{Iwasaki1972}
Iwasaki, H., Miyazawa, S., Koizumi, H., Sugii, K., and Niizeki, N. (1972).
\newblock Ferroelectric and optical properties of {Pb$_5$Ge$_3$O$_{11}$} and its isomorphous compound {Pb$_5$Ge$_2$SiO$_{11}$}.
\newblock {\em J. Appl. Phys.}, 43(12):4907--4915.

\bibitem[Iwasaki and Sugii, 1971]{Iwasaki1971}
Iwasaki, H. and Sugii, K. (1971).
\newblock Optical activity of ferroelectric 5{PbO}{\textperiodcentered}3{GeO$_2$} single crystals.
\newblock {\em Appl. Phys. Lett.}, 19(4):92--93.

\bibitem[Jiang and Zhang, 2019]{Jiang2019}
Jiang, A.~Q. and Zhang, Y. (2019).
\newblock Next-generation ferroelectric domain-wall memories: {P}rinciple and architecture.
\newblock {\em NPG Asia Mater.}, 11(1):2.

\bibitem[John and Mariamma, 2021]{John2021}
John, N. and Mariamma, A.~T. (2021).
\newblock Recent developments in the chiroptical properties of chiral plasmonic gold nanostructures: bioanalytical applications.
\newblock {\em Mikrochim Acta}, 188(12):1--25.

\bibitem[Junquera et~al., 2023]{Junquera2023}
Junquera, J., Nahas, Y., Prokhorenko, S., Bellaiche, L., {\'I}{\~n}iguez, J., Schlom, D.~G., Chen, L.-Q., Salahuddin, S., Muller, D.~A., Martin, L.~W., et~al. (2023).
\newblock Topological phases in polar oxide nanostructures.
\newblock {\em Rev. Mod. Phys.}, 95(2):025001.

\bibitem[Karpov et~al., 2017]{Karpov2017}
Karpov, D., Liu, Z., dos Santos~Rolo, T., Harder, R., Balachandran, P., Xue, D., Lookman, T., and Fohtung, E. (2017).
\newblock Three-dimensional imaging of vortex structure in a ferroelectric nanoparticle driven by an electric field.
\newblock {\em Nat. Commun.}, 8(1):280.

\bibitem[Khomskii, 2009]{Khomskii2009}
Khomskii, D. (2009).
\newblock Classifying multiferroics: {M}echanisms and effects.
\newblock {\em Physics}, 2:20.

\bibitem[Kim et~al., 2022a]{Kim2022}
Kim, K.~T., McCarter, M.~R., Stoica, V.~A., Das, S., Klewe, C., Donoway, E.~P., Burn, D.~M., Shafer, P., Rodolakis, F., Gon{\c{c}}alves, M.~A., et~al. (2022a).
\newblock Chiral structures of electric polarization vectors quantified by {X}-ray resonant scattering.
\newblock {\em Nat. Commun.}, 13(1):1--10.

\bibitem[Kim et~al., 2022b]{Kim2022enantio}
Kim, R.~M., Huh, J.-H., Yoo, S., Kim, T.~G., Kim, C., Kim, H., Han, J.~H., Cho, N.~H., Lim, Y.-C., Im, S.~W., et~al. (2022b).
\newblock Enantioselective sensing by collective circular dichroism.
\newblock {\em Nature}, 612(7940):470--476.

\bibitem[Kimel et~al., 2020]{Kimel2020}
Kimel, A., Kalashnikova, A., Pogrebna, A., and Zvezdin, A. (2020).
\newblock Fundamentals and perspectives of ultrafast photoferroic recording.
\newblock {\em Phys. Rep.}, 852:1--46.

\bibitem[Kirsanov et~al., 2021]{Kirsanov2021}
Kirsanov, N. et~al. (2021).
\newblock Long-distance quantum key distribution based on the physical loss control.
\newblock {\em arXiv preprint arXiv:2105.00035}.

\bibitem[Kittel, 1946]{Kittel1946}
Kittel, C. (1946).
\newblock Theory of the structure of ferromagnetic domains in films and small particles.
\newblock {\em Phys. Rev.}, 70(11-12):965.

\bibitem[Kobayashi et~al., 1971]{Kobayashi1971}
Kobayashi, J., Bouillot, J., and Kinoshita, K. (1971).
\newblock Optical activity of ferroelectric dicalcium strontium propionate.
\newblock {\em Phys. Status Solidi B}, 47(2):619--628.

\bibitem[Kobayashi et~al., 1991]{Kobayashi1991}
Kobayashi, J., Uchino, K., and Asahi, T. (1991).
\newblock Optical properties of rochelle salt.
\newblock {\em Phys. Rev. B}, 43(7):5706.

\bibitem[Kondovych et~al., 2017]{Kondovych2017}
Kondovych, S., Luk’yanchuk, I., Baturina, T.~I., and Vinokur, V.~M. (2017).
\newblock Gate-tunable electron interaction in high-$\kappa$ dielectric films.
\newblock {\em Sci. Rep.}, 7(1):42770.

\bibitem[Kondovych et~al., 2023]{Kondovych2023}
Kondovych, S., Pavlenko, M., Tikhonov, Y., Razumnaya, A., and Lukyanchuk, I. (2023).
\newblock Vortex states in a {PbTiO$_3$} ferroelectric cylinder.
\newblock {\em SciPost Phys.}, 14:056.

\bibitem[Kong et~al., 2020]{Kong2020}
Kong, X.-T., Besteiro, L.~V., Wang, Z., and Govorov, A.~O. (2020).
\newblock Plasmonic chirality and circular dichroism in bioassembled and nonbiological systems: {T}heoretical background and recent progress.
\newblock {\em Adv. Mater.}, 32(41):1801790.

\bibitem[Koralewski and Habry{\l}o, 1983]{Koralewski1983}
Koralewski, M. and Habry{\l}o, S. (1983).
\newblock Optical activity of {TGS} and {DTGS} crystals.
\newblock {\em Ferroelectrics}, 46(1):13--17.

\bibitem[Kornev et~al., 2004]{Kornev2004}
Kornev, I., Fu, H., and Bellaiche, L. (2004).
\newblock Ultrathin films of ferroelectric solid solutions under a residual depolarizing field.
\newblock {\em Phys. Rev. Lett.}, 93(19):196104.

\bibitem[Kumar et~al., 2016]{Kumar2016}
Kumar, J., Thomas, K.~G., and Liz-Marz{\'a}n, L.~M. (2016).
\newblock Nanoscale chirality in metal and semiconductor nanoparticles.
\newblock {\em Chem. Commun.}, 52(85):12555--12569.

\bibitem[Kuwata-Gonokami et~al., 2005]{Kuwata-Gonokami2005}
Kuwata-Gonokami, M., Saito, N., Ino, Y., Kauranen, M., Jefimovs, K., Vallius, T., Turunen, J., and Svirko, Y. (2005).
\newblock Giant optical activity in quasi-two-dimensional planar nanostructures.
\newblock {\em Phys. Rev. Lett.}, 95(22):227401.

\bibitem[Lahoche et~al., 2008]{Lahoche2008}
Lahoche, L., Luk'yanchuk, I., and Pascoli, G. (2008).
\newblock Stability of vortex phases in ferroelectric easy-plane nano-cylinders.
\newblock {\em Integr. Ferroelectr.}, 99(1):60--66.

\bibitem[Landau and Lifshitz, 1935]{Landau1935}
Landau, L.~D. and Lifshitz, E.~M. (1935).
\newblock On the theory of the dispersion of magnetic permeability in ferromagnetic bodies.
\newblock {\em Phys. Z. Sowjetunion}, 8(153):101--114.

\bibitem[Landau and Lifshitz, 2013]{Landau5}
Landau, L.~D. and Lifshitz, E.~M. (2013).
\newblock {\em Statistical physics: {V}olume 5}, volume~5.
\newblock Elsevier.

\bibitem[Landau et~al., 2013]{Landau8}
Landau, L.~D., Pitaevskii, L., and Lifshitz, E. (2013).
\newblock {\em Electrodynamics of continuous media}.
\newblock {E}lsevier.

\bibitem[Lazzeretti, 2017]{Lazzeretti2017}
Lazzeretti, P. (2017).
\newblock Chiral discrimination in nuclear magnetic resonance spectroscopy.
\newblock {\em J. Phys. Condens. Matter}, 29(44):443001.

\bibitem[Lee et~al., 2009]{Lee2009}
Lee, D., Behera, R.~K., Wu, P., Xu, H., Li, Y., Sinnott, S.~B., Phillpot, S.~R., Chen, L., and Gopalan, V. (2009).
\newblock Mixed {B}loch-{N\'e}el-{I}sing character of 180$^\circ$ ferroelectric domain walls.
\newblock {\em Phys. Rev. B}, 80(6):060102.

\bibitem[Lee et~al., 2018]{Lee2018}
Lee, H.-E., Ahn, H.-Y., Mun, J., Lee, Y.~Y., Kim, M., Cho, N.~H., Chang, K., Kim, W.~S., Rho, J., and Nam, K.~T. (2018).
\newblock Amino-acid-and peptide-directed synthesis of chiral plasmonic gold nanoparticles.
\newblock {\em Nature}, 556(7701):360--365.

\bibitem[Leonov et~al., 2016]{Leonov2016}
Leonov, A., Monchesky, T., Romming, N., Kubetzka, A., Bogdanov, A., and Wiesendanger, R. (2016).
\newblock The properties of isolated chiral skyrmions in thin magnetic films.
\newblock {\em New J. Phys.}, 18(6):065003.

\bibitem[Levstik et~al., 1985]{Levstik1985}
Levstik, A., Filipi{\v{c}}, C., Prelov{\v{s}}ek, P., Blinc, R., and Shuvalov, L. (1985).
\newblock Existence of a {L}ifshitz point in incommensurate {RbH$_3$(SeO$_3$)$_2$}.
\newblock {\em Phys. Rev. Lett.}, 54(14):1567.

\bibitem[Li et~al., 2020a]{Li2020}
Li, J., Wang, M., Wu, Z., Li, H., Hu, G., Jiang, T., Guo, J., Liu, Y., Yao, K., Chen, Z., et~al. (2020a).
\newblock Tunable chiral optics in all-solid-phase reconfigurable dielectric nanostructures.
\newblock {\em Nano Lett.}, 21(2):973--979.

\bibitem[Li et~al., 2021]{Li2021}
Li, Q., Stoica, V.~A., Pa{\'s}ciak, M., Zhu, Y., Yuan, Y., Yang, T., McCarter, M.~R., Das, S., Yadav, A.~K., Park, S., et~al. (2021).
\newblock Subterahertz collective dynamics of polar vortices.
\newblock {\em Nature}, 592(7854):376--380.

\bibitem[Li et~al., 2020b]{Li2020b}
Li, Y., Rui, G., Zhou, S., Gu, B., Yu, Y., Cui, Y., and Zhan, Q. (2020b).
\newblock Enantioselective optical trapping of chiral nanoparticles using a transverse optical needle field with a transverse spin.
\newblock {\em Opt. Express}, 28(19):27808--27822.

\bibitem[Li et~al., 2017]{Li2017}
Li, Z., Wang, Y., Tian, G., Li, P., Zhao, L., Zhang, F., Yao, J., Fan, H., Song, X., Chen, D., et~al. (2017).
\newblock High-density array of ferroelectric nanodots with robust and reversibly switchable topological domain states.
\newblock {\em Sci. Adv.}, 3(8):e1700919.

\bibitem[Li et~al., 2023]{Li2023}
Li, Z., Zhang, S., Li, X., Wu, Y., Wang, H., and Liu, H. (2023).
\newblock Center-type topological domain states in ferroelectric nanodots tailored from thin films.
\newblock {\em Phys. Status Solidi Rapid Res. Lett.}, page {2200424}.

\bibitem[Lich et~al., 2023]{Lich2023}
Lich, L.~V., Dang, H.~T., and Dinh, V.-H. (2023).
\newblock Polar toron structure in ferroelectric core-shell nanoparticles.
\newblock {\em Scr. Mater.}, 236:115641.

\bibitem[Lich et~al., 2022]{Lich2022}
Lich, L.~V., Nguyen, T.-G., Dinh, V.-H., and Phan, M.-H. (2022).
\newblock Theoretical studies on controlling the chirality of helical polarization vortices in ferroelectric nanowires: {I}mplications for reconfigurable electronic devices.
\newblock {\em ACS Appl. Nano Mater.}, 5(11):16509--16518.

\bibitem[Lininger et~al., 2022]{Lininger2022}
Lininger, A., Palermo, G., Guglielmelli, A., Nicoletta, G., Goel, M., Hinczewski, M., Strangi, G., et~al. (2022).
\newblock Chirality in light--matter interaction.
\newblock {\em Adv. Mater.}, 2107325.

\bibitem[Liu et~al., 2021]{Liu2021}
Liu, J., Yang, L., Qin, P., Zhang, S., Yung, K. K.~L., and Huang, Z. (2021).
\newblock Recent advances in inorganic chiral nanomaterials.
\newblock {\em Adv. Mater.}, page 2005506.

\bibitem[Liu et~al., 2017]{Liu2017book}
Liu, J.~P., Zhang, Z., and Zhao, G. (2017).
\newblock {\em Skyrmions: {T}opological structures, properties, and applications}.
\newblock CRC Press.

\bibitem[Liu et~al., 2009]{Liu2009}
Liu, N., Liu, H., Zhu, S., and Giessen, H. (2009).
\newblock Stereometamaterials.
\newblock {\em Nat. Photonics}, 3(3):157--162.

\bibitem[Liu et~al., 2022]{Liu2022}
Liu, Y., Watanabe, H., and Nagaosa, N. (2022).
\newblock Emergent magnetomultipoles and nonlinear responses of a magnetic hopfion.
\newblock {\em Phys. Rev. Lett.}, 129(26):267201.

\bibitem[Long et~al., 2020]{Long2020}
Long, G., Sabatini, R., Saidaminov, M.~I., Lakhwani, G., Rasmita, A., Liu, X., Sargent, E.~H., and Gao, W. (2020).
\newblock Chiral-perovskite optoelectronics.
\newblock {\em Nat. Rev. Mater.}, 5(6):423--439.

\bibitem[Louis et~al., 2012]{Louis2012}
Louis, L., Kornev, I., Geneste, G., Dkhil, B., and Bellaiche, L. (2012).
\newblock Novel complex phenomena in ferroelectric nanocomposites.
\newblock {\em J. Phys. Condens. Matter}, 24(40):402201.

\bibitem[Lovesey et~al., 2005]{Lovesey2005}
Lovesey, S., Balcar, E., Knight, K., and {Fernández Rodríguez}, J. (2005).
\newblock Electronic properties of crystalline materials observed in x-ray diffraction.
\newblock {\em Phys. Rep.}, 411(4):233--289.

\bibitem[Lovesey and van~der Laan, 2018]{Lovesey2018}
Lovesey, S.~W. and van~der Laan, G. (2018).
\newblock Resonant x-ray diffraction from chiral electric-polarization structures.
\newblock {\em Phys. Rev. B}, 98(15):155410.

\bibitem[Lu et~al., 2018]{Lu2018}
Lu, L., Nahas, Y., Liu, M., Du, H., Jiang, Z., Ren, S., Wang, D., Jin, L., Prokhorenko, S., Jia, C.-L., et~al. (2018).
\newblock Topological defects with distinct dipole configurations in {PbTiO$_3$/SrTiO$_3$} multilayer films.
\newblock {\em Phys. Rev. Lett.}, 120(17):177601.

\bibitem[Luk'yanchuk et~al., 2018]{Lukyanchuk2018}
Luk'yanchuk, I., Sen{\'e}, A., and Vinokur, V. (2018).
\newblock Electrodynamics of ferroelectric films with negative capacitance.
\newblock {\em Phys. Rev. B}, 98(2):024107.

\bibitem[Lukyanchuk et~al., 2021]{Lukyanchuk2021pat}
Lukyanchuk, I., Tikhonov, Y., Razumnaya, A., and Vinokour, V. (2021).
\newblock Thermal and electromagnetic generation and switching of chirality in ferroelectrics.
\newblock European Patent App. 21199906.5.

\bibitem[Lukyanchuk et~al., 2019]{Lukyanchuk2019}
Lukyanchuk, I., Zaitseva, E., Levashenko, V., Kvassay, M., Kondovych, S., Tikhonov, Y., Baudry, L., and Razumnaya, A. (2019).
\newblock Ferroelectric multiple-valued logic units.
\newblock {\em Ferroelectrics}, 543(1):213--221.

\bibitem[Luk’yanchuk et~al., 2022]{Lukyanchuk2022}
Luk’yanchuk, I., Razumnaya, A., Sen\'e, A., Tikhonov, Y., and Vinokur, V. (2022).
\newblock The ferroelectric field-effect transistor with negative capacitance.
\newblock {\em Npj Comput. Mater.}, 8(1):1--8.

\bibitem[Luk’yanchuk et~al., 2020]{Lukyanchuk2020}
Luk’yanchuk, I., Tikhonov, Y., Razumnaya, A., and Vinokur, V. (2020).
\newblock Hopfions emerge in ferroelectrics.
\newblock {\em Nat. Commun.}, 11(1):1--7.

\bibitem[Luk’yanchuk and Vinokur, 2021]{Lukyanchuk2021}
Luk’yanchuk, I. and Vinokur, V.~M. (2021).
\newblock Dynamics of polarization vortices revealed in a ferroelectric material.
\newblock {\em Nature}, 592(7854):359--360.

\bibitem[Luk’yanchuk et~al., 2009]{Lukyanchuk2009}
Luk’yanchuk, I.~A., Lahoche, L., and Sen{\'e}, A. (2009).
\newblock Universal properties of ferroelectric domains.
\newblock {\em Phys. Rev. Lett.}, 102(14):147601.

\bibitem[Mackinnon, 2019]{Mackinnon2019}
Mackinnon, N. (2019).
\newblock On the differences between helicity and chirality.
\newblock {\em J. Opt.}, 21(12):125402.

\bibitem[Malozemoff and Slonczewski, 2016]{Malozemoff2016}
Malozemoff, A. and Slonczewski, J.~C. (2016).
\newblock {\em Magnetic domain walls in bubble materials: {A}dvances in materials and device research}, volume~1.
\newblock Academic press.

\bibitem[Mangeri et~al., 2017]{Mangeri2017}
Mangeri, J., Espinal, Y., Jokisaari, A., Alpay, S.~P., Nakhmanson, S., and Heinonen, O. (2017).
\newblock Topological phase transformations and intrinsic size effects in ferroelectric nanoparticles.
\newblock {\em Nanoscale}, 9(4):1616--1624.

\bibitem[Manzi et~al., 2023]{Manzi2023}
Manzi, M., Pica, G., De~Bastiani, M., Kundu, S., Grancini, G., and Saidaminov, M.~I. (2023).
\newblock Ferroelectricity in hybrid perovskites.
\newblock {\em J. Phys. Chem. Lett.}, 14(14):3535--3552.

\bibitem[Martelli et~al., 2015]{Martelli2015}
Martelli, P.-W., Mefire, S.~M., and Luk'yanchuk, I.~A. (2015).
\newblock Multidomain switching in the ferroelectric nanodots.
\newblock {\em EPL (Europhys. Lett.)}, 111(5):50001.

\bibitem[Mason, 1984]{Mason1984}
Mason, S.~F. (1984).
\newblock Origins of biomolecular handedness.
\newblock {\em Nature}, 311(5981):19--23.

\bibitem[Masuda et~al., 2019]{Masuda2019}
Masuda, K., Le~Van, L., Shimada, T., and Kitamura, T. (2019).
\newblock Topological ferroelectric nanostructures induced by mechanical strain in strontium titanate.
\newblock {\em Phys. Chem. Chem. Phys.}, 21(40):22420--22428.

\bibitem[Mej{\'\i}a-Salazar and Oliveira~Jr, 2018]{Mejia2018}
Mej{\'\i}a-Salazar, J. and Oliveira~Jr, O.~N. (2018).
\newblock Plasmonic biosensing: {F}ocus review.
\newblock {\em Chem. Rev.}, 118(20):10617--10625.

\bibitem[Mermin, 1979]{Mermin1979}
Mermin, N.~D. (1979).
\newblock The topological theory of defects in ordered media.
\newblock {\em Rev. Mod. Phys.}, 51(3):591.

\bibitem[Mineev, 1998]{Mineev1998}
Mineev, V.~P. (1998).
\newblock {\em Topologically stable defects and solitons in ordered media}, volume~1.
\newblock CRC Press.

\bibitem[Mitra and Basak, 2022]{Mitra2022}
Mitra, S. and Basak, M. (2022).
\newblock Diverse bio-sensing and therapeutic applications of plasmon-enhanced nanostructures.
\newblock {\em Mater. Today}.

\bibitem[Mitsui and Nomura, 1981]{LandoldtOx}
Mitsui, T. and Nomura, S. (1981).
\newblock {\em Ferroelectrics and related substances, subvolume a: {O}xides, in {L}andolt-{B}ornstern numerical data and functional relationships}, volume~16.
\newblock Springer Berlin, Heidelberg, and New York.

\bibitem[Mitsui and Nomura, 1982]{LandoldtNonOx}
Mitsui, T. and Nomura, S. (1982).
\newblock {\em Ferroelectrics and related substances, subvolume b: {N}on-oxides, in {L}andolt-{B}ornstern numerical data and functional relationships}, volume~16.
\newblock Springer Berlin, Heidelberg, and New York.

\bibitem[Moffatt et~al., 1992]{Moffatt1992}
Moffatt, H.~K., Zaslavsky, G., Comte, P., and Tabor, M. (1992).
\newblock {\em Topological aspects of the dynamics of fluids and plasmas}, volume 218.
\newblock Springer Science \& Business Media.

\bibitem[Monastyrsky, 2007]{Monastyrsky2007}
Monastyrsky, M. (2007).
\newblock {\em Topology in molecular biology}.
\newblock Springer.

\bibitem[Moriya, 1960]{Moriya1960}
Moriya, T. (1960).
\newblock Anisotropic superexchange interaction and weak ferromagnetism.
\newblock {\em Phys. Rev.}, 120(1):91.

\bibitem[Morozovska et~al., 2021]{Morozovska2021}
Morozovska, A.~N., Hertel, R., Cherifi-Hertel, S., Reshetnyak, V.~Y., Eliseev, E.~A., and Evans, D.~R. (2021).
\newblock Chiral polarization textures induced by the flexoelectric effect in ferroelectric nanocylinders.
\newblock {\em Phys. Rev. B}, 104(5):054118.

\bibitem[Mun et~al., 2020]{Mun2020}
Mun, J., Kim, M., Yang, Y., Badloe, T., Ni, J., Chen, Y., Qiu, C.-W., and Rho, J. (2020).
\newblock Electromagnetic chirality: from fundamentals to nontraditional chiroptical phenomena.
\newblock {\em Light Sci. Appl.}, 9(1):1--18.

\bibitem[Nafie, 2011]{Nafie2011}
Nafie, L.~A. (2011).
\newblock {\em Vibrational optical activity: {P}rinciples and applications}.
\newblock John Wiley \& Sons.

\bibitem[Nagaosa and Tokura, 2013]{Nagaosa2013}
Nagaosa, N. and Tokura, Y. (2013).
\newblock Topological properties and dynamics of magnetic skyrmions.
\newblock {\em Nat. Nanotechnol.}, 8(12):899--911.

\bibitem[Nahas et~al., 2020a]{Nahas2020b}
Nahas, Y., Prokhorenko, S., Fischer, J., Xu, B., Carr{\'e}t{\'e}ro, C., Prosandeev, S., Bibes, M., Fusil, S., Dkhil, B., Garcia, V., et~al. (2020a).
\newblock Inverse transition of labyrinthine domain patterns in ferroelectric thin films.
\newblock {\em Nature}, 577(7788):47--51.

\bibitem[Nahas et~al., 2015]{Nahas2015}
Nahas, Y., Prokhorenko, S., Louis, L., Gui, Z., Kornev, I., and Bellaiche, L. (2015).
\newblock Discovery of stable skyrmionic state in ferroelectric nanocomposites.
\newblock {\em Nat. Commun.}, 6:8542.

\bibitem[Nahas et~al., 2020b]{Nahas2020}
Nahas, Y., Prokhorenko, S., Zhang, Q., Govinden, V., Valanoor, N., and Bellaiche, L. (2020b).
\newblock Topology and control of self-assembled domain patterns in low-dimensional ferroelectrics.
\newblock {\em Nat. Commun.}, 11(1):1--8.

\bibitem[Naumov et~al., 2004]{Naumov2004}
Naumov, I.~I., Bellaiche, L., and Fu, H. (2004).
\newblock Unusual phase transitions in ferroelectric nanodisks and nanorods.
\newblock {\em Nature}, 432(7018):737.

\bibitem[Neubrech et~al., 2020]{Neubrech2020}
Neubrech, F., Hentschel, M., and Liu, N. (2020).
\newblock Reconfigurable plasmonic chirality: {F}undamentals and applications.
\newblock {\em Adv. Mater.}, 32(41):1905640.

\bibitem[Neumann, 1885]{Neumann1885}
Neumann, F.~E. (1885).
\newblock {\em Vorlesungen {\"u}ber die Theorie der Elasticit{\"a}t der festen K{\"o}rper und des Licht{\"a}thers: gehalten an der Universit{\"a}t K{\"o}nigsberg}, volume~5.
\newblock BG Teubner.

\bibitem[Nguyen et~al., 2023]{Nguyen2023}
Nguyen, K.~X., Jiang, Y., Cao, M.~C., Purohit, P., Yadav, A.~K., Garc{\'\i}a-Fern{\'a}ndez, P., Tate, M.~W., Chang, C.~S., Aguado-Puente, P., {\'I}{\~n}iguez, J., et~al. (2023).
\newblock Transferring orbital angular momentum to an electron beam reveals toroidal and chiral order.
\newblock {\em Phys. Rev. B}, 107(20):205419.

\bibitem[Omari and Hayward, 2014]{Omari2014}
Omari, K. and Hayward, T. (2014).
\newblock Chirality-based vortex domain-wall logic gates.
\newblock {\em Phys. Rev. Appl.}, 2(4):044001.

\bibitem[Paiva-Marques et~al., 2020]{Paiva-Marques2020}
Paiva-Marques, W.~A., Reyes~G{\'o}mez, F., Oliveira~Jr, O.~N., and Mej{\'\i}a-Salazar, J.~R. (2020).
\newblock Chiral plasmonics and their potential for point-of-care biosensing applications.
\newblock {\em Sensors}, 20(3):944.

\bibitem[Pakhomov et~al., 2013]{Pakhomov2013}
Pakhomov, A., Luk'yanchuk, I., and Sidorkin, A. (2013).
\newblock Frequency dependence of the dielectric permittivity in ferroelectric thin films with 180$^\circ$ domain structure.
\newblock {\em Ferroelectrics}, 444(1):177--182.

\bibitem[Patti et~al., 2019]{Patti2019}
Patti, F., Saija, R., Denti, P., Pellegrini, G., Biagioni, P., Iat{\`\i}, M.~A., and Marag{\`o}, O.~M. (2019).
\newblock Chiral optical tweezers for optically active particles in the {T}-matrix formalism.
\newblock {\em Sci. Rep.}, 9(1):1--10.

\bibitem[Pavlenko et~al., 2022]{Pavlenko2022}
Pavlenko, M.~A., Di~Rino, F., Boron, L., Kondovych, S., Sen{\'e}, A., Tikhonov, Y.~A., Razumnaya, A.~G., Vinokur, V.~M., Sepliarsky, M., and Lukyanchuk, I.~A. (2022).
\newblock Phase diagram of a strained ferroelectric nanowire.
\newblock {\em Crystals}, 12(4):453.

\bibitem[Pertsev et~al., 1998]{Pertsev1998PRL}
Pertsev, N., Zembilgotov, A., and Tagantsev, A. (1998).
\newblock Effect of mechanical boundary conditions on phase diagrams of epitaxial ferroelectric thin films.
\newblock {\em Phys. Rev. Lett.}, 80(9):1988.

\bibitem[Pfeifer et~al., 2007]{Pfeifer2007}
Pfeifer, R.~N., Nieminen, T.~A., Heckenberg, N.~R., and Rubinsztein-Dunlop, H. (2007).
\newblock Colloquium: {M}omentum of an electromagnetic wave in dielectric media.
\newblock {\em Rev. Mod. Phys.}, 79(4):1197.

\bibitem[Pinto et~al., 2020]{Pinto2020}
Pinto, M.~M., Fernandes, C., and Tiritan, M.~E. (2020).
\newblock Chiral separations in preparative scale: {A} medicinal chemistry point of view.
\newblock {\em Molecules}, 25(8):1931.

\bibitem[Polimeno et~al., 2018]{Polimeno2018}
Polimeno, P., Magazzu, A., Iati, M.~A., Patti, F., Saija, R., Boschi, C. D.~E., Donato, M.~G., Gucciardi, P.~G., Jones, P.~H., Volpe, G., et~al. (2018).
\newblock Optical tweezers and their applications.
\newblock {\em J. Quant. Spectrosc. Radiat. Transf.}, 218:131--150.

\bibitem[Pour et~al., 2011]{Pour2011}
Pour, S.~O., Bell, S.~E., and Blanch, E.~W. (2011).
\newblock Use of a hydrogel polymer for reproducible surface-enhanced {R}aman optical activity ({SEROA}).
\newblock {\em Chem. Commun.}, 47(16):4754--4756.

\bibitem[Prosandeev and Bellaiche, 2007]{Prosandeev2007nd}
Prosandeev, S. and Bellaiche, L. (2007).
\newblock Characteristics and signatures of dipole vortices in ferroelectric nanodots: {F}irst-principles-based simulations and analytical expressions.
\newblock {\em Phys. Rev. B}, 75(9):094102.

\bibitem[Prosandeev and Bellaiche, 2008]{Prosandeev2008}
Prosandeev, S. and Bellaiche, L. (2008).
\newblock Order parameter in complex dipolar structures: {M}icroscopic modeling.
\newblock {\em Phys. Rev. B}, 77(6):060101.

\bibitem[Prosandeev et~al., 2013]{Prosandeev2013}
Prosandeev, S., Malashevich, A., Gui, Z., Louis, L., Walter, R., Souza, I., and Bellaiche, L. (2013).
\newblock Natural optical activity and its control by electric field in electrotoroidic systems.
\newblock {\em Phys. Rev. B}, 87(19):195111.

\bibitem[Prosandeev et~al., 2019]{Prosandeev2019}
Prosandeev, S., Prokhorenko, S., Nahas, Y., and Bellaiche, L. (2019).
\newblock Prediction of a novel topological multidefect ground state.
\newblock {\em Phys. Rev. B}, 100(14):140104.

\bibitem[Qi et~al., 2021]{Qi2021}
Qi, L., Ruan, S., and Zeng, Y.-J. (2021).
\newblock Review on recent developments in {2D} ferroelectrics: {T}heories and applications.
\newblock {\em Adv. Mater.}, 33(13):2005098.

\bibitem[Ramesh and Schlom, 2019]{Ramesh2019}
Ramesh, R. and Schlom, D.~G. (2019).
\newblock Creating emergent phenomena in oxide superlattices.
\newblock {\em Nat. Rev. Mater.}, 4(4):257--268.

\bibitem[Razumnaya et~al., 2023]{Razumnaya2023}
Razumnaya, A.~G., Tikhonov, Y.~A., Vinokur, V.~M., and Lukyanchuk, I.~A. (2023).
\newblock Ferroelectric topologically configurable multilevel logic unit.
\newblock {\em Neuromorph. Comput. Eng.}, 3(2):024003.

\bibitem[Ren et~al., 2021]{Ren2021}
Ren, Y., Chen, Q., He, M., Zhang, X., Qi, H., and Yan, Y. (2021).
\newblock Plasmonic optical tweezers for particle manipulation: Principles, methods, and applications.
\newblock {\em ACS Nano}, 15(4):6105--6128.

\bibitem[Rijal et~al., 2023]{Rijal2023}
Rijal, S., Nahas, Y., Prokhorenko, S., and Bellaiche, L. (2023).
\newblock Dynamics of polar vortex crystallization.
\newblock {\em arXiv preprint arXiv:2302.07380}.

\bibitem[R{\o}rvik et~al., 2011]{Rorvik2011}
R{\o}rvik, P.~M., Grande, T., and Einarsrud, M.-A. (2011).
\newblock One-dimensional nanostructures of ferroelectric perovskites.
\newblock {\em Adv. Mater.}, 23(35):4007--4034.

\bibitem[Rybakov et~al., 2015]{Rybakov2015}
Rybakov, F.~N., Borisov, A.~B., Bl{\"u}gel, S., and Kiselev, N.~S. (2015).
\newblock New type of stable particle-like states in chiral magnets.
\newblock {\em Phys. Rev. Lett.}, 115(11):117201.

\bibitem[Rybakov et~al., 2022]{Rybakov2022}
Rybakov, F.~N., Kiselev, N.~S., Borisov, A.~B., D{\"o}ring, L., Melcher, C., and Bl{\"u}gel, S. (2022).
\newblock Magnetic hopfions in solids.
\newblock {\em APL Mater.}, 10(11):111113.

\bibitem[Salje and Scott, 2014]{Salje2014}
Salje, E. and Scott, J. (2014).
\newblock Ferroelectric {B}loch-line switching: {A} paradigm for memory devices?
\newblock {\em Appl. Phys. Lett.}, 105(25):252904.

\bibitem[Sawada et~al., 1977]{Sawada1977}
Sawada, A., Ishibashi, Y., and Takagi, Y. (1977).
\newblock Ferroelasticity and the origin of optical activity of {Ca$_2$Sr(C$_2$H$_5$CO$_2$)$_6$} ({DSP}).
\newblock {\em J. Phys. Soc. Japan}, 43(1):195--203.

\bibitem[Saxena et~al., 2020]{Saxena2020}
Saxena, A., Kevrekidis, P.~G., and Cuevas-Maraver, J. (2020).
\newblock Nonlinearity and topology.
\newblock In {\em Emerging Frontiers in Nonlinear Science}, pages 25--54. Springer.

\bibitem[Scott, 2000]{ScottBook}
Scott, J.~F. (2000).
\newblock {\em Ferroelectric Memories}.
\newblock Springer Berlin, Heidelberg, and New York.

\bibitem[Seidel, 2019]{Seidel2019}
Seidel, J. (2019).
\newblock Nanoelectronics based on topological structures.
\newblock {\em Nat. Mater.}, 18(3):188.

\bibitem[Sen{\'e} et~al., 2011]{Sene2011}
Sen{\'e}, A., Baudry, L., Luk’yanchuk, I., Lahoche, L., and El~Amraoui, Y. (2011).
\newblock Field-induced vortices in weakly anisotropic ferroelectrics.
\newblock {\em Superlattices Microstruct.}, 49(3):314--317.

\bibitem[Seul and Andelman, 1995]{Seul1995}
Seul, M. and Andelman, D. (1995).
\newblock Domain shapes and patterns: the phenomenology of modulated phases.
\newblock {\em Science}, 267(5197):476--483.

\bibitem[Shafer et~al., 2018]{Shafer2018}
Shafer, P., Garc{\'\i}a-Fern{\'a}ndez, P., Aguado-Puente, P., Damodaran, A.~R., Yadav, A.~K., Nelson, C.~T., Hsu, S.-L., Wojde{\l}, J.~C., {\'I}{\~n}iguez, J., Martin, L.~W., et~al. (2018).
\newblock Emergent chirality in the electric polarization texture of titanate superlattices.
\newblock {\em Proc. Natl. Acad. Sci}, 115(5):915--920.

\bibitem[Shao et~al., 2023]{Shao2023}
Shao, Y.-T., Das, S., Hong, Z., Xu, R., Chandrika, S., G{\'o}mez-Ortiz, F., Garc{\'\i}a-Fern{\'a}ndez, P., Chen, L.-Q., Hwang, H.~Y., Junquera, J., et~al. (2023).
\newblock Emergent chirality in a polar meron to skyrmion phase transition.
\newblock {\em Nature Commun.}, 14(1):1355.

\bibitem[Shimada et~al., 2016]{Shimada2016}
Shimada, T., Lich, L.~V., Nagano, K., Wang, J.-S., Wang, J., and Kitamura, T. (2016).
\newblock Polar superhelices in ferroelectric chiral nanosprings.
\newblock {\em Sci. Rep.}, 6(1):1--9.

\bibitem[Shuvalov and Ivanov, 1964]{Shuvalov1964}
Shuvalov, L. and Ivanov, N. (1964).
\newblock The change of optical activity of ferroelectrics if they are repoled.
\newblock {\em Sov. Phys. Crystallogr.}, 9:363.

\bibitem[Singh et~al., 2021]{Singh2021}
Singh, A., Dev, K., Siljak, H., Joshi, H.~D., and Magarini, M. (2021).
\newblock Quantum internet - applications, functionalities, enabling technologies, challenges, and research directions.
\newblock {\em IEEE Commun. Surv. Tutor.}, 23(4):2218--2247.

\bibitem[SM, SM]{SM}
SM.
\newblock See supplemental material at link {http://...,} for detailed description of phase-field simulations.

\bibitem[Sohncke, 1879]{SohnckeBook}
Sohncke, L. (1879).
\newblock {\em Entwickelung einer theorie der krystallstruktur}.
\newblock BG Teubner.

\bibitem[Spaldin and Ramesh, 2019]{Spaldin2019}
Spaldin, N.~A. and Ramesh, R. (2019).
\newblock Advances in magnetoelectric multiferroics.
\newblock {\em Nat. Mater.}, 18(3):203--212.

\bibitem[Stachiotti and Sepliarsky, 2011]{Stachiotti2011}
Stachiotti, M. and Sepliarsky, M. (2011).
\newblock Toroidal ferroelectricity in {P}b{T}i{O}$_3$ nanoparticles.
\newblock {\em Phys. Rev. Lett.}, 106(13):137601.

\bibitem[Stengel, 2023]{Stengel2023}
Stengel, M. (2023).
\newblock Macroscopic polarization from nonlinear gradient couplings.
\newblock {\em arXiv preprint arXiv:2304.06613}.

\bibitem[Stengel et~al., 2009]{Stengel2009}
Stengel, M., Vanderbilt, D., and Spaldin, N.~A. (2009).
\newblock First-principles modeling of ferroelectric capacitors via constrained displacement field calculations.
\newblock {\em Phys. Rev. B}, 80(22):224110.

\bibitem[Stephanovich et~al., 2003]{Stephanovich2003}
Stephanovich, V., Luk'yanchuk, I., and Karkut, M. (2003).
\newblock Domain proximity and ferroelectric transition in ferro-paraelectric superlattices.
\newblock {\em Ferroelectrics}, 291(1):169--175.

\bibitem[Stephanovich et~al., 2005]{Stephanovich2005}
Stephanovich, V., Luk’yanchuk, I., and Karkut, M. (2005).
\newblock Domain-enhanced interlayer coupling in ferroelectric/paraelectric superlattices.
\newblock {\em Phys. Rev. Lett.}, 94(4):047601.

\bibitem[Stoica et~al., 2019]{Stoica2019}
Stoica, V., Laanait, N., Dai, C., Hong, Z., Yuan, Y., Zhang, Z., Lei, S., McCarter, M., Yadav, A., Damodaran, A.~R., et~al. (2019).
\newblock Optical creation of a supercrystal with three-dimensional nanoscale periodicity.
\newblock {\em Nat. Mater.}, 18(4):377--383.

\bibitem[Strukov and Levanyuk, 2012]{StrukovBook}
Strukov, B.~A. and Levanyuk, A.~P. (2012).
\newblock {\em Ferroelectric phenomena in crystals: {P}hysical foundations}.
\newblock Springer Science \& Business Media.

\bibitem[Susarla et~al., 2023]{Susarla2023}
Susarla, S., Hsu, S., G{\'o}mez-Ortiz, F., Garc{\'\i}a-Fern{\'a}ndez, P., Savitzky, B.~H., Das, S., Behera, P., Junquera, J., Ercius, P., Ramesh, R., et~al. (2023).
\newblock The emergence of three-dimensional chiral domain walls in polar vortices.
\newblock {\em Nat. Commun.}, 14(1):4465.

\bibitem[Tagantsev et~al., 2001]{Tagantsev2001}
Tagantsev, A.~K., Courtens, E., and Arzel, L. (2001).
\newblock Prediction of a low-temperature ferroelectric instability in antiphase domain boundaries of strontium titanate.
\newblock {\em Phys. Rev. B}, 64(22):224107.

\bibitem[Tagantsev et~al., 2010]{Tagantsev2010book}
Tagantsev, A.~K., Cross, L.~E., and Fousek, J. (2010).
\newblock {\em Domains in ferroic crystals and thin films}, volume~13.
\newblock Springer.

\bibitem[Tan et~al., 2021]{Tan2021}
Tan, C., Dong, Y., Sun, Y., Liu, C., Chen, P., Zhong, X., Zhu, R., Liu, M., Zhang, J., Wang, J., et~al. (2021).
\newblock Engineering polar vortex from topologically trivial domain architecture.
\newblock {\em Nat. Commun.}, 12(1):1--8.

\bibitem[Tang et~al., 2021]{Tang2021}
Tang, Y., Zhu, Y., and Ma, X. (2021).
\newblock Topological polar structures in ferroelectric oxide films.
\newblock {\em J. Appl. Phys.}, 129(20):200904.

\bibitem[Tian et~al., 2021a]{Tian2021}
Tian, G., Yang, W., Gao, X., and Liu, J.-M. (2021a).
\newblock Emerging phenomena from exotic ferroelectric topological states.
\newblock {\em APL Mater.}, 9(2):020907.

\bibitem[Tian et~al., 2021b]{Tian2021chi}
Tian, L., Wang, C., Zhao, H., Sun, F., Dong, H., Feng, K., Wang, P., He, G., and Li, G. (2021b).
\newblock Rational approach to plasmonic dimers with controlled gap distance, symmetry, and capability of precisely hosting guest molecules in hotspot regions.
\newblock {\em J. Am. Chem. Soc.}, 143(23):8631--8638.

\bibitem[Tikhonov et~al., 2020]{Tikhonov2020}
Tikhonov, Y., Kondovych, S., Mangeri, J., et~al. (2020).
\newblock Controllable skyrmion chirality in ferroelectrics.
\newblock {\em Sci. Rep.}, 10(1):8657.

\bibitem[Tikhonov et~al., 2022]{Tikhonov2022}
Tikhonov, Y., Maguire, J.~R., McCluskey, C.~J., McConville, J.~P., Kumar, A., Lu, H., Meier, D., Razumnaya, A., Gregg, J.~M., Gruverman, A., et~al. (2022).
\newblock Polarization topology at the nominally charged domain walls in uniaxial ferroelectrics.
\newblock {\em Adv. Mater.}, 34(45):2203028.

\bibitem[Tokura et~al., 2014]{Tokura2014}
Tokura, Y., Seki, S., and Nagaosa, N. (2014).
\newblock Multiferroics of spin origin.
\newblock {\em Rep. Prog. Phys.}, 77(7):076501.

\bibitem[Torsi et~al., 2013]{Torsi2013}
Torsi, L., Magliulo, M., Manoli, K., and Palazzo, G. (2013).
\newblock Organic field-effect transistor sensors: {A} tutorial review.
\newblock {\em Chem. Soc. Rev.}, 42(22):8612--8628.

\bibitem[Valev et~al., 2013]{Valev2013}
Valev, V.~K., Baumberg, J.~J., Sibilia, C., and Verbiest, T. (2013).
\newblock Chirality and chiroptical effects in plasmonic nanostructures: {F}undamentals, recent progress, and outlook.
\newblock {\em Adv. Mater.}, 25(18):2517--2534.

\bibitem[Van~Lich et~al., 2017]{VanLich2017}
Van~Lich, L., Shimada, T., Wang, J., Dinh, V.-H., Bui, T.~Q., Kitamura, T., et~al. (2017).
\newblock Switching the chirality of a ferroelectric vortex in designed nanostructures by a homogeneous electric field.
\newblock {\em Phys. Rev. B}, 96(13):134119.

\bibitem[Vinegrad et~al., 2018]{Vinegrad2018}
Vinegrad, E., Vestler, D., Ben-Moshe, A., Barnea, A.~R., Markovich, G., and Cheshnovsky, O. (2018).
\newblock Circular dichroism of single particles.
\newblock {\em ACS Photonics}, 5(6):2151--2159.

\bibitem[Volovik et~al., 2019]{Volovik2019}
Volovik, G., Rysti, J., M{\"a}kinen, J., and Eltsov, V. (2019).
\newblock Spin, orbital, {W}eyl and other glasses in topological superfluids.
\newblock {\em J. Low Temp. Phys.}, 196(1):82--101.

\bibitem[Wagni{\`e}re, 2008]{Wagniere2008}
Wagni{\`e}re, G.~H. (2008).
\newblock {\em On chirality and the universal asymmetry: {R}eflections on image and mirror image}.
\newblock John Wiley \& Sons.

\bibitem[Walter et~al., 2016]{Walter2016}
Walter, R., Prokhorenko, S., Gui, Z., Nahas, Y., and Bellaiche, L. (2016).
\newblock Electrical control of chiral phases in electrotoroidic nanocomposites.
\newblock {\em Adv. Electron. Mater.}, 2(1):1500218.

\bibitem[Wang et~al., 2022a]{Wang2022b}
Wang, C., Wang, T., Zhang, W., Jiang, J., Chen, L., and Jiang, A. (2022a).
\newblock Analog ferroelectric domain-wall memories and synaptic devices integrated with {Si} substrates.
\newblock {\em Nano Res.}, 15(4):3606--3613.

\bibitem[Wang et~al., 2023]{Wang2023}
Wang, C., You, L., Cobden, D., and Wang, J. (2023).
\newblock Towards two-dimensional van der waals ferroelectrics.
\newblock {\em Nat. Mater.}, pages 1--11.

\bibitem[Wang et~al., 2018]{Wang2018}
Wang, L., Feng, Q., Kim, Y., Kim, R., Lee, K.~H., Pollard, S.~D., Shin, Y.~J., Zhou, H., Peng, W., Lee, D., et~al. (2018).
\newblock Ferroelectrically tunable magnetic skyrmions in ultrathin oxide heterostructures.
\newblock {\em Nat. Mater.}, 17(12):1087--1094.

\bibitem[Wang et~al., 2020]{Wang2020}
Wang, Y., Feng, Y., Zhu, Y., Tang, Y., Yang, L., Zou, M., Geng, W., Han, M., Guo, X., Wu, B., et~al. (2020).
\newblock Polar meron lattice in strained oxide ferroelectrics.
\newblock {\em Nat. Mater.}, 19(8):881--886.

\bibitem[Wang et~al., 2022b]{Wang2022}
Wang, Y., Tang, Y., Zhu, Y., and Ma, X. (2022b).
\newblock Entangled polarizations in ferroelectrics: {A} focused review of polar topologies.
\newblock {\em Acta Mater.}, page 118485.

\bibitem[Wenzel, 2018]{Wenzel2018}
Wenzel, T.~J. (2018).
\newblock {\em Differentiation of chiral compounds using {NMR} spectroscopy}.
\newblock John Wiley \& Sons.

\bibitem[Wojde{\l} and {\'I}{\~n}iguez, 2014]{Wojdel2014}
Wojde{\l}, J.~C. and {\'I}{\~n}iguez, J. (2014).
\newblock Ferroelectric transitions at ferroelectric domain walls found from first principles.
\newblock {\em Phys. Rev. Lett.}, 112(24):247603.

\bibitem[Wu and Pauly, 2022]{Wu2022}
Wu, W. and Pauly, M. (2022).
\newblock Chiral plasmonic nanostructures: recent advances in their synthesis and applications.
\newblock {\em Mater. Adv.}, 3:186--215.

\bibitem[Wu and Zheng, 2017]{Wu2017}
Wu, Z. and Zheng, Y. (2017).
\newblock Moir{\'e} chiral metamaterials.
\newblock {\em Adv. Opt. Mater.}, 5(16):1700034.

\bibitem[Xia et~al., 2011]{Xia2011}
Xia, Y., Zhou, Y., and Tang, Z. (2011).
\newblock Chiral inorganic nanoparticles: {O}rigin, optical properties and bioapplications.
\newblock {\em Nanoscale}, 3(4):1374--1382.

\bibitem[Xiong et~al., 2022]{Xiong2022}
Xiong, P., Ma, W., Yuan, S., Liu, Y., and Wang, B. (2022).
\newblock Control of the chirality of a vortex in a ferroelectric nanodot by uniform electric fields mediated by inhomogeneous surface screening.
\newblock {\em AIP Adv.}, 12(1):015001.

\bibitem[Xu et~al., 2019]{Xu2019}
Xu, R., Liu, S., Saremi, S., Gao, R., Wang, J., Hong, Z., Lu, H., Ghosh, A., Pandya, S., Bonturim, E., et~al. (2019).
\newblock Kinetic control of tunable multi-state switching in ferroelectric thin films.
\newblock {\em Nat. Commun.}, 10(1):1282.

\bibitem[Yadav et~al., 2016]{Yadav2016}
Yadav, A., Nelson, C., Hsu, S., Hong, Z., Clarkson, J., Schlep{\"u}tz, C., Damodaran, A., Shafer, P., Arenholz, E., Dedon, L., et~al. (2016).
\newblock Observation of polar vortices in oxide superlattices.
\newblock {\em Nature}, 530(7589):198.

\bibitem[Yadav et~al., 2019]{Yadav2019}
Yadav, A.~K., Nguyen, K.~X., Hong, Z., Garc{\'\i}a-Fern{\'a}ndez, P., Aguado-Puente, P., Nelson, C.~T., Das, S., Prasad, B., Kwon, D., Cheema, S., et~al. (2019).
\newblock Spatially resolved steady-state negative capacitance.
\newblock {\em Nature}, 565(7740):468--471.

\bibitem[Ye et~al., 2018]{Ye2018}
Ye, H.-Y., Tang, Y.-Y., Li, P.-F., Liao, W.-Q., Gao, J.-X., Hua, X.-N., Cai, H., Shi, P.-P., You, Y.-M., and Xiong, R.-G. (2018).
\newblock Metal-free three-dimensional perovskite ferroelectrics.
\newblock {\em Science}, 361(6398):151--155.

\bibitem[Ye and Lin, 2017]{Ye2017}
Ye, Q. and Lin, H. (2017).
\newblock On deriving the {M}axwell stress tensor method for calculating the optical force and torque on an object in harmonic electromagnetic fields.
\newblock {\em Eur. J. Phys.}, 38(4):045202.

\bibitem[Yin et~al., 2021]{Yin2021}
Yin, J., Zong, H., Tao, H., Tao, X., Wu, H., Zhang, Y., Zhao, L.-D., Ding, X., Sun, J., Zhu, J., et~al. (2021).
\newblock Nanoscale bubble domains with polar topologies in bulk ferroelectrics.
\newblock {\em Nat. Commun.}, 12(1):1--8.

\bibitem[Yoo and Park, 2015]{Yoo2015}
Yoo, S. and Park, Q.-H. (2015).
\newblock Enhancement of chiroptical signals by circular differential {M}ie scattering of nanoparticles.
\newblock {\em Sci. Rep.}, 5(1):1--8.

\bibitem[Yu et~al., 2018]{Yu2018}
Yu, X., Koshibae, W., Tokunaga, Y., Shibata, K., Taguchi, Y., Nagaosa, N., and Tokura, Y. (2018).
\newblock Transformation between meron and skyrmion topological spin textures in a chiral magnet.
\newblock {\em Nature}, 564(7734):95--98.

\bibitem[Yuan et~al., 2018a]{Yuan2018}
Yuan, S., Chen, W., Liu, J., Liu, Y., Wang, B., and Zheng, Y. (2018a).
\newblock Torsion-induced vortex switching and skyrmion-like state in ferroelectric nanodisks.
\newblock {\em J. Phys. Condens. Matter}, 30(46):465304.

\bibitem[Yuan et~al., 2018b]{Yuan2018switch}
Yuan, S., Chen, W., Ma, L., Ji, Y., Xiong, W., Liu, J., Liu, Y., Wang, B., and Zheng, Y. (2018b).
\newblock Defect-mediated vortex multiplication and annihilation in ferroelectrics and the feasibility of vortex switching by stress.
\newblock {\em Acta Mater.}, 148:330--343.

\bibitem[Yuan et~al., 2023]{Yuan2023}
Yuan, S., Chen, Z., Prokhorenko, S., Nahas, Y., Bellaiche, L., Liu, C., Xu, B., Chen, L., Das, S., and Martin, L.~W. (2023).
\newblock Hexagonal close-packed polar-skyrmion lattice in ultrathin ferroelectric {PbTiO$_3$} films.
\newblock {\em Phys. Rev. Lett.}, 130(22):226801.

\bibitem[Zabalo and Stengel, 2023]{Zabalo2023}
Zabalo, A. and Stengel, M. (2023).
\newblock Natural optical activity from density-functional perturbation theory.
\newblock {\em arXiv preprint arXiv:2304.00048}.

\bibitem[Zafar and Ragusa, 2020]{Zafar2020}
Zafar, M.~S. and Ragusa, A. (2020).
\newblock Chirality at the nanoparticle surface: {F}unctionalization and applications.
\newblock {\em Appl. Sci.}, 10(15):5357.

\bibitem[Zang et~al., 2018]{Zang2018}
Zang, J., Cros, V., and Hoffmann, A. (2018).
\newblock {\em Topology in magnetism}, volume 192.
\newblock Springer.

\bibitem[Zel’{d}ovich, 1957]{Zeldovich1957}
Zel’{d}ovich, I.~B. (1957).
\newblock Electromagnetic interaction with parity violation.
\newblock {\em J. Exp. Theor. Phys.}, 33:1531--1533.

\bibitem[Zhang et~al., 2020]{Zhang2020}
Zhang, H.-Y., Song, X.-J., Chen, X.-G., Zhang, Z.-X., You, Y.-M., Tang, Y.-Y., and Xiong, R.-G. (2020).
\newblock Observation of vortex domains in a two-dimensional lead iodide perovskite ferroelectric.
\newblock {\em J. Am. Chem. Soc.}, 142(10):4925--4931.

\bibitem[Zhang and Xiong, 2023]{Zhang2023bioelectronics}
Zhang, H.-Y. and Xiong, R.-G. (2023).
\newblock Ferroelectric polymers take a step toward bioelectronics.
\newblock {\em Science}, 381(6657):484--485.

\bibitem[Zhang et~al., 2011]{Zhang2011}
Zhang, Q., Herchig, R., and Ponomareva, I. (2011).
\newblock Nanodynamics of ferroelectric ultrathin films.
\newblock {\em Phys. Rev. Lett.}, 107(17):177601.

\bibitem[Zhang et~al., 2017]{Zhang2017}
Zhang, Q., Xie, L., Liu, G., Prokhorenko, S., Nahas, Y., Pan, X., Bellaiche, L., Gruverman, A., and Valanoor, N. (2017).
\newblock Nanoscale bubble domains and topological transitions in ultrathin ferroelectric films.
\newblock {\em Adv. Mater.}, 29(46):1702375.

\bibitem[Zhang et~al., 2023]{Zhang2023}
Zhang, X., Chen, H., Tian, G., Yang, W., Fan, Z., Hou, Z., Chen, D., Zeng, M., Qin, M., Gao, J., et~al. (2023).
\newblock Creation and erasure of polar bubble domains in {PbTiO$_3$} films by mechanical stress and light illuminations.
\newblock {\em J. Materiomics}.

\bibitem[Zhang et~al., 2021]{Zhang2021}
Zhang, Y., Min, C., Dou, X., Wang, X., Urbach, H.~P., Somekh, M.~G., and Yuan, X. (2021).
\newblock Plasmonic tweezers: {F}or nanoscale optical trapping and beyond.
\newblock {\em Light Sci. Appl.}, 10(1):1--41.

\bibitem[Zhang et~al., 2014]{Zhang2014}
Zhang, Y., Oka, T., Suzuki, R., Ye, J., and Iwasa, Y. (2014).
\newblock Electrically switchable chiral light-emitting transistor.
\newblock {\em Science}, 344(6185):725--728.

\bibitem[Zhang et~al., 2022]{Zhang2022}
Zhang, Y., Yu, S., Han, B., Zhou, Y., Zhang, X., Gao, X., and Tang, Z. (2022).
\newblock Circularly polarized luminescence in chiral materials.
\newblock {\em Matter}.

\bibitem[Zhao et~al., 2021]{Zhao2021}
Zhao, H.~J., Chen, P., Prosandeev, S., Artyukhin, S., and Bellaiche, L. (2021).
\newblock Dzyaloshinskii--{M}oriya-like interaction in ferroelectrics and antiferroelectrics.
\newblock {\em Nat. Mater.}, 20(3):341--345.

\bibitem[Zhao et~al., 2016]{Zhao2016}
Zhao, Y., Saleh, A.~A., and Dionne, J.~A. (2016).
\newblock Enantioselective optical trapping of chiral nanoparticles with plasmonic tweezers.
\newblock {\em ACS Photonics}, 3(3):304--309.

\bibitem[Zhao et~al., 2017]{Zhao2017}
Zhao, Y., Saleh, A.~A., Van De~Haar, M.~A., Baum, B., Briggs, J.~A., Lay, A., Reyes-Becerra, O.~A., and Dionne, J.~A. (2017).
\newblock Nanoscopic control and quantification of enantioselective optical forces.
\newblock {\em Nat. Nanotechnol.}, 12(11):1055--1059.

\bibitem[Zhao et~al., 2004]{Zhao2004}
Zhao, Z., Buscaglia, V., Viviani, M., Buscaglia, M.~T., Mitoseriu, L., Testino, A., Nygren, M., Johnsson, M., and Nanni, P. (2004).
\newblock Grain-size effects on the ferroelectric behavior of dense nanocrystalline {BaTiO$_3$} ceramics.
\newblock {\em Phys. Rev. B}, 70(2):024107.

\bibitem[Zheng et~al., 2018]{Zheng2018}
Zheng, F., Rybakov, F.~N., Borisov, A.~B., Song, D., Wang, S., Li, Z.-A., Du, H., Kiselev, N.~S., Caron, J., Kov{\'a}cs, A., et~al. (2018).
\newblock Experimental observation of chiral magnetic bobbers in {B}20-type {FeGe}.
\newblock {\em Nat. Nanotechnol.}, 13(6):451--455.

\bibitem[Zheng et~al., 2023]{Zheng2023}
Zheng, W., Wang, X., Zhang, X., Chen, B., Suo, H., Xing, Z., Wang, Y., Wei, H.-L., Chen, J., Guo, Y., et~al. (2023).
\newblock Emerging halide perovskite ferroelectrics.
\newblock {\em Adv. Mater.}, 35(21):2205410.

\bibitem[Zheng and Chen, 2017]{Zheng2017}
Zheng, Y. and Chen, W. (2017).
\newblock Characteristics and controllability of vortices in ferromagnetics, ferroelectrics, and multiferroics.
\newblock {\em Rep. Prog. Phys.}, 80(8):086501.

\bibitem[Zubko et~al., 2012]{Zubko2012}
Zubko, P., Jecklin, N., Torres-Pardo, A., Aguado-Puente, P., Gloter, A., Lichtensteiger, C., Junquera, J., St{\'e}phan, O., and Triscone, J.-M. (2012).
\newblock Electrostatic coupling and local structural distortions at interfaces in ferroelectric/paraelectric superlattices.
\newblock {\em Nano Lett.}, 12(6):2846--2851.

\end{thebibliography}

\end{document}